\documentclass[11pt,twoside,openright]{book}
\pdfoutput=1
\usepackage{widebar}
\usepackage[colorlinks=true]{hyperref}
\usepackage{morewrites}
\usepackage[paperwidth=8.5in, paperheight=11in,bindingoffset=.75in]{geometry}

\usepackage[english]{babel}
\usepackage[exerciseaslist=true,copyexercisesinsolutions=true,exercisesfontsize=small]{exsol}
\usepackage{amsmath}
\usepackage{amssymb}
\usepackage{slashed}
\usepackage{array}

\usepackage{tikz}
\usetikzlibrary{arrows, intersections,calc,shapes}
\usetikzlibrary{positioning}
\usetikzlibrary{decorations.pathmorphing}
\usetikzlibrary{decorations.markings}

\tikzset{
particle/.style={thick,draw=black, postaction={decorate},
    decoration={markings,mark=at position .7 with {\arrow[black]{triangle 45}}}},
photon/.style={decorate, draw=black,
    decoration={coil,aspect=0}},
gluon/.style={decorate, draw=black,
    decoration={coil,aspect=0.3,segment length=5pt,amplitude=3pt}}
}

\usepackage[shortlabels]{enumitem} 
\usepackage{mathtools} 
\usepackage{float} 
\usepackage[stable]{footmisc} 

\makeatletter
\newsavebox\myboxA
\newsavebox\myboxB
\newlength\mylenA
\newcommand*\xoverline[2][0.75]{%
    \sbox{\myboxA}{$\m@th#2$}%
    \setbox\myboxB\null
    \ht\myboxB=\ht\myboxA%
    \dp\myboxB=\dp\myboxA%
    \wd\myboxB=#1\wd\myboxA
    \sbox\myboxB{$\m@th\overline{\copy\myboxB}$}
    \setlength\mylenA{\the\wd\myboxA}
    \addtolength\mylenA{-\the\wd\myboxB}%
    \ifdim\wd\myboxB<\wd\myboxA%
       \rlap{\hskip 0.5\mylenA\usebox\myboxB}{\usebox\myboxA}%
    \else
        \hskip -0.5\mylenA\rlap{\usebox\myboxA}{\hskip 0.5\mylenA\usebox\myboxB}%
    \fi}
\makeatother

\newcommand\Htilde{\widetilde{H}}
\newcommand{\fracup}[2]{\frac{\raisebox{0.25ex}{$\displaystyle #1$}}{#2}}

\title{TASI-2013 Lectures on Flavor Physics}
\author{Benjam\'\i{}n Grinstein}
\date{October 2014} 

\begin{document}
\frontmatter 

\let\cleardoublepage\clearpage

\maketitle
\tableofcontents

\chapter{Preface}
This ``book'' is based on lectures I gave at TASI during the summer of
2013. This document is not intended as a reference work. It is
certainly not encyclopedic. It is not even complete.  This is because these are lectures for  graduate students, and I aimed at pedagogy. 
So don't look here for a complete list of topics, nor for a complete
set of references. My hope is that a physics student who has taken
some courses on Quantum Field Theory and has been exposed to the Standard Model of electroweak interactions and has heard of pions and $B$ mesons will be able to learn a lot of flavor physics. 

The experienced reader may disagree with my choices. Heck, I may
disagree with my choice if I teach this again. In spite of years of
experience, I almost always find in retrospect that the approaches
opted for on teaching a course for the first time are far from
optimal. The course is just a crude approximation to one that is an
evolving project. If only I got to teach this a few more times, I would get
really good at it.

But my institution, and most institutions in the USA can't afford to spend Professors' time teaching various advanced and technical courses; the demand is largest for large undergraduate service courses, for pre-meds and engineering majors,  where the Distinguished Professor's  unique expertise is,  frankly, irrelevant and useless. Still, it is what pays the bills.  

That's where TASI comes in to fill a tremendous need (fill a hunger, would be even more appropriate) of the students of theoretical particle physics. I feel privileged and honored  that I have been given the opportunity to present these lectures on Flavor Physics and hope that the writeup of these lectures can be of use to many current and future students that may not have the good fortune of attending a TASI.

Being lectures, there are lots of exercises that go with these. The exercises are not collected at the back, not even at then end of each chapter or section. They are interspersed in the material. The problems tend to expand or check on one point and I think it's best for a student to solve the exercises in context.  I have many ideas for additional exercises, but only limited time. I hope to add some more in time. Some day I will publish the solutions.  Some are already typed into the TeX source and I hope to keep adding to it. You should be able to find the typeset solutions as an ancillary file in the arXiv submission. 

No one is perfect and I am certainly far from it. I would appreciate
alert readers to send me any typos, errors or any other needed
corrections they may find. Suggestions for any kind of improvement are
welcome. I will be indebted if you'd  send them to me at bgrinstein@ucsd.edu

\bigskip

{\raggedleft Benjam\'\i{}n Grinstein\\
 San Diego, June 2014\\
}
\mainmatter
\chapter{Flavor Theory}

\section{Introduction: What/Why/How?}

\paragraph{WHAT:} There are six different types of quarks: $u$ (``up''), $d$ (``down''),
$s$ (``strange''), $c$ (``charm''), $b$ (``bottom'') and $t$
(``top''). Flavor physics is the study of different types of quarks,
or ``flavors,'' their spectrum and the transmutations among
them. More generally different types of leptons, ``lepton flavors,''
can also be included in this topic, but in this lectures we
concentrate on quarks and the  hadrons that contain them. 

\paragraph{WHY:} Flavor physics is very rich. You should have a copy of the \href{http://pdg.lbl.gov}{PDG},
or at least a bookmark to \href{http://pdg.lbl.gov}{pdg.lbl.gov} on your computer. A quick inspection
of the PDG reveals that a great majority of content gives transition
rates among hadrons with different quark content, mostly decay
rates. This is all flavor physics. We aim at understanding this wealth
of information in terms of some simple basic principles. That we may
be able to do this is striking endorsement of the validity of our theoretical model of
nature, and gives stringent constraints on any new model of nature you
may invent. Indeed, many models you may have heard about, in fact many
of the most popular models, like gauge mediated SUSY breaking and
extended technicolor, were invented to address the strong constraints
imposed by flavor physics. Moreover, all observed CP violation (CPV)
in nature is tied to flavor changing interactions, so understanding of
this fundamental phenomenon is the domain of flavor physics. 

\paragraph{HOW:} The richness of  flavor physics comes at a price: while flavor
transitions occur intrinsically at the quark level, we only observe
transitions among hadrons. Since quarks are bound in hadrons by the
strong interactions we face the problem of confronting theory with
experiment in the context of mathematical models that are not
immediately amenable to simple analysis, like perturbation
theory. Moreover, the physics of flavor more often than not involves
several disparate time (or energy) scales, making even dimensional analysis
somewhere between difficult and worthless. Many tools have been
developed to address these issues, and these lectures will cover many
of them. Among these:
\begin{itemize}
\item Symmetries allow us to relate different processes and sometimes
  even to predict the absolute rate of a transition.
\item Effective Field Theory (EFT) allows to systematically
disentangle the effects of disparate scales. Fermi theory is an EFT
for electroweak interactions at low energies. Chiral Lagrangians
encapsulate the information of symmetry relations  of  transitions
among pseudo-Goldstone bosons. Heavy Quark Effective Theory (HQET)
disentangles the scales associated with the masses of heavy quarks
from the scale associated with hadron dynamics and makes explicit spin
and heavy-flavor symmetries. And so on.
\item Monte-Carlo simulations of strongly interacting quantum field theories on the lattice
can be used to compute some quantities of basic interest that cannot be computed using perturbation theory.  
\end{itemize}

\section{Flavor in the Standard Model}
Since the Standard Model of Strong and Electroweak interactions (SM) works so well, we will adopt it as our standard (no pun intended) paradigm. All alternative theories that are presently studied build on the SM; we refer to them collectively as Beyond the SM (BSM). Basing our discussion on the SM is very useful:
\begin{itemize}
\item It will allow us to introduce concretely the methods used to think about and quantitatively analyze Flavor physics. It should be straightforward to extend the  techniques introduced in the context of the SM to specific BSM models. 
\item Only to the extent that we can make precise calculations in the SM and confront them with comparably precise experimental results can we meaningfully study effects of other (BSM) models. 
\end{itemize}
So let's review the SM. At the very least, this allows us to agree on
notation. The SM is a gauge theory, with gauge group $SU(3)\times
SU(2) \times U(1)$. The $SU(3)$ factor models the strong interactions of ``colored'' quarks and gluons,
$SU(2)\times U(1)$ is the famous Glashow-Weinberg-Salam model of the
electroweak interactions. Sometimes we will refer to these as
$SU(3)_\text{c}$ and $SU(2)_\text{W}\times U(1)_\text{Y}$ to
distinguish them from other transformations with the same groups. The matter content of the model consists of
color triplet quarks: left handed spinor doublets $q^i_L$ with $U(1)$
``hypercharge'' $Y=1/6$ and  right handed spinor singlets $u^i_R$ and
$d^i_R$ with $Y=2/3$ and $Y=-1/3$. The color ($SU(3)$),
weak ($SU(2)$), and Lorentz-transformation indices are implicit. The
``$i$'' index runs over $i=1,2,3$ accounting for three copies, or ``generations.'' A more concise description is $q^i_L=(3,2)_{1/6}$, meaning that $q^i_L$ transforms as a $\mathbf{3}$ under $SU(3)$, a $\mathbf{2}$ under $SU(2)$ and has $Y=1/6$ (the $U(1)$ charge). Similarly, $u^i_R=(3,1)_{2/3}$ and $d^i_R=(3,1)_{-1/3}$. The leptons are color singlets: $\ell^i_L=(1,2)_{-1/2}$ and $e^i_R=(1,1)_{-1}$. 

We give names to the quarks in different generations:
\begin{equation}
q^i_L=\left(\begin{pmatrix}u_L\\d_L\end{pmatrix}, \begin{pmatrix}c_L\\s_L\end{pmatrix}, \begin{pmatrix}t_L\\b_L\end{pmatrix}\right), \qquad u^i_R=(u_R, c_R, t_R), \qquad d^i_R=(d_R,s_R,b_R).
\end{equation}
Note that we have used the same symbols, ``$u$'' and ``$d$,'' to denote the collection of quarks in a generation and the individual elements in the first generation. When the superscript $i$ is explicit this should give rise to no confusion. But soon we will want to drop the superscript to denote collectively the generations as vectors $q_L$, $u_R$ and $d_R$, and then we will have to rely on the context to figure out whether it is the collection or the individual first element that we are referring to. For this reason some authors use the capital letters $U_R$ and $D_R$ to denote the vectors in generation space. But I want to reserve $U$ for unitary transformations, and I think you should have no problem figuring out what we are talking about from context.

Similarly, for leptons we have
\begin{equation}
\ell^i_L=\left(\begin{pmatrix}\nu_{eL}\\e_L\end{pmatrix}, \begin{pmatrix}\nu_{\mu L}\\\mu_L\end{pmatrix}, \begin{pmatrix}\nu_{\tau L}\\\tau_L\end{pmatrix}\right), \qquad e^i_R=(e_R, \mu_R,\tau_R).
\end{equation}

The last ingredient of the SM is the Brout-Englert-Higgs (BEH) field, $H$, a collection of complex scalars transforming as $(1,2)_{1/2}$. The BEH field has an expectation value, which we take to be
\begin{equation}
\label{eq:higgsvev}
\langle H\rangle =\frac1{\sqrt2}\begin{pmatrix}0\\v\end{pmatrix}.
\end{equation}
The hermitian conjugate field $\Htilde=i\sigma^2 H^*$ transforms as
$(1,2)_{-1/2}$ and is useful in constructing Yukawa interactions
invariant under the electroweak group.
The covariant derivative is
\begin{equation}
D_\mu =\partial_\mu+ig_s T^a A^a_\mu+i g_2\frac{\sigma^j}2 W^j_\mu +i g_1YB_\mu.
\end{equation}
Here we have used already the Pauli $\sigma^i$ matrices as generators
of $SU(2)$, since the only fields which are non-singlets under this
group are all doublets (and, of course, one should replace zero for
$\sigma^j$ above in the case of singlets). It should also be clear
that we are using the generalized Einstein convention: the repeated
index $a$ is summed over $a=1,\ldots, N_c^2-1$, where $N_c=3$ is the
number of colors, and $j$ is summed over $j=1,2,3$. The generators
$T^a$ of
$SU(3)$ are normalized so that in the fundamental representation
$\text{Tr}(T^aT^b)=\tfrac12 \delta^{ab}$. With this we see that
$\langle H\rangle $ is invariant under $Q=\tfrac12\sigma^3+Y$, which
we identify as the generator of an unbroken $U(1)$ gauge group,
electromagnetic charge. The field strength tensors for $A_\mu$,
$W_\mu$ and $B_\mu$ are denoted as $G_{\mu\nu}$, $W_{\mu\nu}$,  and
$B_{\mu\nu}$, respectively, and that of electromagnetism by
$F_{\mu\nu}$. 

The Lagrangian of the SM is the most general combination of monomials constructed out of these fields constrained by (i) Lorentz invariance, (ii) Gauge invariance, and (iii) renormalizability. This last one implies that this monomials, or ``operators,'' are of dimension no larger than four. Field redefinitions by linear transformations that preserve Lorentz and gauge invariance bring the kinetic terms to canonical form. The remaining terms are potential energy terms, either Yukawa interactions or BEH-field self-couplings. The former are central to our story: 
\begin{equation} 
  -\mathcal{L}_{\text{Yuk}}=\sum_{i,j}\left[\lambda_{U}{}^i_j\Htilde  \widebar q_{Li} u_R^j+\lambda_D{}^i_{j} H \widebar q_{Li} d_R^j +\lambda_E{}^i_{j} H \widebar \ell_{Li} e_R^j+\text{h.c.}  \right]
\end{equation}
We will mostly avoid explicit index notation from here on. The reason for upper and lower indices will become clear below. The above equation can be written more compactly  as  
\begin{equation} 
  \label{eq:yuk} -\mathcal{L}_{\text{Yuk}}=\Htilde \widebar q_{L}\lambda_{U} u_R+ H \widebar q_{L}\lambda_D d_R +H \widebar \ell_{L}\lambda_E e_R+\text{h.c.} 
\end{equation}

\paragraph{Flavor ``symmetry.''} In the absence of Yukawa interactions
({\it i.e.}, setting $\lambda_{U}=\lambda_{D}=\lambda_{E}=0$ above)
the SM Lagrangian has a large global symmetry. This is because the
Lagrangian is just the sum of covariantized kinetic energy therms,
$\sum_n\widebar\psi_n i\slashed{D}\psi_n$, with the sum running over all
the fields in irreducible representations of the the SM gauge group, and
one can make linear unitary transformations among the fields in a
given SM-representation  without altering the
Lagrangian:
\[
q_L\to U_q\; q_L\;, \quad u_R\to U_u\; u_R \;,\quad\ldots \quad e_R\to U_e\; e_R~,
\]
where $U^\dagger_qU_q^{\phantom{\dagger}}=\cdots=U^\dagger_eU_e^{\phantom{\dagger}}=1$. 
 Since there are
$N_f=3$ copies of each SM-representation this means these are $N_f\times
N_f $ matrices, so that for each SM-representation the redefinition freedom is by elements of the group  $U(N_f)$.  Since there are five distinct SM-representations (3
for quarks and 2 for leptons), the full symmetry group is
$U(N_f)^5=U(3)^5$.\footnote{Had we kept indices explicitly we would have written \(
q_L^i\to U_q{}^i_j\; q_L^j\;,  u_R^i\to U_u{}^i_j\; u_R^j \;,\ldots ,  e_R^i\to U_e{}^i_j\; e_R^j
\). The fields transform in the fundamental representation of $SU(N_f)$. We use upper indices for this. Objects, like the hermitian conjugate of the fields, that transform in the anti-fundamental representation, carry lower indices. The transformation matrices have one upper and one  lower indices, of course. } In the quantum theory each of the $U(1)$ factors
(corresponding to a redefinition of the $N_f$ fields in a given
SM-representation by multiplication by a common phase) is anomalous, so
the full symmetry group is smaller. One can make non-anomalous 
combinations of these $U(1)$'s, most famously $B-L$, a symmetry that
rotates quarks and leptons simultaneously, quarks with $-1/3$ the
phase of leptons. For our purposes it is the non-abelian factors that
are most relevant, so we will be happy to restrict our attention to
the symmetry group $SU(N_f)^5$.

The flavor symmetry is broken explicitly by the Yukawa
interactions. We can keep track of the pattern of symmetry breaking by
treating the Yukawa couplings as ``spurions,'' that is, as constant
fields. For example, under $SU(N_f)_q\times SU(N_f)_u$ the first term in 
\eqref{eq:yuk} is invariant if we declare that $\lambda_U$ transforms as  a bi-fundamental, 
$\lambda_U\to U_q\lambda_U U_u^\dagger$; check:   
\[
\widebar q_{L}\lambda_{U} u_R \to  \widebar q_{L}U_q^\dagger (U_q^{\phantom{\dagger}}\lambda_{U}U_u^\dagger)U_u u_R= \widebar q_{L}\lambda_{U} u_R.
\]
So this, together with  $\lambda_D\to U_q^{\phantom{\dagger}}\lambda_D U_d^\dagger$  and $\lambda_E\to U_\ell^{\phantom{\dagger}} \lambda_E U_e^\dagger$  renders the whole Lagrangian invariant. 

Why do we care? As we will see, absent tuning or large parametric suppression, {\it new interactions that break this  ``symmetry'' tend to produce rates of flavor transformations that are inconsistent with observation.} This is not an absolute truth, rather a statement about the generic case. 

In these lectures we will be mostly concerned with hadronic flavor, so  from here on
we focus on the $G_F\equiv SU(3)^3$ that acts on quarks. 

\section{The CKM matrix and the KM model of CP-violation}
Replacing  the BEH field by its VEV, Eq.~\eqref{eq:higgsvev},  in the Yukawa terms  in \eqref{eq:yuk} 
we obtain  mass terms for  quarks and leptons:
\begin{equation}
 \label{eq:qmass} 
-\mathcal{L}_{\text{m}}=\frac{v}{\sqrt2} \widebar u_{L}\lambda_{U} u_R+\frac{v}{\sqrt2}    \widebar d_{L}\lambda_D d_R +\frac{v}{\sqrt2}  \widebar e_{L}\lambda_E e_R+\text{h.c.} 
\end{equation}
For simpler computation and interpretation of the model it is best to make further field redefinitions that render the  mass terms diagonal while maintaining the canonical form of the kinetic terms (diagonal, with unit normalization). The field redefinition must be linear (to maintain explicit renormalizability of the model) and commute with the Lorentz group and the part of the gauge group that is unbroken by the electroweak VEV (that is, the $U(1)\times SU(3)$ of electromagnetism and color). This means the linear transformation can act to mix only quarks with the same handedness and electric charge (and the same goes  for leptons):
\begin{equation}
\label{eq:quarkV}
u_R \to V_{u_R} u_R,\quad u_L \to V_{u_L} u_L,\quad  d_R \to V_{d_R} d_R,\quad  d_L \to V_{d_L} d_L.
\end{equation}
Finally, the linear transformation will preserve the form of the kinetic terms, say, $\widebar u_L i \slashed{\partial} u_L
\to 
(\widebar u_L V_{u_L}^\dagger)  i \slashed{\partial} (V_{u_L} u_L)=\widebar u_L (V_{u_L}^\dagger V_{u_L}) i \slashed{\partial} u_L$, if $V_{u_L}^\dagger  V_{u_L} =1$, that is, if it  is unitary. 

Now, choose to make a redefinition by matrices that diagonalize the mass terms, 
\begin{equation}
\label{eq:massDtransf}
V_{u_L}^\dagger \lambda_{U} V_{u_R}^{\phantom{\dagger}}= \lambda_{U}^\prime, \quad V_{d_L}^\dagger \lambda_{D} V_{d_R}^{\phantom{\dagger}}= \lambda_{D}^\prime\;.
\end{equation}
Here the matrices with a prime, $\lambda_{U}^\prime$ and $\lambda_{D}^\prime$, are diagonal, real and positive. 

\begin{exercises}
\begin{exercise}
Show that this can always be done. That is, that an arbitrary matrix $M$ can be transformed  into a real, positive diagonal matrix $M'=P^\dagger MQ$ by a pair of unitary matrices, $P$ and $Q$. 
\end{exercise}
\begin{solution}
I'll give you a physicist's proof. If you want to be a mathematician, and use Jordan Normal forms, be my guest. Consider the matrices $M^\dagger M $ and $ M M^\dagger$. They are both hermitian so they can each be diagonalized by a unitary transformation. Moreover, they both obviously have real non-negative  eigenvalues. And they  have the same eigenvalues: using the properties of the determinant you can see that the characteristic polynomial is the same, $\det(M^\dagger M -x)=\det(MM^\dagger-x)$. So we have matrices $P$ and $Q$ such that $P^\dagger ( MM^\dagger) P=Q^\dagger ( M^\dagger M) Q=D=$ real, non-negative, diagonal. We can rewrite $D= P^\dagger ( MM^\dagger) P= ( P^\dagger  M Q)( P^\dagger  M Q)^\dagger= XX^\dagger$, where $X=  P^\dagger  M Q$. Similarly, we also have $D=X^\dagger X$, and multiplying this by X on the left we combine the two results into $XD=DX$. Let's assume all the entries in $D$ are all different and non-vanishing  (I will leave out the special cases, you can feel in the details). Then $DX-XD=0$ means, in components $(D_{ii}-D_{jj})X_{ij}=0$ which means that $X_{ij}=0$ for $j\ne i$. So $X$ is diagonal, with $|X_{ii}| = \sqrt{D_{ii}}$. We can always take $ P^\dagger  M Q=\sqrt{D}=M'$, by further transformation by a diagonal unitary matrix on the  left or right.
\end{solution}
\end{exercises}

Then from 
\begin{equation}
-\mathcal{L}_{\text{m}}=\frac{v}{\sqrt2} \Big( \widebar u_{L}\lambda'_{U} u_R+ \widebar d_{L}\lambda'_D d_R +  \widebar e_{L}\lambda_E e_R+\text{h.c.}\Big)
=\frac{v}{\sqrt2}\Big(\widebar u\lambda'_{U} u+   \widebar d\lambda'_D d +  \widebar e\lambda_E e\Big)
\end{equation}
we read off the diagonal mass matrices, $m_U= v \lambda'_{U}/\sqrt2$, $m_D=v \lambda'_D /\sqrt2$ and $m_E=v \lambda_E/\sqrt2 $. 

Since the field redefinitions in \eqref{eq:quarkV} do not commute with the electroweak  group, it is not guaranteed that the Lagrangian is independent of the  matrices $V_{u_L}, \ldots, V_{d_R}$. We did choose the transformations to leave the kinetic terms in  canonical form. We now check the effect of \eqref{eq:quarkV}  on the gauge interactions. Consider first the singlet fields $u_R$. Under the field redefinition we have
\[
\widebar u_R \,( g_s\slashed{A}^a T^a+ \tfrac23 g_1 \slashed{B}) u_R \to 
\widebar u_R V^\dagger_{u_R} \, ( g_s\slashed{A}^a T^a+\tfrac23 g_1 \slashed{B}) V_{u_R} u_R = \widebar u_R \, ( g_s\slashed{A}^a T^a+\tfrac23 g_1 \slashed{B}) u_R~.
\]
It remains unchanged. Clearly the same happens with the $d_R$ fields. The story gets more interesting with the left handed fields, since they form doublets. First let's look at the terms that are diagonal in the doublet space:
\begin{multline*}
\widebar q_L ( g_s\slashed{A}^a
T^a+\tfrac12g_2\slashed{W}^3\sigma^3+\tfrac16 g_1\slashed{B})q_L \\
= \widebar u_L ( g_s\slashed{A}^a T^a+\tfrac12g_2\slashed{W}^3+\tfrac16 g_1\slashed{B}) u_L+
\widebar d_L ( g_s\slashed{A}^a T^a-\tfrac12g_2\slashed{W}^3+\tfrac16 g_1\slashed{B})d_L
\end{multline*}
This is clearly invariant under \eqref{eq:quarkV}. Finally we have the off-diagonal terms. For these let us introduce
\[\sigma^\pm = \frac{\sigma^1\pm i\sigma^2}{\sqrt2}, \quad
\text{and}\quad W^\pm= \frac{W^1\mp i W^2}{\sqrt2}
\]
so that $\sigma^1W^1+\sigma^2 W^2 = \sigma^+W^++\sigma^- W^-$ and $(\sigma^+)_{12}=\sqrt2$,   $(\sigma^-)_{21}=\sqrt2$, and all other elements vanish. It is now easy to expand:
\begin{multline}
\label{eq:chargedCurr}
\widebar q_L\tfrac12 g_2(\sigma^1W^1+\sigma^2 W^2 ) q_L = \tfrac{1}{\sqrt2}g_2\widebar u_L\slashed{W}^+ d_L +\tfrac{1}{\sqrt2}g_2
 \widebar d_L\slashed{W}^- u_L\\ \to \tfrac{1}{\sqrt2}g_2\widebar
 u_L(V_{u_L}^\dagger V_{d_L}^{\phantom{\dagger}})\slashed{W}^+ d_L + \tfrac{1}{\sqrt2}g_2 \widebar d_L(V_{d_L}^\dagger V_{u_L}^{\phantom{\dagger}})\slashed{W}^- u_L
\end{multline}
A relic of our field redefinitions has remained in the form of the unitary matrix $V=V_{u_L}^\dagger V_{d_L}$. We call this the Cabibbo-Kobayashi-Maskawa (CKM) matrix. 

A general unitary $3\times 3$ matrix has $3^2$ complex entries, constrained by $3$ complex plus $3$ real conditions. So the CKM matrix is in general parametrized by 9 real entries.  But not all are of physical consequence. 
We can perform further transformations of the form of
\eqref{eq:quarkV} that leave the mass matrices in
\eqref{eq:massDtransf}  diagonal and
non-negative if the unitary matrices are diagonal with $V_{u_L}=V_{u_R}=\text{diag}(e^{i\alpha_1},e^{i\alpha_2},e^{i\alpha_3}) $ and  $V_{d_L}=V_{d_R}=\text{diag}(e^{i\beta_1},e^{i\beta_2},e^{i\beta_3})$.  $V$ is redefined by $V_{ij}\to e^{i(\beta_j-\alpha_i)}V_{ij}$. These five independent phase differences reduce the number of independent parameters in $V$ to $9-5=4$. It can be shown that this can in general be taken to be 3 rotation angles and one complex phase. It will be useful to label the matrix elements by the quarks they connect:
\[
V= \begin{pmatrix}
V_{ud} & V_{us} & V_{ub}\\
V_{cd} & V_{cs} & V_{cb}\\
V_{td} & V_{ts} & V_{tb}
\end{pmatrix}~.
\]

\smallskip

Observations:
\begin{enumerate}
\item That there is one irremovable phase in $V$ impies that CP is not a symmetry of the SM Lagrangian. It is broken by the terms $\widebar u_L V \slashed {W}^+ d_L + \widebar d_L V^\dagger \slashed{W}^- u_L$. To see this, recall that under CP 
$\widebar u_L \gamma^\mu  d_L \rightarrow -\widebar d_L \gamma_\mu  u_L$ and $W^{+\mu}\to -W^-_\mu$. Hence CP invariance requires $V^\dagger =V$. 
\begin{exercises}
\begin{exercise}
In QED, charge conjugation is $\widebar e\gamma^\mu e\to -\widebar e \gamma^\mu e$ and $A^\mu \to - A^\mu$.  So $\widebar e \slashed {A} e$ is invariant under $C$.\\
So what about QCD? Under charge conjugation  $\widebar q T^a \gamma^\mu q \to \widebar q (-T^a)^T \gamma^\mu q$, but $ (-T^a)^T=(-T^a)^*$ does not equal $-T^a$ (nor $T^a$). So what does charge conjugation mean in QCD? How does the gluon field, $A^a_\mu$, transform?
\end{exercise}
\begin{solution}
I have never seen this discussed in a textbook, or elsewhere. Maybe one of the readers will write a nice article for AJP (don't forget to include me!). If you think of the ``transformation arrow'' more properly as the action by a unitary operator on the Hilbert space, $C$, 
so that $\widebar e\gamma^\mu e\to -\widebar e \gamma^\mu e$ really means $C(\widebar e\gamma^\mu e)C^{-1}= -\widebar e \gamma^\mu e$, then it is clear that $T^a$ is not changed since it is a $c$-number that commutes with $C$. What we need is
$A_\mu^a T^a\to A_\mu^a (-T^a)^T$. This is accomplished by a transformation $ A_\mu^a\to R^{ab}A_\mu^b$ with a real matrix $R$ that must take $T^a$ into minus its transpose: $R^{ba} T^b = -T^{aT}$. Since the matrices $T^a$ are in the fundamental representation of $SU(3)$ we have $R^{ca}=2\text{Tr}[T^c(R^{ba} T^b)] =-2\text{Tr}( T^c T^{aT})$. $R$ is indeed real:
$(R^{ca})^* =-2\text{Tr}( T^{c*} T^{a\dagger})$, then using  $T^{a\dagger}=T^a$, $\text{Tr}(A^T)=\text{Tr}(A)$ and cyclicity of trace, it follows that $R$ is a real symmetric matrix. Notice that $R^2=1$ for consistency (the negative transpose of the negative transpose is the identity). You can check this using the identity $2T^a_{ij}T^a_{mn}=\delta_{in}\delta_{mj}-\frac13\delta_{ij}\delta_{mn}$.

In physical terms this means that under charge conjugation the, say, blue-antigreen gluon is transformed into minus the green-antiblue gluon, and so on. 

I have seen in places  an explanation for charge conjugation in QCD along these lines: first take the quark field $q$ and rewrite in terms of a left- and a right-handed fields, $q_L$ and $q_R$. Then replace $q_R$ by its charge-conjugate, which is also a left-handed field, $q^c_L$. Now $q_L$ is a triplet under color while $q_L^c$ is an antitriplet under color. So charge conjugation is simply $q_L\leftrightarrow q^c_L$. This is incomplete (and therefore wrong). If you were to ignore the transformation of the gluon field the resulting Lagrangian would not be gauge invariant since now the covariant  derivative acting on $q_L$ has a generator for an anti-triplet, $-T^{aT}$, while the covariant derivative acting on $q_L^c$ has generator $T^a$ appropriate for a triplet. It is only after you transform the gluon field that everything works as it should!
\end{solution}
\begin{exercise}
If two entries  in $m_U$ (or in $m_D$) are equal show that $V$ can be brought into a real matrix and hence is an  orthogonal transformation (an element of $O(3)$). 
\end{exercise}
\begin{solution}
Without loss of generality we may assume the first two entries in
$m_U$ are equal. This means that the remnant freedom to redefine
quark fields without changing neither the kinetic nor the mass terms
is not just by individual phases on all flavors but also by a
$2\times2$ unitary matrix acting on the degenerate quarks. Let
$u_{L,R}\to Uu_{L,R} $, then $U$ is of the form
\[\left(\begin{array}{c|c}A & 0\\ \hline \\[-2.5ex] 0 & e^{i\alpha_3}\end{array}
\right)\]
where $A$ is a $2\times2$ unitary matrix and ``0'' stands for a
2-component zero vector. 
Let also $W$ be the diagonal matrix with entries $e^{i\beta_i}$,
$i=1,2,3$, and redefine $d_{L,R}\to W d_{L,R}$. This has the effect of
redefining $V\to U^\dagger VW$. To see what is going on let's write
$V$ in terms of a $2\times2$ submatrix, $X$, two 2-component column
vectors, $\psi$ and $\eta$, and a complex number, $z$:
\[\left(\begin{array}{c|c}X & \psi\\ \hline \\[-2.5ex]\eta^T & z\end{array}.
\right)\]
Then $V$ is transformed into
\begin{equation}
\label{eqsol:vtrans}
V=\left(\begin{array}{c|c}A^\dagger X\begin{pmatrix}e^{i\beta_1}&0\\0& e^{i\beta_2}\end{pmatrix}  & e^{i\beta_3}A^\dagger\psi\\[1.8ex]
    \hline \\[-1.9ex] e^{-i\alpha_3}\eta^T\begin{pmatrix}e^{i\beta_1}&0\\0& e^{i\beta_2}\end{pmatrix} & e^{i(\beta_3-\alpha_3)}z\end{array}
\right).\end{equation}
Now we can choose $A^\dagger$ so that $e^{i\beta_3}A^\dagger\psi$ has
vanishing lower component and real upper component. This still leaves
freedom in $A$ to make a rotation by a phase of the (vanishing) lower
component. So we may take
\[
\psi= \begin{pmatrix}|\psi|\\0\end{pmatrix}\qquad\text{and}\qquad
e^{i\beta_3}A^\dagger=
\begin{pmatrix}1 &0\\0&e^{i(\gamma+\beta_3)}\end{pmatrix}.
\]
At this point it is worth making the trivial observation that for
fixed $\beta_3$ one can make the third row of the new $V$ matrix in
\eqref{eqsol:vtrans} real by choosing $\beta_1,\beta_2$
and~$\alpha_3$. We are left with the $2\times2$ block,
\[
A^\dagger X\begin{pmatrix}e^{i\beta_1}&0\\0& e^{i\beta_2}\end{pmatrix} =\begin{pmatrix}e^{-i\beta_3} &0\\0&e^{i\gamma}\end{pmatrix}X\begin{pmatrix}e^{i\beta_1}&0\\0& e^{i\beta_2}\end{pmatrix} 
\]
Now choose $\beta_3$ and $\gamma$ to make real the first column. This
means that the only entries of $V$ with a phase are the top two
entries of the second column, $V_{21}$ and $V_{22}$. But unitarity of
$V$ requires $V_{2i}V_{3i}^*=0$. Since $V_{32}=0$ this can only be
satisfied if $V_{21}$ is real. Then $V_{2i}V_{1i}^*=0$ can only be
satisfied if $V_{22}$ is real. Hence all elements in $V$ are real.
\end{solution}
\end{exercises}
\item Precise knowledge of the elements of $V$ is necessary to  constrain new physics (or to test the validity of the SM/CKM theory). We will describe below how well we know them and how. But for now it is useful to have a sketch that gives a rough order of magnitude of the magnitude of the elements in $V$:
\begin{equation}
\label{eq:Veps}
V\sim \begin{pmatrix}\epsilon^0 & \epsilon^1& \epsilon^3\\\epsilon^1 & \epsilon^0& \epsilon^2\\\epsilon^3 & \epsilon^2& \epsilon^0\end{pmatrix}, \qquad\text{with $\epsilon\sim 10^{-1}$.}
\end{equation}
\item Since $VV^\dagger=V^\dagger V=1$ the rows as well as the columns  of $V$ are orthonormal vectors. In particular,  $\sum_k V^{\phantom{*}}_{ik}V^*_{jk}=0$ for $j\ne i$. Three complex numbers that  sum to zero are represented on the complex plane as a triangle. As the following table shows, the resulting triangles are very different in shape. Two of them are very squashed, with one side much smaller than the other two, while the third one has all sides of comparable size. As we shall see, this will play a role in understanding when CP asymmetries can be sizable. 

\medskip

\begin{tabular}[c]{l|l|c|m{4cm}}
$ij$ & $\sum V_{ik}^{\phantom{*}}V_{jk}^*=0$ & $\sim \epsilon^n$& \parbox[t]{5cm}{shape\\ (normalized to unit base)}\\ \hline\hline
12 & $V_{ud}^{\phantom{*}}V_{cd}^*+V_{us}^{\phantom{*}}V_{cs}^*+V_{ub}^{\phantom{*}}V_{cb}^*=0$ & $\epsilon + \epsilon +\epsilon^5=0$ & \parbox[c][1cm]{5cm}{\begin{tikzpicture}
\draw (0,0) -- (4,0) -- node[right]{$\epsilon^4$} (4,0.2) -- (0,0);
\end{tikzpicture}} \\ \hline
23 & $V_{cd}^{\phantom{*}}V_{td}^*+V_{cs}^{\phantom{*}}V_{ts}^*+V_{cb}^{\phantom{*}}V_{tb}^*=0$ & $\epsilon^4 + \epsilon^2 +\epsilon^2=0$ & \parbox[c][1.2cm]{5cm}{\begin{tikzpicture}
\draw (0,0) -- (4,0) -- node[right]{$\epsilon^2$} (4,0.5) -- (0,0);
\end{tikzpicture}} \\ \hline
13 & $V_{ud}^{\phantom{*}}V_{td}^*+V_{us}^{\phantom{*}}V_{ts}^*+V_{ub}^{\phantom{*}}V_{tb}^*=0$ & $\epsilon^3 + \epsilon^3 +\epsilon^3=0$ & \parbox[c][2.5cm]{5cm}{\begin{tikzpicture}
\draw (0,0) -- (4,0) -- node[right]{$1$} (3,2) -- (0,0);
\end{tikzpicture}} \\\hline
\end{tabular}

\bigskip

These are called ``unitarity triangles.'' The most commonly discussed is in the 1-3 {\it columns}, 
\[
V_{ud}^{\phantom{*}}V_{ub}^*+V_{cd}^{\phantom{*}}V_{cb}^*+V_{td}^{\phantom{*}}V_{tb}^*=0\quad\Rightarrow\quad \text{ \parbox[c][2.5cm]{5cm}{\begin{tikzpicture}[scale=0.5]
\draw (0,0) --node[below]{1}  (4,0) -- node[right]{$\sim\!1$} (1.6,2) -- node[left]{$\sim\!1$}(0,0);
\end{tikzpicture}} }
\]
Dividing by the middle term we can be more explicit as to what we mean by the unit base unitarity triangle:
\[
\frac{V_{ud}^{\phantom{*}}V_{ub}^*}{V_{cd}^{\phantom{*}}V_{cb}^*}+1+\frac{V_{td}^{\phantom{*}}V_{tb}^*}{V_{cd}^{\phantom{*}}V_{cb}^*}=0
\]
We drew this on the complex plane and introduced some additional notation:  the complex plane is $z=\widebar\rho + i \widebar \eta$
and the internal angles of the triangle are\footnote{This convention is popular in the US, while in Japan  a different convention is more common:   $\phi_1=\beta$, $\phi_2=\alpha$ and $\phi_3=\gamma$.}  $\alpha$, $\beta$ and $\gamma$; see Fig.~\ref{fig:UTsketch}.

\begin{figure}
\begin{center}
\begin{tikzpicture}[scale=1.7,>=triangle 90]
\coordinate (A) at (0,0);
\coordinate (B) at  (4,0); 
\coordinate (C) at  (40:3cm) ;

\draw[->,>=stealth,thick] ($(A)+(-0.7,0)$) -- ($(B)+(1,0)$) node[below right]{$\widebar\rho$};
\draw[->,>=stealth,thick] ($(A)+(0,-0.7)$) -- +(0,3.3) node[left]{$\widebar\eta$};

\draw[very thick]   (A) -- (B) -- node[above right]{$\displaystyle\left|\frac{V_{td}^{\phantom{*}}V_{tb}^*}{V_{cd}^{\phantom{*}}V_{cb}^*}\right|$} (C)  node[below=0.9cm]{ $\alpha$} -- node[above left]{$\displaystyle\left|\frac{V_{ud}^{\phantom{*}}V_{ub}^*}{V_{cd}^{\phantom{*}}V_{cb}^*}\right|$}(0,0);

\draw[black]  ($(B)+(155:0.5cm)$)  node[left]{$\beta$};

\draw[->] ($(A)+(0.5,0)$) arc (0:40:0.5cm)  ;
\draw  ($(A)+(20:0.5cm)$)  node[right]{$\gamma$};

\draw[->] ($(C)+(220:0.5)$) arc (220:310:0.5);
\draw[->] ($(B)+(130:0.5)$) arc (130:180:0.5);

\end{tikzpicture}
\end{center}
\caption{\label{fig:UTsketch} Unitarity triangle in the $\xoverline\rho$-$\xoverline\eta$ plane. The base is of unit length. The sense of the angles is indicated by arrows.}
\end{figure}
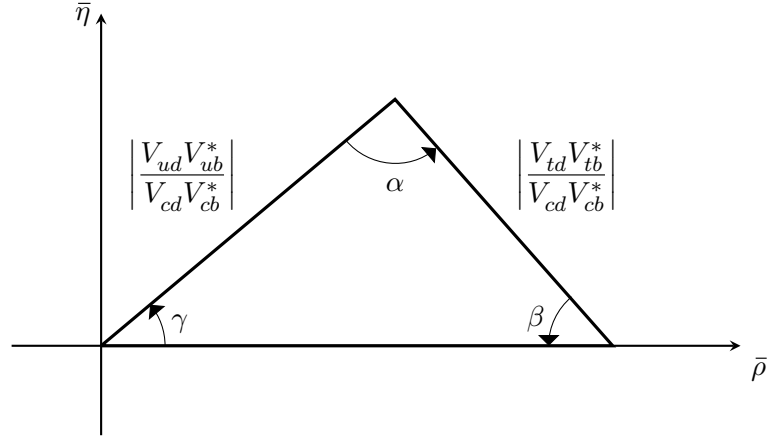

The angles of the unitarity triangle, of course, are completely determined by the CKM matrix, as you will now explicitly show:

\begin{figure}
\begin{center}
\includegraphics[width=4in]{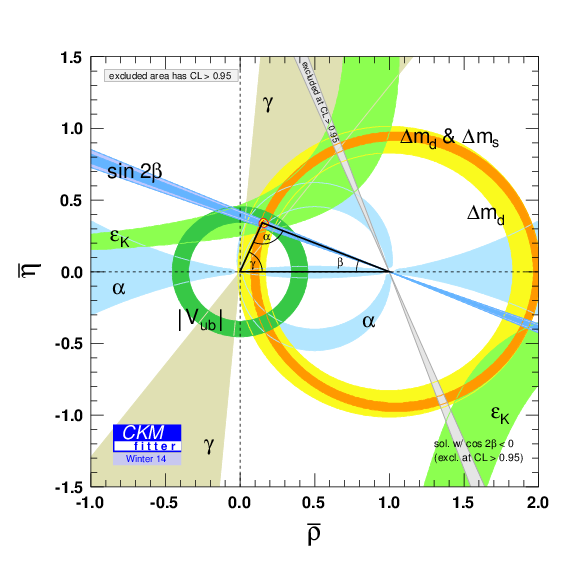}\hfill
\includegraphics[width=4in]{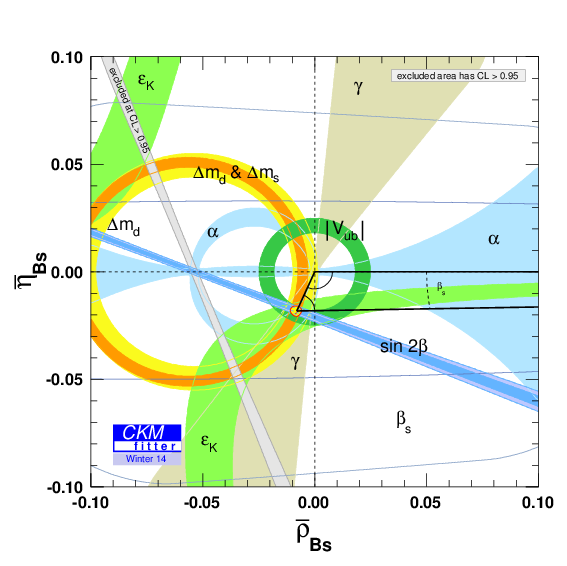}
\end{center}
\caption{\label{fig:CKMfitterTRIANGLES}  Experimentally determined unitarity triangles~\cite{ckmfitter}. Upper pane:  ``fat''  1-3  columns triangle.  Lower pane: ``skinny''  2-3 columns triangle.}
\end{figure}

\begin{exercises}
\begin{exercise}
\label{eqsol:alphabetagamma}
Show that 
\begin{enumerate}[(i)]
\item $\displaystyle \beta= \text{arg}\left(- \fracup{V_{cd}^{\phantom{*}}V_{cb}^*}{V_{td}^{\phantom{*}}V_{tb}^*}\right)$, 
 $\displaystyle \alpha= \text{arg}\left(- \fracup{V_{td}^{\phantom{*}}V_{tb}^*}{V_{ud}^{\phantom{*}}V_{ub}^*}\right)$ and 
 $\displaystyle \gamma= \text{arg}\left(- \fracup{V_{ud}^{\phantom{*}}V_{ub}^*}{V_{cd}^{\phantom{*}}V_{cb}^*}\right)$.
\item These are invariant under phase redefinitions of quark fields (that is, under the remaining arbitrariness). Hence these are candidates for observable quantities.  
\item The area of the triangle is $-\frac12\,\text{Im}\frac{\raisebox{0.45ex}{$\scriptstyle V_{ud}^{\phantom{*}}V_{ub}^*$}}{V_{cd}^{\phantom{*}}V_{cb}^*}
=-\frac12\,\frac1{|V_{cd}^{\phantom{*}}V_{cb}^*|^2}\text{Im}\left(V_{ud}^{\phantom{*}}V_{cd}^*V_{cb}^{\phantom{*}}V_{ub}^*\right)$. 
\item The product $J=
\text{Im}\left(V_{ud}^{\phantom{*}}V_{cd}^*V_{cb}^{\phantom{*}}V_{ub}^*\right)$ (a ``Jarlskog invariant'') is invariant under re-phasing of quark fields. 
\end{enumerate}
Note that
$\text{Im}\left(V_{ij}^{\phantom{*}}V_{kl}^{\phantom{*}}V_{il}^*V_{kj}^*\right)=J(\delta_{ij}\delta_{kl}-\delta_{il}\delta_{kj})$
is the common area of all the un-normalized triangles. The area of a
normalized triangle is $J$ divided by the square of the magnitude of the side that is normalized to unity. 
\end{exercise}
\begin{solution}
\begin{enumerate}[(i)]
\item Take the equation that defines the triangle
\[
\frac{V_{ud}^{\phantom{*}}V_{ub}^*}{V_{cd}^{\phantom{*}}V_{cb}^*}+1+\frac{V_{td}^{\phantom{*}}V_{tb}^*}{V_{cd}^{\phantom{*}}V_{cb}^*}=0
\]
and depict it as a triangle in the complex plane:
\begin{center}
\begin{tikzpicture}[scale=1.7,>=triangle 90]
\coordinate (A) at (0,0);
\coordinate (B) at  (4,0); 
\coordinate (C) at  (40:3cm) ;

\draw[->,>=stealth,thick] ($(A)+(-0.7,0)$) -- ($(B)+(1,0)$) node[below right]{$\widebar\rho$};
\draw[->,>=stealth,thick] ($(A)+(0,-0.7)$) -- +(0,3.3) node[left]{$\widebar\eta$};

\draw[->,very thick]   (A) -- (B);
\draw[->,very thick] (B) -- node[above right]{$\displaystyle\frac{V_{td}^{\phantom{*}}V_{tb}^*}{V_{cd}^{\phantom{*}}V_{cb}^*}$} (C);
\draw[->,very thick] (C)  node[below=0.9cm]{ $\alpha$} -- node[above left]{$\displaystyle\frac{V_{ud}^{\phantom{*}}V_{ub}^*}{V_{cd}^{\phantom{*}}V_{cb}^*}$}(0,0);

\draw[black]  ($(B)+(155:0.5cm)$)  node[left]{$\beta$};
\draw[->] ($(A)+(0.5,0)$) arc (0:40:0.5cm)  ;
\draw  ($(A)+(20:0.5cm)$)  node[right]{$\gamma$};
\draw[->] ($(C)+(220:0.5)$) arc (220:310:0.5);
\draw[->] ($(B)+(130:0.5)$) arc (130:180:0.5);
\end{tikzpicture}
\end{center}
Note that the vector from the origin to $(\bar\rho,\bar\eta)$ is the opposite of $\frac{V_{ud}^{\phantom{*}}V_{ub}^*}{V_{cd}^{\phantom{*}}V_{cb}^*}$, so the angle $\gamma$ is the argument of minus this, $ \gamma= \text{arg}\left(- \frac{V_{ud}^{\phantom{*}}V_{ub}^*}{V_{cd}^{\phantom{*}}V_{cb}^*}\right)$. Next, the angle that $\frac{V_{td}^{\phantom{*}}V_{tb}^*}{V_{cd}^{\phantom{*}}V_{cb}^*}$ makes with the $\bar\rho$ axis is $\pi-\beta= \text{arg}\left( \frac{V_{td}^{\phantom{*}}V_{tb}^*}{V_{cd}^{\phantom{*}}V_{cb}^*}\right)$, from which $\beta= \text{arg}\left(- \frac{V_{cd}^{\phantom{*}}V_{cb}^*}{V_{td}^{\phantom{*}}V_{tb}^*}\right)$ follows. $\alpha$ is most easily obtained from $\alpha+\beta+\gamma=\pi$ using the two previous results and the fact that $\text{arg}(z_1)+\text{arg}(z_2)=\text{arg}(z_1z_2)$.
\item In the numerator or denominator of these expressions, the re-phasing of the charge-$+\frac23$ quarks cancel; for example, $V_{td}^{\phantom{*}}V_{tb}^*\to
(e^{i\phi}V_{td}^{\phantom{*}})(e^{i\phi}V_{tb}^{\phantom{*}})^*=V_{td}^{\phantom{*}}V_{tb}^*$. The re-phasing of the charge-$-\frac13$ quarks cancel between numerator and denominator; for example, for the $d$ quark  $\frac{V_{ud}^{\phantom{*}}V_{ub}^*}{V_{cd}^{\phantom{*}}V_{cb}^*}\to \frac{e^{i\phi}V_{ud}^{\phantom{*}}V_{ub}^*}{e^{i\phi}V_{cd}^{\phantom{*}}V_{cb}^*}=\frac{V_{ud}^{\phantom{*}}V_{ub}^*}{V_{cd}^{\phantom{*}}V_{cb}^*}$.
\item From question (i) we see that $\bar\eta=-\text{Im}\frac{V_{ud}^{\phantom{*}}V_{ub}^*}{V_{cd}^{\phantom{*}}V_{cb}^*}$, and this is the height of the triangle of unit base. The area is $1/2$ base time height from which the first result follows. The second expression is obtained from the first by multiplying  by
$1=\frac{V_{cd}^*V_{cb}^{\phantom{*}}}{V_{cd}^*V_{cb}^{\phantom{*}}}$.
\item In $J=
\text{Im}\left(V_{ud}^{\phantom{*}}V_{cd}^*V_{cb}^{\phantom{*}}V_{ub}^*\right)$ each $V_{ix}$ appears with one, and only one, other factor of $V_{iy}^*$, and one, and only one, factor of $V_{jx}^*$.
\end{enumerate}
\end{solution}
\end{exercises}

\item {\it Parametrization of $V$:} Since there are only four independent parameters in the matrix that contains $3\times3$ complex entries, it is useful to have a completely general parametrization in terms of four parameters. The standard parametrization can be understood as a sequence of rotations about the three axes, with the middle rotation incorporating also a phase transformation:
\begin{gather*}
V=CBA, \\
\intertext{where}
A=\begin{pmatrix} c_{12}& s_{12}&0\\  -s_{12}& c_{12}&0\\ 0& 0&1\end{pmatrix},\quad
B=\begin{pmatrix} c_{13}& 0& s_{13}e^{-i\delta}\\  0& 1&0\\ -s_{13}e^{i\delta}&0& c_{13} \end{pmatrix},\quad
C=\begin{pmatrix} 1&0& 0\\0& c_{23}& s_{23}\\ 0& -s_{23}& c_{23}\end{pmatrix}.
\end{gather*}
Here we have used the shorthand, $c_{ij}=\cos\theta_{ij}$, $s_{ij}=\sin\theta_{ij}$, where the angles $\theta_{ij}$ all lie on the first quadrant. From the phenomenologically observed rough order of magnitude of elements in $V$ in~\eqref{eq:Veps} we see that the angles $\theta_{ij}$ are all small. But the phase $\delta$ is large, else all triangles would be squashed.

An alternative and popular parametrization is due to Wolfenstein. It follows from the above by introducing parameters $A$, $\lambda$, $\rho$ and $\eta$ according to
\begin{equation}
\label{eq:Wpar}
s_{12}=\lambda, \quad s_{23}=A\lambda^2,\quad s_{13}e^{i\delta}=A\lambda^3(\rho+i\eta)
\end{equation}
The advantage of this parametrization is that if $\lambda$ is of the order of  $\epsilon$, while the other parameters are of order one, 
then the CKM has the rough order in~\eqref{eq:Veps}. It is easy to see
that $\rho$ and $\eta$ are very close to, but not quite, the coordinates of the apex of the unitarity triangle in Fig.~\ref{fig:UTsketch}. One can adopt the alternative, but tightly related parametrization in terms of $A$, $\lambda$, $\widebar\rho$ and~$\widebar\eta$:
\[
s_{12}=\lambda, \quad s_{23}=A\lambda^2,\quad s_{13}e^{i\delta}=A\lambda^3(\widebar\rho+i\widebar \eta)\frac{\sqrt{1-A^2\lambda^4}}{\sqrt{1-\lambda^2}[1-A^2\lambda^4(\widebar\rho+i\widebar \eta)]} .
\]

\begin{exercises}
\begin{exercise}
\begin{enumerate}[(i)]
\item Show that  \[
\widebar\rho+i\widebar \eta = - \fracup{V_{ud}^{\phantom{*}}V_{ub}^*}{V_{cd}^{\phantom{*}}V_{cb}^*},
\]
hence $\widebar\rho$ and $\widebar \eta$ are indeed the coordinates of the apex of the unitarity triangle and are invariant under quark phase redefinitions. 
\item Expand in $\lambda\ll1$ to show
\[
V=\begin{pmatrix} 1-\tfrac12\lambda^2 &\lambda& A\lambda^3(\rho-i\eta)\\ -\lambda&  1-\tfrac12\lambda^2 & A\lambda^2\\
A\lambda^3(1-\rho-i\eta)& - A\lambda^2 &1\end{pmatrix} +\mathcal{O}(\lambda^4)
\]

\end{enumerate}
\end{exercise}
\begin{solution}
\begin{enumerate}[(i)]
\item  We are not looking for a graphical representation solution, as was done in  Exercise \ref{eqsol:alphabetagamma}. Instead, we want to show this form the definitions of the parameters $\lambda, A, \bar\rho$ and~$\eta$.  This is just plug in and go. First, 
\[V=
\begin{pmatrix}
 c_{12} c_{13} & c_{13} s_{12} & s_{13}e^{-i\delta} \\
 -c_{12} s_{23} s_{13}e^{i\delta}-c_{23}
   s_{12} & c_{12} c_{23}-s_{12} s_{23}
   s_{13}e^{i\delta} & c_{13} s_{23} \\
 s_{12} s_{23}-c_{12} c_{23} s_{13}e^{i\delta}
   & -c_{12} s_{23}-c_{23} s_{12}
   s_{13}e^{i\delta} & c_{13} c_{23}
\end{pmatrix}.
\]
You can break the computation into
  smaller steps. For example,
  $V_{ud}=c_{12}c_{13}=\sqrt{1-\lambda^2}c_{13}$ and
  $V_{c,b}=c_{13}s_{23}=A\lambda^2 c_ {13}$ so that 
\[ \frac{V_{ud}}{V^*_{cb}}=\frac{\sqrt{1-\lambda^2}}{A\lambda^2}.\]
Similarly,
\[ \frac{V^*_{ub}}{V_{cd}}=-\frac{A\lambda^2\sqrt{1-A^2\lambda^4}z}{\sqrt{(1-\lambda^2)(1-A^2\lambda^4)}},\]
where $z=\bar\rho+i\bar\eta$. The result follows.
\item Again plug in and go. But you can be clever about it. For
  example, since $s_{13}\sim\lambda^3$, we have
  $c_{13}=\sqrt{1-s_{13}^2}=1+\mathcal{O}(\lambda^6)$. Similarly
  $c_{23}=1+\mathcal{O}(\lambda^4)$ and
  $c_{12}=1-\frac12\lambda^2+\mathcal{O}(\lambda^4)$.  Plugging these,
  and \eqref{eq:Wpar} into the explicit form of $V$ above the result follows.
  and the 
\end{enumerate}
\end{solution}
\end{exercises}
\end{enumerate}

\section{Once more on Flavor Symmetry}
\label{Sec:FlavorOps}
Suppose we extend the SM by adding terms (local,\footnote{By ``local''
  we mean a product of fields all evaluated at the same spacetime
  point.} Lorentz and gauge invariant) to the Lagrangian. Since the SM
already includes all possible monomials (``operators'')  of dimension
4 or smaller, we consider adding operators of dim~$\ge5$. We are going
to impose an additional constraint, and we will investigate its
consequence. We will require that these operators be invariant under
the flavor transformations, comprising the group $G_F$:
\begin{equation}
\label{eq:mvf-trans}
q_L\to U_q\; q_L\;, \quad u_R\to U_u\; u_R \;, \quad d_R\to U_d\; d_R~,\quad \lambda_U\to U_q^{\phantom{\dagger}}\lambda_U U_u^\dagger, \quad\lambda_D\to U_q^{\phantom{\dagger}}\lambda_D U_d^\dagger.
\end{equation}
We add some terms to the Lagrangian
\[
\mathcal{L}\to\mathcal{L}+\Delta\mathcal{L}, \qquad \Delta\mathcal{L}=\sum_i c_iO_i
\]
with, for example,
\begin{align*}
O_1&= G^a_{\mu\nu} H \widebar u_R T^a\sigma^{\mu\nu}\lambda_U q_L,\\
O_2 &= \widebar q_L \gamma^\mu  \lambda_U^\dagger\lambda_U^{\phantom{\dagger}}q_L\, \widebar d_R \gamma_\mu
\lambda_D^{\phantom{\dagger}}\lambda_D^\dagger d_R.
\end{align*}
Here $ G^a_{\mu\nu}$ is the  field strength for the $SU(3)_c$ gauge field (which is quite irrelevant for our discussion, so don't be distracted).  Consider these operators when we rotate to the basis in which the mass matrices are diagonal. Start with the first:
\begin{align*}
O_1 &\to G^a_{\mu\nu} H \widebar u_R T^a\sigma^{\mu\nu}  V_{u_R}^\dagger\lambda_U\begin{pmatrix}V_{u_L}u_L\\V_{d_L}d_L\end{pmatrix}\\
&= G^a_{\mu\nu} H \widebar u_R T^a\sigma^{\mu\nu}  (V_{u_R}^\dagger\lambda_UV_{u_L}^{\phantom{\dagger}})\begin{pmatrix}u_L\\V_{u_L\phantom{d_L\!\!\!\!\!\!\!\!\!}}^\dagger V_{d_L}^{\phantom{\dagger}}d_L\end{pmatrix}\\
&= G^a_{\mu\nu} H \widebar u_R T^a\sigma^{\mu\nu}  \lambda_U^\prime \begin{pmatrix}u_L\\ V d_L\end{pmatrix}
\end{align*}
We see that the only flavor-changing interaction is governed by the off-diagonal components of $\lambda_U^\prime V$. Similarly
\[
O_2\to 
\widebar q'_L \gamma^\mu  (\lambda_U^\prime)^2 q'_L\, \widebar d_R \gamma_\mu
(\lambda_D^{\prime})^2 d_R, \quad\text{where}\quad q'_L=\begin{pmatrix}u_L\\ V d_L\end{pmatrix}.
\]

\newpage
\begin{exercises}
\begin{exercise}
Had we considered an operator like $O_1$ but with $\Htilde \widebar d_R$ instead of $ H \widebar u_R$ the flavor off-diagonal terms would have been governed by $\lambda_D V^\dagger$.   
Show this is generally true, that is, that flavor change in any operator is governed by $V$ and powers of $\lambda'$. 
\end{exercise}
\begin{solution}
In any operator use the inverse of \eqref{eq:massDtransf} to write $\lambda_{U,D}$ in terms of $\lambda'_{U,D}$ and the matrices $V_{u_{L,R}}$ and $V_{d_{L,R}}$. Now rotate quarks to go to the mass-diagonal basis. This would be a flavor symmetry transformation if $V_{u_{L}}=V_{d_{L}}=U_q$, so it fails to be a symmetry only because $V=V_{u_{L}}^\dagger V_{d_{L}}^{\phantom{\dagger}}\ne1$, which may appear in these operators. This is the only parameter that is off-diagonal in flavor space. 
\end{solution}
\begin{exercise}
Exhibit examples of operators of dimension 6 that produce flavor
change without involving $\lambda_{U,D}$. Can these be such that only
quarks of charge $+2/3$ are involved? (These would correspond to Flavor
Changing Neutral Currents; see Sec.~\ref{sec:fcnc} below).
\end{exercise}
\begin{solution}
The question is phrased loosely: the answer depends on whether we impose the flavor symmetry \eqref{eq:mvf-trans}. If we don't, then we can simply take an operator like $O_1$ but without the spurion $\lambda_U$ sandwiched between quarks. So, for example, the operator $G^a_{\mu\nu} H \widebar u_R T^a\sigma^{\mu\nu}\kappa q_L$, where $\kappa$ is some arbitrary matrix, when expressed in the mass eigenstate basis gives
\[
 G^a_{\mu\nu} H \widebar u_R T^a\sigma^{\mu\nu}  V_{u_R}^\dagger\kappa\begin{pmatrix}V_{u_L}u_L\\V_{d_L}d_L\end{pmatrix}
\]
Consider, instead, the case when we insist on the symmetry \eqref{eq:mvf-trans}. Now quark bilinears can only be of one SM-representation with itself, $\widebar q_L\gamma^\mu q_L$,  $\widebar q_L\tau^j \gamma^\mu q_L$,  $\widebar u_R\gamma^\mu u_R$  and $\widebar d_R\gamma^\mu d_R$. Of these, only  $\widebar q_L\tau^j \gamma^\mu q_L$ fails to be invariant under the transformation that takes the quarks to the mass eigenstate basis, and then only the terms involving $\tau^\pm$.  So, in the absence of factors of $\lambda_{U,D}$ we can only get charge changing flavor changing interactions. A simple example is the four quark operator $\widebar q_L\tau^j \gamma^\mu q_L\,\widebar q_L\tau^j \gamma_\mu q_L$.
\end{solution}
\end{exercises}

This construction, restricting the higher dimension operators by the flavor symmetry with the Yukawa couplings treated  as spurions, goes by the name of the principle of Minimal Flavor Violation (MFV). Extensions of the SM in which the  only breaking of $G_F$ is by $\lambda_U$ and $\lambda_D$ automatically satisfy  MFV. As we will see they are much less constrained by flavor changing and CP-violating observables than models with generic breaking of $G_F$. Let's consider some examples:

\begin{enumerate}
\item {\it The supersymmetrized SM.} I am not calling this the MSSM,
  because the discussion applies as well to the zoo of models in which the BEH sector has been extended, {\it e.g.}, the NMSSM. In the absence of SUSY breaking this model satisfies the principle of MFV. The Lagrangian is 
\[
\mathcal{L}
=\int \!\! d^4\theta\;\left[\widebar Q e^VQ+\widebar Ue^V U +\widebar D e^V D\right]+\text{gauge \& H kinetic terms}+\int \!\! d^2\theta\,W +\text{h.c.}
\]
with superpotential 
\[
W=H_1 U y_UQ +H_2 D y_D Q+\text{non-quark-terms}
\]
Here $V$ stands for the vector superfields\footnote{Since I will not make explicit use of vector superfields, there should be no confusion with the corresponding symbol for the the CKM matrix, which is used ubiquitously in these lectures.} and $Q$, $D$, $U$, $H_1$ and $H_2$ are  chiral superfields with the following quantum numbers:
\begin{equation*}
\begin{aligned}
Q & \sim (3,2)_{1/6}\\
U & \sim (\widebar 3,1)_{-2/3}\\
D & \sim (\widebar 3,1)_{1/3}\\
\end{aligned}\qquad
\begin{aligned}
H_1 & \sim (1,2)_{1/2}\\
H_2 & \sim (1,2)_{-1/2}\\
\end{aligned}
\end{equation*}
The fields on the left column come in three copies, the three generations we call flavor. We are again suppressing that index (as well as the gauge and Lorentz indices). Unlike the SM case, this Lagrangian is not the most general one for these fields  once renormalizability, Lorentz and gauge invariance are imposed. In addition one needs to impose, of course, supersymmetry. But even that is not enough. One has to impose an $R$-symmetry to forbid dangerous baryon number violating renormalizable interactions. 

When the Yukawa couplings are neglected, $y_U=y_D=0$, this theory has a $SU(3)^3$ flavor symmetry. The symmetry is broken only by the couplings and we can keep track of this again by treating the couplings as spurions. Specifically, under $SU(3)^3$, 
\[
Q\to U_q Q,\quad U \to S_UU , \quad D\to S_D D,\quad y_U\to S_U^*y_U U_q^\dagger, \quad y_D\to S_D^*y_D U_q^\dagger
\]
Note that this has both quarks and squarks transforming together. 
The transformations on quarks may look a little different than the transformation in the SM, Eq.~\eqref{eq:mvf-trans}. But they are the same, really. The superficial difference is that here the quark fields are all written as left-handed fields, which are obtained by charge-conjugation from the right handed ones in the standard representation of the SM. So in fact, the couplings are related by 
$y_U=\lambda_U^\dagger$ and $y_D=\lambda_D^\dagger$, and the transformations on the right handed fields by $S_U=U_u^*$ and $S_D=U_d^*$. While the relations are easily established, it is worth emphasizing that we could have carried out the analysis in the new basis without need to connect to the SM basis. All that matters is the way in which symmetry considerations restrict  certain interactions. 

Now let's add soft SUSY breaking terms. By ``soft'' we mean operators of dimension less than 4. Since we are focusing on flavor, we only keep terms that include fields that carry flavor:
\begin{multline}
\label{eq:SUSYbkg}
\Delta\mathcal{L}_{\text{SUSY-bkg}}=\phi_q^*\mathcal{M}^2_q\phi_q+\phi_u^*\mathcal{M}^2_u\phi_u+\phi_d^*\mathcal{M}^2_d\phi_d\\
+(\phi_{h_1}\phi_u g_U \phi_q +\phi_{h_2}\phi_d g_D \phi_q +\text{h.c.})
\end{multline}
Here $\phi_X$ is the scalar SUSY-partner of the quark $X$. 
This breaks the flavor symmetry unless
$\mathcal{M}^2_{q,u,d}\propto\mathbf{1}$ and $g_{U,D}\propto y_{U,D}$
(see, however, Exercise~\ref{ex:susy-bkg}). And unless these conditions are satisfied new flavor changing interactions are generically present and large. The qualifier ``generically'' is because the effects can be made small by lucky coincidences (fine tunings) or if the masses of scalars are large. 

This is the motivation  for gauge mediated SUSY-breaking~\cite{Dine:1993yw}:

\bigskip

\begin{centering}
\begin{tikzpicture}[scale=1]
\coordinate (A) at (-4,0);
\coordinate (B) at  (4,0); 

\path node at (A) [shape=ellipse,draw] {\parbox[c]{3cm}{\begin{centering}SUSY\\ breaking sector\\\end{centering}}}
node at (B)[shape=ellipse,draw] {SUSY SM} ;

\draw ($(A)+(2.5,0)$) -- node{\parbox[c]{3cm}{\begin{centering}gauge \\ interaction\\\end{centering}}} ($(B)+(-1.5,0)$);

\end{tikzpicture}
\end{centering}

\bigskip

The gauge interactions, {\it e.g.}, $\widebar Q e^V Q$, are diagonal
in flavor space. In theories of supergravity mediated supersymmetry
breaking the flavor problem is severe. To repeat, this is why gauge mediation and its variants were invented. 

\item {\it  MFV Fields.} Recently CDF and D0 reported a larger than
  expected forward-backward asymmetry in $t\widebar t$ pairs
  produced in $p\widebar p$ collisions~\cite{Aaltonen:2011kc}.  Roughly speaking, define
  the forward direction as the direction in which the protons move,
  and classify the outgoing particles of a collision according to
  whether they move in the forward or backward direction. You can be
  more careful and define this relative to the CM of the colliding
  partons, or better yet in terms of rapidity, which is invariant
  under boosts along the beam direction. But we need not worry about
  such subtleties: for our purposes we want to understand how flavor
  physics plays a role in this process that one would have guessed is
  dominated  by  SM interactions~\cite{Kuhn:2011ri}. Now, we take this as an educational example, but I should warn you that by the time you read this the reported effect may have evaporated. In fact, since the lectures were given D0 has revised its result and the deviation from the SM expected asymmetry is now much smaller~\cite{Leone:2014gwa}. 

There are two types of BSM models that explain this asymmetry, classified according to the the type of new particle exchange that produces the asymmetry:

\begin{enumerate}[(i)]
\item $s$-channel. For example an ``axi-gluon,'' much like a gluon but massive and coupling to axial currents of  quarks. 
The interference between vector and axial currents, 
\begin{tikzpicture}[scale=0.7]
\coordinate[label=left:$u$] (u);
\coordinate[label=right:$t$,right=3 cm of u] (t);
\coordinate[label=left:$\widebar u$,below=1.25cm of u] (baru);
\coordinate[label=right:$\widebar t$,below=1.25cm of t] (bart);
\coordinate (v1) at ($0.5*(u)+0.5*(baru)+(0.5,0)$);
\coordinate (v2) at ($0.5*(t)+0.5*(bart)+(-0.5,0)$);

\draw[particle] (u) -- (v1);
\draw[particle] (v1) -- (baru);
\draw[particle] (bart) -- (v2);
\draw[particle] (v2) -- (t);

\draw[gluon] (v1) -- node[above=2pt]{$g$} (v2);

\node at ($(v2)+(1.4,0)$) {$+$};

\begin{scope}[xshift=6cm]
\coordinate[label=left:$u$] (u);
\coordinate[label=right:$t$,right=3 cm of u] (t);
\coordinate[label=left:$\widebar u$,below=1.25cm of u] (baru);
\coordinate[label=right:$\widebar t$,below=1.25cm of t] (bart);
\coordinate (v1) at ($0.5*(u)+0.5*(baru)+(0.5,0)$);
\coordinate (v2) at ($0.5*(t)+0.5*(bart)+(-0.5,0)$);

\draw[particle] (u) -- (v1);
\draw[particle] (v1) -- (baru);
\draw[particle] (bart) -- (v2);
\draw[particle] (v2) -- (t);

\draw[photon] (v1) -- node[above=2pt]{$a$} (v2);
\end{scope}
\end{tikzpicture} 
produces a FB-asymmetry. It turns out that it is best to have the sign of the axigluon coupling to $t$-quarks be opposite that of the coupling to $u$ quarks,  in order to get the correct sign of the FB-asymmetry without violting  constraints from direct detection at the LHC. But different couplings to $u$ and $t$ means flavor symmetry violation and by now you should suspect that any complete model will be subjected to severe constraints from flavor physics. 
\item $t$-channel: for example, one may exchange a scalar, and the amplitude now looks like this:
\begin{tikzpicture}[scale=0.7]
\coordinate[label=left:$u$] (u);
\coordinate[label=right:$t$,right=3 cm of u] (t);
\coordinate[label=left:$\widebar u$,below=1.25cm of u] (baru);
\coordinate[label=right:$\widebar t$,below=1.25cm of t] (bart);
\coordinate (v1) at ($0.5*(u)+0.5*(baru)+(0.5,0)$);
\coordinate (v2) at ($0.5*(t)+0.5*(bart)+(-0.5,0)$);

\draw[particle] (u) -- (v1);
\draw[particle] (v1) -- (baru);
\draw[particle] (bart) -- (v2);
\draw[particle] (v2) -- (t);

\draw[gluon] (v1) -- node[above=2pt]{$g$} (v2);

\node at ($(v2)+(1.4,0)$) {$+$};

\begin{scope}[xshift=6cm]
\coordinate[label=left:$u$] (u);
\coordinate[label=right:$t$,right=3 cm of u] (t);
\coordinate[label=left:$\widebar u$,below=1.25cm of u] (baru);
\coordinate[label=right:$\widebar t$,below=1.25cm of t] (bart);
\coordinate (v1) at ($0.5*(u)+0.5*(t)$);
\coordinate (v2) at ($0.5*(baru)+0.5*(bart)$);

\draw[particle] (u) -- (v1);
\draw[particle] (v1) -- (t);
\draw[particle] (bart) -- (v2);
\draw[particle] (v2) -- (baru);

\draw[dashed] (v1) -- node[right]{$\phi$} (v2);
\end{scope}
\end{tikzpicture} 

This model has introduced a scalar $\phi$ with a coupling $\phi \widebar t u$ (plus its hermitian conjugate). 
This clearly violates flavor symmetry. Not only we expect that the
effects of  this flavor violating coupling would be directly
observable but, since the coupling is introduced in the mass
eigenbasis, we suspect there are also other couplings involving the
charge-$+2/3$  quarks, as in $\phi \widebar c u$ and $\phi
\widebar t u$  and flavor diagonal ones. This is because even if we
started with only one coupling  in some generic basis of fields, when we rotate the fields to go the mass eigenstate basis we will generate all the other couplings.  Of course this does not have to happen, but it will, generically, unless there is some underlying reason, like a symmetry. Moreover, since couplings to a scalar involve both right and left handed quarks, and the left handed quarks are in doublets of the electroweak group, we may also have flavor changing interactions involving the charge-$(-1/3)$ quarks in these models. 
\end{enumerate}

One way around these difficulties is to build the model so that it satisfies the principle of MFV, by design. Instead of having only a single scalar field, as above, one may include a multiplet of scalars transforming in some representation of $G_F$. So, for example,  
one can have a charged scalar multiplet $\phi$ transforming in the
$(\mathbf{3},\mathbf{\widebar{3}}, 1)$ representation of  $SU(3)_q\times
SU(3)_u\times SU(3)_d$, with gauge quantum numbers $(1,2)_{-1/2}$ and with interaction term 
\[
\lambda \widebar q_L\phi u_R\qquad\text{with}\quad \phi\to U_{q_L} \phi\, U^\dagger_{u_R}\,.
\]
Note that the coupling $\lambda$ is a single number (if we want invariance under flavor). This actually works! See \cite{Grinstein:2011yv}.
\begin{exercises}
\begin{exercise}
\protect\label{ex:susy-bkg}Below Eq.~\protect\eqref{eq:SUSYbkg} we said, ``This breaks the flavor
symmetry unless $\mathcal{M}^2_{q,u,d}\propto\mathbf{1}$ and
$g_{U,D}\propto y_{U,D}$.'' This is not strictly correct (or, more
bluntly, it is a lie). While not correct it is the simplest
choice. Why? Exhibit alternatives, that is, other forms for $\mathcal{M}^2_{q,u,d}$ and
$g_{U,D}$ that respect the symmetry. {\it Hint: Read below. See \protect\eqref{eq:lambdacube}.}
\end{exercise}
\begin{solution}
Flavor symmetry requires that $\mathcal{M}^2_q\to U_q\mathcal{M}^2_q U_q^\dagger$, $\mathcal{M}^2_u\to S_U\mathcal{M}^2_u S_U^\dagger$,  $\mathcal{M}^2_d\to S_D\mathcal{M}^2_d S_D^\dagger$, $g_U\to S_U^*g_U U_q^\dagger$ and $y_D\to S_D^*y_D U_q^\dagger$. 
\end{solution}
\begin{exercise}
Classify all possible dim-4 interactions of Yukawa form in the SM. To
this end list all possible Lorentz scalar combinations you can form
out of pairs of SM quark fields. Then give explicitly the
transformation properties of the scalar field, under the gauge and
flavor symmetry groups, required to make the Yukawa interaction
invariant. Do this first without including the SM Yukawa couplings as
spurions  and then including also one power of the SM Yukawa couplings.
\end{exercise}
\end{exercises}
\end{enumerate}

\section{FCNC}
\label{sec:fcnc}
This stands for {\bf F}lavor {\bf C}hanging {\bf N}eutral {\bf
  C}urrents, but it is used more generally to mean Flavor Changing
Neutral transitions, not necessarily ``currents.''  By this we mean an
interaction that changes flavor but does not change electric
charge. For example, a transition from a $b$-quark to an $s$- or
$d$-quarks would be flavor changing neutral, but not  so a transition from a $b$-quark to a $c$- or $u$-quark.  Let's review flavor changing transitions in the SM:
\begin{enumerate}
\item Tree level. Only interactions with the charge vector bosons
  $W^\pm$ change flavor; {\it cf.} \eqref{eq:chargedCurr}. The photon and $Z$ coupe diagonally in flavor space, so these ``neutral currents'' are flavor conserving.   \\
\begin{tikzpicture} 
\coordinate[label=left:$d$] (d);
\coordinate[label=right:$u$,right=3 cm of d] (u);

\coordinate (v1) at ($0.5*(u)+0.5*(d)+(0,-0.5)$);
\coordinate (v2) at ($(v1)+ (2,-0.5)$);
\coordinate[label=right:$\widebar \nu$] (barnu) at ($(v2)+(1.5,0.5)$);
\coordinate[label=right:$e$] (e) at  ($(v2)+(1.5,-0.5)$);

\draw[particle] (d) -- (v1);
\draw[particle] (v1) -- (u);
\draw[particle] (barnu) -- (v2);
\draw[particle] (v2) -- (e);

\draw[photon] (v2) -- node[below]{$W$} (v1);

\node at ($(v1)+(-5.5,0)$) {For example, $n\to p e\widebar\nu$ is };

\end{tikzpicture}
 \item 1-loop. Can we have FCNCs at 1-loop? Say, $b\to s\gamma$? Answer: YES. Here is a diagram:
\begin{tikzpicture} 
\coordinate[label=left:$b$] (b) at (-3,0);
\coordinate[label=right:$s$] (s) at (3,0);

\coordinate (v1) at (-1,0);
\coordinate (v2) at (1,0);
\coordinate (v3) at (0,-1);

\coordinate[label=right:$\gamma$] (g) at ($(v3)+(2,-0.3)$);

\draw[particle] (b) -- (v1);
\draw[particle]  (v1) arc (180:270:1) node[left=0.7cm]{$u,c,t$};
\draw[particle] (v3) arc (270:360:1);
\draw[particle] (v2) -- (s);

\draw[photon] (g) --  (v3);
\draw[photon] (v1) --  node[above]{$W$} (v2);

\end{tikzpicture}\\
Hence, FCNC are suppressed in the SM by a 1-loop factor of $\displaystyle \sim \frac{g_2^2}{16\pi^2}\sim\frac{\alpha}{4\pi c^2_W}$ relative to the flavor changing charged currents. 
\end{enumerate}
\begin{exercises}
\begin{exercise}
Just in case you have never computed the $\mu$-lifetime, verify that 
\[\tau^{-1}_\mu\approx\Gamma(\mu\to e\nu_\mu\widebar\nu_e) = \frac{G_F^2m_\mu^5}{192\pi^3}\]
neglecting $m_e$, at lowest order in perturbation theory.
\end{exercise}
\begin{exercise}
Compute the amplitude for $Z\to b\widebar s$ in the SM to lowest
order in perturbation theory (in the strong and electroweak
couplings).  Don't bother to compute integrals explicitly, just make
sure they are finite (so you could evaluate them numerically if need
be). Of course, if you can express the result in closed analytic form,
you should. See~Ref.~\cite{Clements:1982mk}.
\end{exercise}
\end{exercises}

\section{GIM-mechanism: more suppression of FCNC in SM}
\subsection{Old GIM}
\label{ssec:oldgim}
Let' s imagine a world with a light top and a hierarchy  $m_u< m_c <
m_t \ll M_W$. Just in case you forgot, the real world is not like
this, but rather it has $m_u\ll m_c \ll M_W \approx \tfrac12 m_t$. We can make a lot of progress towards the computation of the Feynman graph for $b\to s\gamma$ discussed previously without computing any integrals explicitly:\\[0.5cm]
\parbox[c]{6cm}{\begin{tikzpicture}[scale=0.7]
\coordinate[label=left:$b$] (b) at (-3,0);
\coordinate[label=right:$s$] (s) at (3,0);
\coordinate (v1) at (-1,0);
\coordinate (v2) at (1,0);
\coordinate (v3) at (0,-1);
\coordinate (g) at ($(v3)+(2,-0.3)$);
\draw[particle] (b) -- (v1);
\draw[particle]  (v1) arc (180:270:1) node[left=0.7cm]{$u,c,t$};
\draw[particle] (v3) arc (270:360:1);
\draw[particle] (v2) -- (s);
\draw[photon] (g) node[right] {$\gamma(q,\epsilon)$} --(v3);
\draw[photon] (v1) --  node[above]{$W$} (v2);
\end{tikzpicture}
}
$\displaystyle = e q_\mu\epsilon_\nu\widebar u(p_s)\sigma^{\mu\nu}{\textstyle \left(\frac{1+\gamma_5}{2}\right)}u(p_b)\frac{m_b}{M_W^2}\,\frac{g_2^2}{16\pi^2}\cdot I$\\
where
\[I=\sum_{i=u,c,t}V_{ib}^{\phantom{*}}V_{is}^*F({\textstyle\frac{m_i^2}{M_W^2}})
 \]
and $F(x)$ is some function that results form doing the integral
explicitly, and we expect it to be of order 1. The coefficient of this
unknown integral can be easily understood. First, it has the obvious
loop factor ($g_2^2/16\pi^2$), photon coupling constant ($e$) and CKM
factors $V_{ib}^{\phantom{*}}V_{is}^*$ from the charged curent
interactions. 
Next, in order to produce a real (on-shell) photon the interaction has
to be
of the transition magnetic-moment form, $F_{\mu\nu}\widebar
s\sigma^{\mu\nu} b$, which translates into the Dirac spinors $u(p)$
for the quarks combining with the photon's momentum $q$ and
polarization vector ($\epsilon$) through $ q_\mu\epsilon_\nu\widebar
u(p_s)\sigma^{\mu\nu}u(p_b)$.\footnote{The other possibility, that the
  photon field $A_\mu$ couples to a flavor changing current, $A_\mu
  \widebar b \gamma^\mu s$, is forbidden by electromagnetic gauge
  invariance. If you don't like this argument, here is an
  alternative: were you to expand the amplitude in powers of $q/M_Z$
  you would find the lowest order contribution, $\epsilon^\mu
  \widebar u(p_s)\gamma^{\mu}u(p_b)$ is absent by gauge invariance,
  and the leading contribution is linear in momentum, as exhibited. } Finally, notice that the external quarks interact with
the rest of the diagram through a weak interaction, which involves
only left-handed fields. This would suggest getting an amplitude
proportional to $\widebar
u(p_s)\left(\frac{1+\gamma_5}{2}\right)\sigma^{\mu\nu}{\textstyle
  \left(\frac{1-\gamma_5}{2}\right)}u(p_b)$ which, of course,
vanishes. So we need one or the other of the external quarks to flip
its  chirality,  and only then interact. A chirality flip produces a factor of the mass of the quark and we have chosen to flip the chirality of the $b$ quark because $m_b\gg m_s$. This explain both the factor of $m_b$ and the projector $\frac{1+\gamma_5}{2}$ acting on the spinor for the $b$-quark. The correct dimensions are made up by the factor of $1/M_W^2$. 

Now, since we are pretending $m_u< m_c < m_t \ll M_W$, let's expand in a Taylor series, $F(x)=F(0)+xF'(0)+\cdots$ 
\[
I=\left(\sum_{i=u,c,t} V_{ib}^{\phantom{*}}V_{is}^*\right) F(0)+\left(\sum_{i=u,c,t} V_{ib}^{\phantom{*}}V_{is}^*\frac{m_i^2}{M_W^2}\right) F'(0)+\cdots
\]
Unitarity of the CKM matrix gives $\sum_{i=u,c,t} V_{ib}^{\phantom{*}}V_{is}^*=0$ so the first term vanishes. Moreover, we can rewrite the unitarity relation as giving one term as a combination of the other two, for example, 
\[ V_{tb}^{\phantom{*}}V_{ts}^* =- \sum_{i=u,c} V_{ib}^{\phantom{*}}V_{is}^*
\]
giving us
\[
I\approx  -F'(0) \sum_{i=u,c} V_{ib}^{\phantom{*}}V_{is}^*\frac{m_t^2-m_i^2}{M_W^2}
\]
We have uncovered additional FCNC suppression factors. Roughly, 
\[
I\sim V_{ub}^{\phantom{*}}V_{us}^*\frac{m_t^2-m_u^2}{M_W^2} + V_{cb}^{\phantom{*}}V_{cs}^*\frac{m_t^2-m_c^2}{M_W^2}\sim \epsilon^4\frac{m_t^2}{M_W^2} +\epsilon^2\frac{m_t^2}{M_W^2} .
\]
So in addition the 1-loop suppression, there is a mass suppression ($m_t^2/M_W^2$) and a mixing angle suppression ($\epsilon^2$).  This combination of suppression factors was uncovered by Glashow, Iliopoulos and Maiani (hence ``GIM'') back in the days when we only knew about the existence of three flavors, $u$, $d$ and $s$. They studied neutral kaon mixing, which involves a FCNC for $s$ to $d$ transitions and realized that theory would grossly over-estimate the mixing rate unless a fourth quark existed (the charm quark $c$) that would produce the above type of cancellation (in the 2-generation case). Not only did they explain kaon mixing and predicted the existence of charm, they even gave a rough upper bound for the mass of the charm quark, which they could do since the contribution to the FCNC grows rapidly with the mass, as shown above. We will study kaon mixing in some detail later, and we will see that the top quark contribution to mixing is roughly as large as that of the charm quark: Glashow, Iliopoulos and Maiani were a bit lucky, the parameters of the SM-CKM could have easily favored top quark mediated dominance in kaon mixing and their bound could have been violated. As it turns out, the charm was discovered shortly after their work, and the mass turned out to be close to their upper bound.

\subsection{Modern GIM}
We have to revisit the above story, since $m_t\ll M_W$ is not a good approximation. Consider our example above, $b\to s\gamma$. The function $F(x)$ can not be safely Taylor expanded when the argument is the top quark mass. However, $I$ is invariant under $F(x)\to F(x)+ $~constant, so we may choose without loss of generality $F(0)=0$. Then  
\begin{align*}
I&= -V_{cb}^{\phantom{*}}V_{cs}^*\left(F({\textstyle\frac{m_t^2}{M_W^2}})-F'(0)\frac{m_c^2}{M_W^2}\right)
-V_{ub}^{\phantom{*}}V_{us}^*\left(F({\textstyle\frac{m_t^2}{M_W^2}})-F'(0)\frac{m_u^2}{M_W^2}\right)+\cdots\\
&=F({\textstyle\frac{m_t^2}{M_W^2}})V_{tb}^{\phantom{*}}V_{ts}^*
+F'(0)\sum_{i=u,c} V_{ib}^{\phantom{*}}V_{is}^*\frac{m_i^2}{M_W^2}+\cdots\\
&\sim \epsilon^2 F({\textstyle\frac{m_t^2}{M_W^2}})
\end{align*}
We expect $F(x)$ to be order 1. This is indeed the case, $F(x)$ is a slowly increasing function of $x$ that is of order $1$ at the top quark mass. The contributions from $u$ and $c$ quarks to $I$ are completely negligible, and virtual top-quark exchange dominates this amplitude. 

\begin{exercises}
\begin{exercise}
Consider $s\to d\gamma$. Show that the above type of analysis suggests that virtual top quark exchange no longer dominates, but that in fact the charm and top contributions are roughly equally important. {\it Note: For this you need to know the mass of charm relative to $M_W$. If you don't, look it up!}
\end{exercise}
\begin{solution}
For  $s\to d\gamma$ we now have\\[0.5cm]
\parbox[c]{6cm}{\begin{tikzpicture}[scale=0.7]
\coordinate[label=left:$s$] (b) at (-3,0);
\coordinate[label=right:$d$] (s) at (3,0);
\coordinate (v1) at (-1,0);
\coordinate (v2) at (1,0);
\coordinate (v3) at (0,-1);
\coordinate (g) at ($(v3)+(2,-0.3)$);
\draw[particle] (b) -- (v1);
\draw[particle]  (v1) arc (180:270:1) node[left=0.7cm]{$u,c,t$};
\draw[particle] (v3) arc (270:360:1);
\draw[particle] (v2) -- (s);
\draw[photon] (v3) --  (g) node[right] {$\gamma(q,\epsilon)$};
\draw[photon] (v1) --  node[above]{$W$} (v2);
\end{tikzpicture}
}
$\displaystyle = e q_\mu\epsilon_\nu\widebar u(p_d)\sigma^{\mu\nu}{\textstyle \left(\frac{1+\gamma_5}{2}\right)}u(p_d)\frac{m_s}{M_W^2}\,\frac{g_2^2}{16\pi^2}\cdot I$\\
where
\[I=\sum_{i=u,c,t}V_{is}^{\phantom{*}}V_{id}^*F({\textstyle\frac{m_i^2}{M_W^2}})
 \]
We still have 
\[
I=F({\textstyle\frac{m_t^2}{M_W^2}})V_{ts}^{\phantom{*}}V_{td}^*
+F'(0)\sum_{i=u,c} V_{is}^{\phantom{*}}V_{id}^*\frac{m_i^2}{M_W^2}+\cdots
\]
But now the counting of powers of $\epsilon$ is a bit different: $|V_{ts}^{\phantom{*}}V_{td}^*|\sim \epsilon^5$ while $|V_{is}^{\phantom{*}}V_{id}^*|\sim\epsilon$ for either $i=c$ or $i=u$. Since $m_u\ll m_c$ we neglect the $u$-quark contribution. Using $F\sim1$ at the top, then the ratio of top to charm contributions is $\sim \epsilon^5/(\epsilon m_c^2/M_W^2)=(\epsilon^2 M_W/m_c)^2$. Using $M_W/m_c\approx 80/1.5$ and $\epsilon\approx 0.1$ the ratio os 0.3, and we have every right to expect the two contributions are comparable in magnitude. 
\end{solution}
\end{exercises}

\subsection{Bounds on New Physics, GIM and MFV}
Now let's bring together all we have learned. Let's stick to the
process $b\to s \gamma$, which in fact places some of the most
stringent constraints on models of new physics (NP). Let's model the
contribution of NP by adding a dimension 6 operator to the
Lagrangian,\footnote{The field strength should be the one for weak hypercharge, and the coupling constant should be $g_1$. This is just a distraction and does not affect the result; in the interest of pedagogy I have been intentionally sloppy.} 
\[
\Delta \mathcal{L}=\frac{C}{\Lambda^2}e F_{\mu\nu} H \widebar q_L \sigma^{\mu\nu} b_R =
\frac{evC}{\sqrt{2}\Lambda^2} F_{\mu\nu} \widebar s_L \sigma^{\mu\nu} b_R+\cdots
\]
I have assumed the left handed doublet belongs in the second
generation and have gone to unitary gauge. The coefficient of the
operator is $C/\Lambda^2$: $C$ is dimensionless and we assume it is of
order 1, while $\Lambda$ has dimensions of mass and indicates the
energy scale of the NP.  It is easy to compute this term's
contribution to the amplitude. It is even easier to roughly compare it to that of the SM, 
\[
\frac{\mathcal{A}_{\text{NP}}}{\mathcal{A}_{\text{SM}}}\sim \frac{\frac{vC}{\sqrt{2}\Lambda^2}}{|V_{tb}^{\phantom{*}}V_{ts}^*|\frac{\alpha}{4\pi s^2_W}\frac {m_b}{M_W^2}}
\]
Require this ratio be less than, say, 10\%, since the SM prediction agrees at that level with the measurement. This gives,
\[
C^{-1}\Lambda^2 \gtrsim \frac{vM_W^2s_W^2}{\sqrt{2}m_b|V_{tb}^{\phantom{*}}V_{ts}^*|\frac{\alpha}{4\pi}}\cdot\frac1{0.1}
\quad\Rightarrow\quad \Lambda\gtrsim 70~\text{TeV}.
\]
This bound is extraordinarily strong. The energy scale of 70~TeV is  much higher than that of any existing or planned particle physics accelerator facility. 

In the numerical bound above we have taken $C\sim1$, but clearly a small coefficient would help bring the scale of NP closer to experimental reach. The question is what would make the coefficient smaller. One possibility is that the NP is weakly coupled and the process occurs also at 1-loop but with NP mediators in the loop. Then we can expect $C\sim \alpha /4\pi s_W^2$, which brings the bound on the scale of new physics down to about 4~TeV. 

Now let's consider the effect of the principle of MFV. Instead of a single operator we take a collection of operators and make a flavor invariant (when we include spurions). Our first attempt is
\[
\Delta \mathcal{L}=\frac{C}{\Lambda^2}e F_{\mu\nu} H \widebar q_L \lambda_D \sigma^{\mu\nu} d_R
\]
which has the same form, as far as flavor is concerned, as the mass
term in the Lagrangian and therefore it gives no flavor changing
interaction when we go to the field basis that diagonalizes the mass
matrices. This is not a surprise, we have seen this before, in
Sec.~\ref{Sec:FlavorOps}. To get around this we need to construct an
operator which either contains more fields, which will give a loop
suppression in the amplitude plus an additional suppression by powers of
$\Lambda$, or additional factors of spurions. We try the latter. Consider, then 
\[
\Delta \mathcal{L}=\frac{C}{\Lambda^2}e F_{\mu\nu} H \widebar q_L \lambda_U^{\phantom{\dagger}}\lambda_U^\dagger \lambda_D^{\phantom{\dagger}} \sigma^{\mu\nu} d_R.
\]
When you rotate the fields to diagonalize the mass matrix you get, for the charge neutral quark bi-linear, 
\begin{equation}
\label{eq:lambdacube}
\lambda_U^{\phantom{\dagger}}\lambda_U^\dagger \lambda_D^{\phantom{\dagger}} \to V_{d_L}^\dagger \lambda_U^{\phantom{\dagger}}\lambda_U^\dagger \lambda_D^{\phantom{\dagger}} V_{d_R}^{\phantom{\dagger}} =
V_{d_L}^\dagger V_{u_L}^{\phantom{\dagger}}  (\lambda'_U)^2V_{u_L}^\dagger V_{d_L}^{\phantom{\dagger}}   \lambda'_D=V^\dagger (\lambda'_U)^2 V  \lambda'_D,
\end{equation}
our estimate of the NP amplitude is suppressed much like in the SM, by
the mixing angles and the square of the ``small'' quark masses. Our  bound now reads
\[
C^{-1}\Lambda^2 \gtrsim \frac{M_W^2s_W^2}{\sqrt{2}\frac{\alpha}{4\pi}}\cdot\frac1{0.1}
\quad\Rightarrow\quad C^{-1/2}\Lambda\gtrsim 4~\text{TeV}
\]
This is within the reach of the LHC (barely), even if $C\sim1$ which
should correspond to a strongly coupled NP sector. If for a weakly
coupled sector $C$ is one loop suppressed, $\Lambda$ could be
interpreted as a mass $M_{\text{NP}}$ of the NP particles in the loop,
and the analysis gives $M_{\text{NP}}\gtrsim 200$~GeV. The moral is
that if you want to build a NP model to explain putative new phenomena
at the Tevatron or the LHC you can get around constraints from flavor
physics if your model incorporates the principle of MFV or some other
mechanism that suppresses FCNC.

\begin{figure}
\begin{center}
\includegraphics[width=0.8\textwidth]{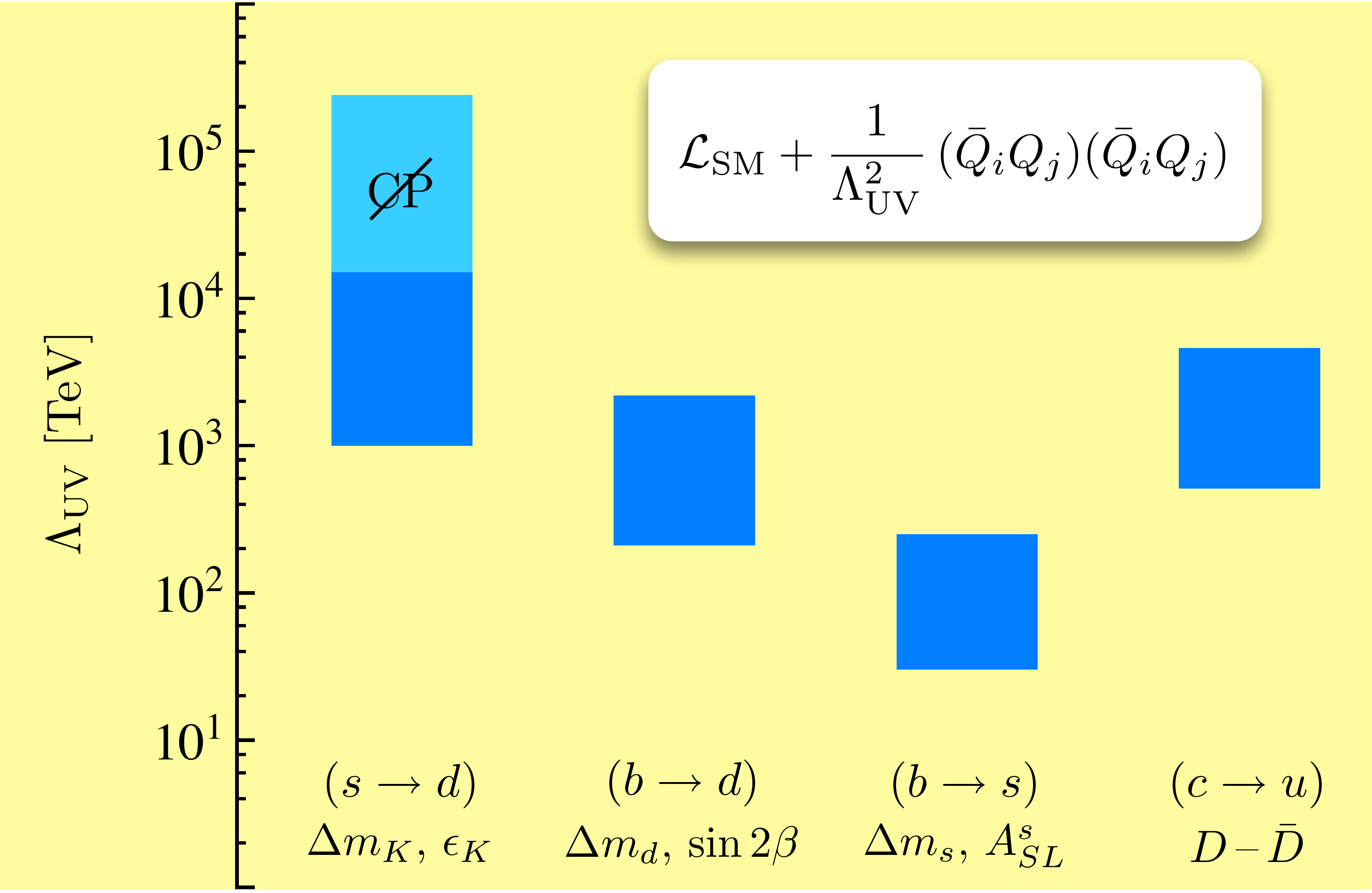}
\end{center}
\caption{\label{fig:neubertEPS2011}Bounds on the scale of NP scale from various processes. The NP is modeled as  dimension 6  operators. No accidental suppression of the coefficient (as in MFV) is included. The $b\to s$ case is consistent with the explicit $b\to s\gamma$ example worked out in these notes. The figure is taken from M.~Neubert's talk at EPS 2011. }
\end{figure}

\begin{exercises}
\begin{exercise}
  Determine how much each of the bounds in
  Fig.~\ref{fig:neubertEPS2011} is weakened if you assume MFV. You may
  not be able to complete this problem if you do not have some idea of
  what the symbols $\Delta M_K$, $\epsilon_K$, etc, mean or what type
  of operators contribute to each process; in that case you should
  postpone this exercise until that material has been covered later in these
  lectures.
\end{exercise}
\end{exercises}

\section{Determination of CKM Elements}
Fig.~\ref{fig:CKMfitterTRIANGLES}  shows the state of the art in our knowledge  of the angles of the unitarity triangles for the 1-3 and 2-3 columns of the CKM. How are these determined? More generally, how are CKM elements measured? Here we give a tremendously compressed description. 

The relative phase between elements of the CKM is associated with possible CP violation. So measurement of rates for processes that are dominated by one entry in the CKM are insensitive to the relative phases. Conversely, CP asymmetries directly probe relative phases:

\begin{enumerate}[(i)]
\item $|V_{ud}|$ is measured through allowed nuclear transitions. The theory is fairly well understood (even if it is nuclear physics) because the transition matrix elements are constrained by symmetry considerations. 
\item $|V_{us}|$, $|V_{cd}|$, $|V_{cs}|$, $|V_{ub}|$, $|V_{cb}|$, are
  primarily probed through semi-leptonic decays of mesons, $M\to M'
  \ell\nu$ ({\it e.g.}, $K^+\to \pi^0 e^+\nu$).  The  theoretical
  difficulty is to produce a reliable estimate of the rate, in terms
  of the CKM matrix elements, in light of the quarks being strongly
  bound in a meson. To appreciate the theoretical challenge  consider
  the decay of a pseudoscalar meson to another pseudoscalar meson. The
  weak interaction couples to a ``$V-A$'' ($V$=vector, $A$=axial)  hadronic current, $\widebar \psi' (\gamma^\mu-\gamma^\mu\gamma_5) \psi$, and a corresponding leptonic current. The latter, being excluded from the strong interactions, offers no difficulty and we can immediately  compute its contribution to the amplitude. The contribution to the amplitude from the hadronic side then involves
\[
\langle p'|V^\mu | p\rangle = f_+(q^2)(p+p')^\mu + f_-(q^2)q^\mu.
\]
The bra and ket stand for the meson final and initial states,
characterized only by their momentum and internal quantum numbers,
which are implicit in the formula. The   matrix element is to be
computed non-perturbatively with regard to the strong
interactions. Only the vector current (not the axial) contributes, by
parity symmetry of the strong interactions.  The matrix element must
be a vector, by Lorentz covariance, and it is written as a linear
combination of the only two vectors, $p$ and $p'$. In the 3-body decay, $p=p'+q$ so $q$ is the sum of the momentum of the lepton pair. The coefficients of
the expansion, or ``form factors,'' are functions of the invariants we
may form out of these vectors. There is only one kinematic variable
one can form, $p\cdot p'$, because $p^2$ and $p^{\prime2}$ are just the fixed square-masses of the mesons. It is conventional to write the form factors as functions of $q^2$. When the term $f_-(q^2) q^\mu$  is contracted with the leptonic current one gets a negligible contribution, $q\cdot (V-A)\sim m_\ell$, at least when $\ell=e$ or $\mu$. So the central problem is to determine $f_+$. Symmetry considerations can produce good estimates of $f_+$ at specific kinematic points, which is sufficient for the determination of the magnitude of the CKM. Alternatively one may determine the form factor using Monte Carlo simulations of QCD on the lattice. 

\begin{exercises}
\begin{exercise}
Show  that $q\cdot (V-A)\sim m_\ell$ for the leptonic charged
current. Be more
precise than ``$\sim$.''
 \end{exercise}
\end{exercises}

For example, if both states are pions then the vector current is the approximately conserved current associated with isospin symmetry. 
This gives $f_+(0)\approx1$. One can repeat this for kaons and pions,
where the symmetry now is Gell-Mann's $SU(3)$, in which the $u$, $d$
and $s$ quarks form a triplet. The pions and kaons, together with the
eta particle form an octet of $SU(3)$. In the symmetry limit one then
still has $f_+(0)=1$, but now the symmetry is not as good as in the
isospin case. Since the largest source of symmetry breaking is the
mass of the strange quark mass, one expects corrections
$f_+(0)-1\propto m_s/\Lambda$,  with $\Lambda$ a hadronic scale,
presumably $\Lambda\sim 1$~GeV. This seems like bad news, an
uncontrolled 10\% correction. Fortunately,  by a theorem of Ademolo
and Gatto, the symmetry breaking parameter appears at second order,
$f_+(0)-1\propto (m_s/\Lambda)^2\sim1$\%. We cannot extend this to the
heavier quarks because then $m_c/\Lambda>1$ is a bad expansion
parameter. Remarkably, for transitions among heavy quarks there is
another symmetry, dubbed ``Heavy Quark Symmetry,'' that allows
similarly successful predictions; for a basic introduction see~\cite{Grinstein:1992ss}. For heavy to light transitions one requires other methods, like lattice QCD, to determine the remaining CKMs. 

The green ring in Fig.~\ref{fig:CKMfitterTRIANGLES} shows the region of the $\bar\rho$-$\bar\eta$ plane allowed by the determination of $|V_{ub}|$. More precisely, note that  $\sqrt{\rho^2+\eta^2}=|V_{ub}/V_{us}V_{cb}|$ so that the ring requires the determination of the three CKM elements. It is labeled ``$|V_{ub}|$'' because this is the least accurately determined of the three CKM elements required. 

\item Neutral Meson Mixing. Next chapter is devoted to this. It
  gives, for example, $V_{tb}^{\phantom{*}} V_{td}^*$ in the case of
  $B_d$ mixing and $V_{tb}^{\phantom{*}} V_{ts}^*$ for $B_s$
  mixing. And the case of $K^0$ mixing is, as we will see,
  fascinatingly subtle and complex. The yellow (``$\Delta m_d$'') and
  orange (``$\Delta m_d$ \& $\Delta m_s$'')  circular rings
  centered at $(1,0)$ in Fig.~\ref{fig:CKMfitterTRIANGLES} are
  determined by the rate of $B_d$ mixing and by the ratio of rates of
  $B_d$ and $B_s$ mixing, respectively. The ratio is used because in
  it some uncertainties cancel, hence yielding a thiner ring. The bright green region labeled
  $\varepsilon_K$ is determined by CP violation in
  $K^0$-$\widebar{K}^0$ mixing.

\item CP asymmetries. Decay asymmetries, measuring the difference in
  rates of a process and the CP conjugate process, directly probe
  relative phases of CKM elements, and in particular the unitarity
  triangle angles $\alpha$, $\beta$ and $\gamma$. We will also study
  these, with particular attention to the  poster boy, the
  determination of $\sin(2\beta)$ from $B_d\to \psi K_S$, which is largely free from hadronic uncertainties. In Fig.~\ref{fig:CKMfitterTRIANGLES}  the blue and brown wedges labeled $\sin2\beta$ and $\gamma$, respectively, and the peculiarly shaped light blue region labeled $\alpha$ are all obtained from various CP asymmetries in decays of $B_d$ mesons.   
\end{enumerate}

\chapter{Neutral Meson Mixing and CP Asymmetries}
\section{Why Study This?}
Yeah, why? In particular why bother with an old subject like neutral-$K$ meson mixing? I offer you an incomplete list of perfectly good reasons:
\begin{enumerate}[(i)]
\item CP violation  was discovered in neutral-$K$ meson mixing. 
\item Best constraints on NP from flavor physics are from meson mixing. Look at Fig.~\ref{fig:neubertEPS2011}, where the best constraint is from CP violation in neutral-$K$ mixing. In fact, other than $A^s_{SL}$, all of the other observables in the figure involve mixing. 
\item It's a really neat phenomenon (and that should be sufficient reason for wanting to learn about it, I hope you will agree).
\item It's an active field of research both in theory and in experiment. I may be just stating the obvious, but the LHCb collaboration has been very active and extremely successful, and even CMS and ATLAS have performed flavor physics analysis. And, of course, there are also several non-LHC experiments ongoing or planned; see, {\it e.g.}, \cite{Venditti:2008zz}.
\end{enumerate}

But there is another reason you should pay attention to this, and more generally to the ``phenomenology'' (as opposed to ``theory'' or ``model building'') part of these lectures.  Instead of playing with Lagrangians and symmetries we will use these to try to understand  dynamics, that is, the actual physical phenomena the Lagrangian and symmetries describe. If you are a model builder you can get by without an understanding of this. 
Sort of. There are enough resources today where you can plug in the data from your model and obtain a prediction that can be tested against experiment. Some of the time. And all of the time without understanding what you are doing. You may get it wrong, you may miss effects. As a rule of thumb, if you are doing something good and interesting, it is novel enough that you may not want to rely on calculations you don't understand and therefore don't know if applicable. Besides, the more you know the better equipped you are to produce interesting physics. 

\section{The $\epsilon$ parameter}
We start our discussion of mixing by concentrating on the neutral $K$ system. It will be straightforward to carry the formalism over to the other neutral meson systems, $D^0$, $B^0=B_d$ and $B_s$. Although they are all based on the same physics, each has its own peculiarities. Moreover it is a historical inconvenience that the notation and conventions used by the different communities that study these mesons  differ unnecessarily. So we have to start somewhere, and we choose the historically important $K^0$-$\widebar K^0$ system. 

Consider the ``weak'' eigenstates $K^0$-$\widebar K^0$, These are really flavor eigenstates in the sense that $K^0$ has the quantum numbers of a $d$ quark and an $s$-antiquark, $ (\widebar s d)$, and $\widebar K^0 =   (\widebar d s)$. Note the peculiar choice the strange meson, $K^0$ contains an anti-strange quark, and carries strangeness $-1$. By ``strangeness'' we mean a  $U(1)$ group that rotates $s_L$ and $s_R$ (in the diagonal mass basis) by a common phase. These states are related by CP, charge conjugation changes particles into anti-particles and parity turns a pseudoscalar into minus itself:
\[ CP| K^0\rangle = -| \widebar K^0\rangle \qquad CP| \widebar K^0\rangle = -| K^0\rangle \]
Now we want to study the time evolution of these one particle states. We can always go to the rest frame,  and since we do not involve many-particle states regardless of their quantum numbers we can model the time evolution by a two-state Schrodinger equation with a $2\times2$ Hamiltonian. 
Of course, since these one particle states may evolve into states that are not accounted for in the two state Hamiltonian, the evolution will not be unitary and the Hamiltonian will not be Hermitian. Keeping this in mind we write, for this effective hamiltonian
\begin{equation}
\label{eq:CPK0}
\mathbf{H}=\mathbf{M}-\frac{i}{2}\mathbf{\Gamma}=\begin{pmatrix}
M-\frac{i}2\Gamma& M_{12}-\frac{i}2\Gamma_{12}\\
M_{12}^*-\frac{i}2\Gamma_{12}^*& M-\frac{i}2\Gamma\end{pmatrix}
\end{equation}
where $\mathbf{M}^\dagger = \mathbf{M}$ and   $ \mathbf{\Gamma}^\dagger =  \mathbf{\Gamma}$. 
\newpage
\begin{exercises}
\begin{exercise}
Show that CPT implies $H_{11}=H_{22}$. Note: If you want to test CPT you relax this constraint. See Ref.~\cite{Huet:1994kr}.
\end{exercise}
\begin{solution}
Let $\Omega=CPT$. We have to be a bit careful in that this is an
anti-unitary operator. The bra-ket notation is somewhat confusing for
anti-linear operators, so we use old fashioned inner product notation
$(\psi,\eta)$ for $\langle\psi|\eta\rangle$. Anti-unitarity means
$(\Omega\psi,\Omega\eta)=(\eta,\omega)$, and anti-linearity means
$\Omega(a\psi+b\eta)= a^*\psi+b^*\eta$, where $a,b$ are constants and
$\psi,\eta$ wave-functions.  Now, the CPT theorem gives $\Omega H
\Omega^{-1}=H^\dagger$. So 
\begin{align*}
(\psi,H\eta)&=(\psi,H\Omega^{-1}\Omega\eta) && \\
&=(\Omega H\Omega^{-1}\Omega\eta, \Omega\psi) && \text{by
  anti-unitarity of $\Omega$}\\
&=( H^\dagger\Omega\eta, \Omega\psi) && \text{by CPT theorem}\\
&=( \Omega\eta, H \Omega\psi) && \text{by definition of adjoint of operator}
\end{align*}
The action of $\Omega$ on the one particle states at rest is just like
that of CP, $\Omega | K^0\rangle = -| \widebar K^0\rangle$ and 
$\Omega| \widebar K^0\rangle = -| K^0\rangle $. So taking $\psi$ and
$\eta$ above to be $| \widebar K^0\rangle $, we have
$H_{22}=(\psi,H\eta)=( \Omega\eta, H \Omega\psi)=H_{11}$.
Note that for $\psi=| K^0\rangle $ and $\eta= | \widebar K^0\rangle$
the same relation gives $H_{12}=H_{12}$. 
\end{solution}
\end{exercises}

CP invariance requires $M_{12}^*=M_{12}$ and $
\Gamma_{12}^*=\Gamma_{12}$. Therefore {\it either}
$\text{Im}M_{12}\ne0$ {\it or} $\text{Im} \Gamma_{12}\ne0$ signal that CP
is violated. Now, to study the time evolution of the system we solve
Schrodinger equation. To this end we first solve the eigensystem for
the effective Hamiltonian. We define the eigenvalues to be
\[
M_{K\!_{L\atop S}}-\tfrac{i}2\Gamma_{K\!_{L\atop S}}=M-\tfrac{i}2\Gamma\pm\tfrac12(\Delta M-\tfrac{i}2\Delta\Gamma)
\]
 and the corresponding eigenvectors are 
\begin{equation}\label{KLSdefd}
|K\!_{L\atop S}\rangle =\frac1{\sqrt{2(1+|\epsilon|^2)}}\left[(1+\epsilon)| K^0\rangle\pm (1-\epsilon)| \widebar K^0\rangle\right]
\end{equation}
If $\epsilon=0$ then these are  $CP$-eigenstates: $CP |K_{L}\rangle=-|K_L\rangle$ and $CP |K_{S}\rangle=|K_S\rangle$. Since $CP |\pi\pi\rangle_{\ell=0}=|\pi\pi\rangle_{\ell=0}$ and  $CP |\pi\pi\pi\rangle_{\ell=0}=-|\pi\pi\pi\rangle_{\ell=0}$ we see that if CP were a good symmetry the decays $K_L\to\pi\pi\pi$ and $K_S\to\pi\pi$ are allowed, but not so the decays $K_L\to\pi\pi$ and $K_S\to\pi\pi\pi$. Barring CP violation in the decay amplitude, observation of $K_L\to\pi\pi$ or $K_S\to\pi\pi\pi$
indicates $\epsilon\ne0$, that is, CP-violation in mixing. 

This is very close to what is observed:
\begin{align}
\label{eq:KpipiBrs}
\text{Br}(K_S\to\pi\pi)&=100.00\pm0.24\%\nonumber\\
\text{Br}(K_L\to\pi\pi)&=0.297\pm0.023\%\\
\text{Br}(K_L\to\pi\pi\pi)&=33.9\pm1.2\%\nonumber
\end{align}
Hence, we conclude {\it (i)} $\epsilon$ is small, and {\it (ii)} CP is not a symmetry. Notice that $3m_\pi\sim 3(140)~\text{MeV}=420~\text{MeV}$ while $m_K\sim490~\text{MeV}$, leaving little phase space for the decays $K\to\pi\pi\pi$. This explains why $K_L$ is much longer lived than $K_S$; the labels ``L'' and ``S'' stand for ``long'' and ``short,'' respectively:
\begin{align*}
\tau_{K_S} &= 0.59\times10^{-10}~\text{s}\\
\tau_{K_L} &= 5.18\times10^{-8}~\text{s}
\end{align*}
For the $D^0$, $B^0=B_d$ and $B_s$ mesons the approximate CP eigenstates have a large number of decay channels available, many consisting of two particle states with much lower masses than the decaying particle and phase space suppression is negligible. The widths of the eigenstates are comparable so it makes no sense to call them ``long'' and ``short.'' Instead they are commonly referred to as heavy and light, $X_H$ and $X_L$ (with $X=D, B_d$ or $B_s$), even though the mass differences are very small too.

\begin{exercises}
\begin{exercise}
In \protect\eqref{eq:KpipiBrs} the branching fraction $\text{Br}(K_S\to\pi\pi)$ is nearly 100\% while $\text{Br}(K_L\to\pi\pi\pi)$ is only about 34\%. Explain. {\it Hint: Consult the \protect\href{http://pdg.lbl.gov}{PDG.}}
\end{exercise}
\end{exercises}

\paragraph{Measuring $\epsilon$.} The semileptonic decay CP-asymmetry is 
\[\delta =\frac{\Gamma(K_L\to\pi^-e^+\nu)-
\Gamma(K_L\to\pi^+e^-\widebar\nu)}{\Gamma(K_L\to\pi^-e^+\nu)+
\Gamma(K_L\to\pi^+e^-\widebar\nu)}
=\frac{|1+\epsilon|^2-|1-\epsilon|^2}{|1+\epsilon|^2+|1-\epsilon|^2}\approx2\text{Re}\epsilon
\]
The first equal sign defines it, the second one computes it (assuming CP symmetry in the decay amplitude) and the last one approximates it ($|\epsilon|\ll1$). Experimental measurement gives $\delta_{\text{exp}}=0.330\pm0.012\%$, from which $\text{Re}\epsilon\simeq1.65\times10^{-3}$.

\subsection{Formulas for $\epsilon$}
Eventually we will want to connect this effective $2\times2$  hamiltonian to the underlying fundamental physics we are studying. This can be done using perturbation theory (in the weak interactions) and is an elementary exercise in Quantum Mechanics (see, {\it e.g.}, Messiah's textbook, p.994 -- 1001 \cite{Messiah}). With $| K^0\rangle =| 1\rangle $ and $| \widebar{K}^0\rangle =| 2\rangle $ one has
\begin{align}
\label{eq:MfromTheory}
M_{ij}&=M\delta_{ij}+\langle i|H|j\rangle+{\sum_n}'\text{PP}\frac{\langle i|H|n\rangle\langle n|H|j\rangle}{M-E_n}
+\cdots \\
\label{eq:GammafromTheory}
\Gamma_{ij}&=2\pi{\sum_n}'\delta(M-E_n)  \langle i|\mathcal{H}|n\rangle\langle n|H|j\rangle 
+\cdots
\end{align}
Here the prime in the summation sign means that the states $ |
1\rangle $ and $| 2\rangle $  are excluded and PP stands for
``principal part.''  Beware the states are normalized by $\langle
i|j\rangle=\delta_{ij}$ rather than $\langle p'|
p\rangle=\frac{E}{m}\delta^{(3)}(p-p')$ (let alone $\langle p'|
p\rangle=2E\delta^{(3)}(p-p')$). Also, $H$ is a Hamiltonian, not a
Hamiltonian density $\mathcal{H} $; $H=\int d^3x\,\mathcal{H} $. It is
the part of the SM Hamiltonian that can produce flavor changes. In the
absence of $H$ the states $| K^0\rangle =| 1\rangle $ and $|
\widebar{K}^0\rangle =| 2\rangle $ would be stable eigenstates of the
Hamiltonian and their time evolution would be by a trivial phase. It
is assumed that this flavor-changing interaction is weak, while there
may be other much stronger interactions (like the strong one that
binds the quarks together). The perturbative expansion is in powers of
the weak interaction while the matrix elements are computed
non-perturbatively with  respect to the remaining interactions. Of
course the weak flavor changing interaction is, well, the Weak
interaction of the electroweak model, and below we denote the Hamiltonian by
$H_w$.

To make the full connection with more fundamental physics, we need a formula for $\epsilon$ in terms of the effective hamiltonian. We start with an exercise:

\begin{exercises}
\begin{exercise}
Show that $\epsilon$ is given by (the solution to)
\[
\frac{1+\epsilon}{1-\epsilon}=2\frac{M_{12}-\frac{i}2\Gamma_{12}}{\Delta M-\frac{i}2\Delta\Gamma}=
\frac12\frac{\Delta M-\frac{i}2\Delta\Gamma}{M^*_{12}-\frac{i}2\Gamma^*_{12}}
\]
where $\Delta M$ and $\Delta \Gamma$ can be themselves determined from (the solution to)
\[(\Delta M)^2-\frac14(\Delta\Gamma)^2=
4|M_{12}|^2-|\Gamma_{12}|^2\quad\text{and}\quad
\Delta M\Delta\Gamma=
4\text{Re}(M_{12}^{\phantom{*}}\Gamma^*_{12})\]
\end{exercise}
\begin{exercise}
If CP is conserved  show that $\Delta M=2\,\text{Re}M_{12}$ and $\Delta \Gamma=2\,\text{Re}\Gamma_{12}$.
\end{exercise}
\begin{solution}
This means  $\epsilon=0$ and $\text{Im}M_{12}=\text{Im}\Gamma_{12}=0$. Form the previous exercise , $\epsilon=0$ gives  $\Delta M-\frac{i}2\Delta\Gamma= 2(M_{12}-\frac{i}2\Gamma_{12})$ which is trivially solved once we use reality of $M_{12}$ and $\Gamma_{12}$ (keep in mind $\Delta M$ and $\Delta \Gamma$ are real, by definition.
\end{solution}
\end{exercises}

As we have seen,  empirically,  $\epsilon$ is not vanishing but, still, it  is small,  so it is natural to assume 
\[\text{Im}M_{12}\ll \text{Re}M_{12}\qquad\text{and}\qquad   
\text{Im}\Gamma_{12}\ll \text{Re}\Gamma_{12}.\]
Then, to linear order in $\epsilon$
\[\epsilon\approx i\frac{ \text{Im}M_{12}-\tfrac{i}2 \text{Im}\Gamma_{12}}{\Delta M -\tfrac{i}2\Delta \Gamma} \]
We will see shortly that $\text{Im}\Gamma_{12}\ll  \text{Im}M_{12} $. Also, empirically $\Delta \Gamma\approx-2\Delta M$; more precisely, $\text{arg}(\Delta M+\frac{i}2\Delta\Gamma)=46.2^\circ$ which we will approximate as $\pi/4$. We finally arrive at the semi-empirical formula
\begin{equation}
\label{eq:epislonEmpirical}
\epsilon\approx e^{i\pi/4}\frac{\text{Im}M_{12}}{\sqrt2\Delta M}
\end{equation}
This looks like a peculiar formula, mixing derived quantities with more fundamental ones (after using empirical input!). So an explanation is in order. We will explain in great detail that the imaginary part of $M_{12}$ is dominated by short distance physics (I will explain what is meant by that) and one can derive nice closed form formulas for it. On the other hand $\Delta M\approx 2\,\text{Re}M_{12}$ is hard to compute, is complicated by long distance physics,  cares little about CP violation, and is measured accurately, so we can just use the value. 

\section{The $\epsilon'$ parameter.}
$\epsilon$ measures the amount of CP admixture of the CP even and odd would-be CP-eigenstates: $K_L$ can decay into $\pi\pi$ because it contains a small ``contamination'' of the CP-even component. But $K_L$ can also decay into $\pi\pi$ even in the absence of this contamination if there is CP violation in the decay process. If this were absent, so all the CP violation would be from mixing, we would expect
\[
\frac{K_L\to\pi^+\pi^-}{K_L\to\pi^0\pi^0}=\frac{K_S\to\pi^+\pi^-}{K_S\to\pi^0\pi^0}.
\]
In other words, both $K_L$ and $K_S$ decays are from the common  CP-even component. But if the CP-odd component can also decay to $\pi\pi$ then the equality is not guaranteed, the CP-even and CP-odd components may decay into the charge and neutral pions with different relative rates. Violations of this relation are measured by the parameter $\epsilon'$. Let
\[
\eta_{+-}\equiv\frac{\langle\pi^+\pi^-|\mathcal{H}_w|K_L\rangle}{\langle\pi^+\pi^-|\mathcal{H}_w|K_S\rangle} \qquad \text{and}\qquad
\eta_{00}\equiv\frac{\langle\pi^0\pi^0|\mathcal{H}_w|K_L\rangle}{\langle\pi^0\pi^0|\mathcal{H}_w|K_S\rangle}.
\]

It is standard practice to parametrize the decay amplitudes in terms of  2-pion states with definite isospin. 
Define
\begin{equation}
\label{eq:AIdefd}
 \prescript{}{_{\text{out}}}\langle \pi\pi(I)|\mathcal{H}_w|K^0 \rangle_{\text{in}} = A_I e^{i\delta_I}
\end{equation}
where $\delta_I$ are the (measured) final states phase shifts for the $s$-wave 2-pion states of isospin $I$, 
$|\pi\pi(I)\rangle_{\text{{out}}}=e^{2i\delta_I}|\pi\pi(I)\rangle_{\text{{in}}}$. $A_I$ is in general complex, but it is conventional to redefine $|K^0\rangle \to e^{i\alpha} |K^0\rangle $ to absorb the phase in $A_0$. So we take $\text{Im}(A_0)=0$. 
\begin{exercises}
\begin{exercise}
Show that if CP is respected by the decay amplitude, then $\eta_{+-}=\eta_{00}=\epsilon$. 
\end{exercise}
\begin{solution}
\[
\eta_{+-}=\frac{\langle\pi^+\pi^-|\mathcal{H}_w|K_L\rangle}{\langle\pi^+\pi^-|\mathcal{H}_w|K_S\rangle} 
=\frac{(1+\epsilon)\langle\pi^+\pi^-|\mathcal{H}_w|K^0\rangle+(1-\epsilon)\langle\pi^+\pi^-|\mathcal{H}_w|\widebar
  K^0\rangle}{(1+\epsilon)\langle\pi^+\pi^-|\mathcal{H}_w|K^0\rangle-(1-\epsilon)\langle\pi^+\pi^-|\mathcal{H}_w|\widebar
  K^0\rangle}
\]
Now that the decay amplitude is CP invariant means that the
hamiltonian that mediates the decay (although not the mixing) is CP
invariant. So we can replace  $(CP)^{-1}\mathcal{H}_w(CP)$ for
$\mathcal{H}_w$ and use $CP|\pi\pi\rangle=|\pi\pi$ and
Eq.~\eqref{eq:CPK0} to show $\langle\pi^+\pi^-|\mathcal{H}_w|\widebar
  K^0\rangle=-\langle\pi^+\pi^-|\mathcal{H}_w| K^0\rangle$. The  the
  amplitudes cancel in the ratio and we obtain
\[\eta_{+-}=\frac{(1+\epsilon)-(1-\epsilon)}{(1+\epsilon)+(1-\epsilon)}=\epsilon
\]
\end{solution}
\begin{exercise}
Show that CPT invariance implies $\prescript{}{{\text{out}}}\langle \pi\pi(I)|H_w|\widebar{K}^0 \rangle_{\text{in}} = -A_I^* e^{i\delta_I}$. 
\end{exercise}
\begin{solution}
Applying CPT to \eqref{eq:AIdefd}, keeping in mind that $T$ is anti-unitary, 
\begin{equation}
- \prescript{}{_{\text{out}}}\langle \widebar{K}^0 |H_w| \pi\pi(I)\rangle_{\text{in}} = A_I e^{i\delta_I}
\end{equation}
Now take the complex conjugate of this and recall that for one particle states one conventionally takes the same phase for `in' and `out' states, $|\widebar{K}^0\rangle_{\text{out}}=|\widebar{K}^0\rangle_{\text{in}}$, while for the 2-particle states we can use $|\pi\pi(I)\rangle_{\text{{out}}}=e^{2i\delta_I}|\pi\pi(I)\rangle_{\text{{in}}}$. The result follows.
\end{solution}
\begin{exercise}
Show that for $s$-wave states
\begin{align}
\tfrac1{\sqrt2}(|\pi^+\pi^-\rangle+|\pi^-\pi^+\rangle)&=
\tfrac1{\sqrt3}|\pi\pi(I=2)\rangle+\sqrt{\tfrac23}|\pi\pi(I=0)\rangle\\
|\pi^0\pi^0\rangle&=
\sqrt{\tfrac23}|\pi\pi(I=2)\rangle-\tfrac1{\sqrt3}|\pi\pi(I=0)\rangle
\end{align}
\end{exercise}
\begin{solution}
This is pure group theory. You can look up the Clebsch-Gordan coefficients. Here, instead, we construct them from scratch. Use the adjoint representation of $SU(2)$:
\begin{equation}
T^1=\begin{pmatrix}
0&1&0\\ -1&0&0\\ 0&0&0\end{pmatrix},\qquad
T^2=\begin{pmatrix}
0&0&-1\\ 0&0&0\\ 1&0&0\end{pmatrix},\qquad
T^3=\begin{pmatrix}
0&0&0\\ 0&0&1\\ 0&-1&0\end{pmatrix},
\end{equation}
These satisfy the commutation relations, $[T^i,T^j]=2i\epsilon^{ijk}T^k$, from which it follows that
\begin{equation}
[T^\pm,T^3]=\mp T^\pm,\qquad [T^+,t^-]=T^3,
\end{equation}
where, as usual, $T^\pm\equiv \frac12(T^1\pm iT^2)$. We want states in a triplet (adjoint) satisfying $T^+|\pi^+\rangle=0$ (highest weight), so that $|\pi^0\rangle=T^-|\pi^+\rangle$ and 
$|\pi^-\rangle=T^-|\pi^0\rangle$. It is easy then to write the normalized states explicitly,
\begin{equation}
\pi^+=\frac1{\sqrt2}\begin{pmatrix}0\\1\\-i\end{pmatrix},\qquad
\pi^0=\begin{pmatrix}i\\0\\0\end{pmatrix}\qquad\text{and}\qquad
\pi^-=\frac1{\sqrt2}\begin{pmatrix}0\\1\\i\end{pmatrix}.
\end{equation}
Then 
\begin{equation}
\pi^+\pi^+\xrightarrow[T^-]{}\frac{\pi+\pi^0+\pi^0\pi^+}{\sqrt2}
\xrightarrow[T^-]{}\frac{2\pi^0\pi^0+\pi+\pi^-+\pi^-\pi^+}{\sqrt6}\,\cdots
\end{equation}
gives the normalized $I=2$ neutral state
\begin{equation}
\pi\pi(I=2)\frac1{\sqrt6}(2\pi^0\pi^0+\pi+\pi^-+\pi^-\pi^+).
\end{equation}
The $I=0$ state can be found by requiring that its variation vanishes. Taking an arbitrary linear combination
$\pi\pi(I=0) a\pi^0\pi^0+b\pi+\pi^-+c\pi^-\pi^+$, the condition is $b=c=-a$, and normalizing
gives the normalized $I=2$ neutral state
\begin{equation}
\pi\pi(I=0)\frac1{\sqrt3}(-\pi^0\pi^0+\pi+\pi^-+\pi^-\pi^+).
\end{equation}
The solution follows from inverting these.
\end{solution}
\end{exercises}
Combining we find
\begin{equation}
\eta_{+-}=\frac{e^{i(\delta_2-\delta_0)} (i\text{Im}A_2+\epsilon \text{Re} A_2)+\sqrt2 \epsilon A_0}{
e^{i(\delta_2-\delta_0)} (\text{Re}A_2+i\epsilon \text{Re} A_2)+\sqrt2  A_0}.
\end{equation}
These can be further simplified using empirical information (data) and approximations. The ``$\Delta I=1/2$ rule''  is the observation that $|A_2|/A_0\sim 1/20$. Using the approximations $|A_2|/A_0\ll 1$ and $|\epsilon|\ll 1$ above, and writing $\eta_{+-}=\epsilon+\epsilon'$ we obtain
\begin{equation}
\label{eq:epsprime}
\epsilon'\approx i\frac{\text{Im}A_2}{\sqrt2 A_0}e^{i(\delta_2-\delta_0)}.
\end{equation}
Moreover, $\eta_{00}=\epsilon-2\epsilon'$ so that 
\begin{equation}
\frac{\eta_{+-}}{\eta_{00}}-1\approx 3\frac{\epsilon'}{\epsilon}.
\end{equation}
Experimentally it is found that $\text{Re}\,\epsilon'/\epsilon=(1.66\pm0.23)\times10^{-3}$, so that indeed the approximations made are valid and CP-violation in the decay amplitudes is small. 

Incidentally, we can now show we were justified in assuming
$\text{Im}\Gamma_{12}\ll\text{Re}\Gamma_{12}$. From Eq.~\eqref{eq:GammafromTheory}
\[
\Gamma_{12}=(2\pi){\sum_n}'{}\delta(M-E_n)\langle{K^0}|H_w|n\rangle\,\langle n |H_w|\widebar{K}^0\rangle.
\]
Note that since $\text{Br}(K_S\to\pi\pi)$ is nearly 100\%, $\Gamma$ is dominated by $\pi\pi$ intermediate states. Since $A_0\gg|A_2|$, $\Gamma$ is dominated by $\pi\pi(I=0)$. In our convention $\text{Im}A_0=0$ and therefore $\text{Im}\Gamma_{12}$ does not get a contribution from  $\pi\pi(I=0)$.

\section{Sketch of SM accounting for $\epsilon$ and
  $\epsilon'/\epsilon$.}
\subsection{$\epsilon$.}
We would like to determine the SM prediction for  $\epsilon$. Rather
than giving a full computation from  first principles, we will use the
semi-empirical formula \eqref{eq:epislonEmpirical}. You may wonder, is
this a cheat, since we used data, {\it e.g.,}
$\Delta\Gamma\approx2\Delta M$,  to derive this formula? The point is
that the emphasis is
in accounting for CPV:
\begin{enumerate}[(i)]
\item In principle we do not need to use empirical input.
\item It is useful already this way. For example, as we will see this is often sufficient in constraining NP  that enters only through short distance effects. 
\end{enumerate}

So our task is to estimate $\text{Im} M_{12}$. From \eqref{eq:MfromTheory},
\begin{equation}
\label{eq:M12fromTheory}
M_{12}=\langle K^0|H_w|\widebar{K}^0\rangle+{\sum_n}\text{PP}\frac{\langle K^0|{H}_w|n\rangle\langle n|{H}_w|\widebar{K}^0\rangle}{m_{K^0}-E_n}
+\cdots \\
\end{equation}
We use this below. Beware this still has the peculiar non-relativistic
normalization of states. For the first term this is easy: replace the
hamiltonian density $\mathcal{H}_w$ for the Hamiltonian $H_w$ and let
the states be normalized a la Bjorken and Drell, as $\langle p'|
p\rangle=\frac{E}{m}\delta^{(3)}(p-p')$. One better: then divide by
$2m_K$ so the states are now relativistically normalized, $\langle p'|
p\rangle=2E\delta^{(3)}(p-p')$.

\paragraph{Short Distance {\it vs} long distance contributions.} So
what is $H_w=\int d^3x\,\mathcal{H}_w$? We could simply use $\mathcal{H}_w=g_2W^+_\mu
J^\mu+\text{h.c.}$, where $J^\mu$ is the charged hadronic (quark)
current (no need to include the neutral current coupling to $Z^\mu$ because it does not change flavor). But then we would need to go to high orders in the above
formula. Diagrammatically the $\Delta S=2$ process is from (to lowest
order):
\begin{equation}
\label{eq:fig:KKbarmix}
\hbox to 5.2cm{\hskip-0.9cm\vbox to 1.2cm {\begin{tikzpicture} 
\coordinate (A) at (0,0);
\coordinate (x) at (1,0);
\coordinate (y) at (0,-1);

\coordinate (v1) at ($(A)+(x)$);
\coordinate (v2) at  ($(v1)+(x)$);
\coordinate (v3) at  ($(v2)+(x)$);
\coordinate (v4) at ($(A)+(y)$);
\coordinate (v5) at  ($(v4)+(x)$);
\coordinate (v6) at  ($(v5)+(x)$);
\coordinate (v7) at  ($(v6)+(x)$);

\draw[particle] (A) -- node[above=0.1cm]{$s$} (v1);
\draw[particle] (v1) -- node[above]{$u,c,t$} (v2);
\draw[particle] (v2) -- node[above=0.1cm]{$d$} (v3);
\draw[particle] (v7) -- node[below=0.1cm]{$s$} (v6);
\draw[particle] (v6) -- node[below]{$u,c,t$} (v5);
\draw[particle] (v5)  -- node[below=0.0cm]{$d$} (v4);

\draw[photon] (v1) -- node[left]{$W$} (v5);
\draw[photon] (v2) -- node[right]{$W$} (v6);

\draw[fill] ($(A)+0.5*(y)$) ellipse (0.1cm and 0.5cm) ;
\draw[fill] ($(A)+0.5*(y)+3*(x)$) ellipse (0.1cm and 0.5cm) ;

\draw[line width=3pt] ($(A)+0.5*(y)$) --
+($-0.5*(x)$)node[left]{$\widebar K^0$} ;
\draw[line width=3pt] ($(A)+0.5*(y)+3*(x)$) -- +($0.5*(x)$)node[right]{${K}^0$} ;

\end{tikzpicture} 
}}
 \end{equation}
\vskip0.7cm
\noindent This is fourth order in $H_w$. But since $m_K\ll M_W$ we expect that Fermi theory, in which we replace
the $W$-propagator mediated interaction by a local 4-fermion vertex,
is a good approximation, {\it i.e.}
\begin{equation}
\label{eq:fig4vertex}
\hbox to 5.2cm{\hskip-0.9cm\vbox to 1.3cm {\begin{tikzpicture}[scale=0.7] 
\coordinate (A) at (0,0);
\coordinate (x) at (1,0);
\coordinate (y) at (0,1);

\coordinate (v1) at ($(A)-(x)+(y)$);
\coordinate (v2) at  ($(A)+(x)+(y)$);
\coordinate (v3) at  ($(A)-(x)-(y)$); 
\coordinate (v4) at ($(A)+(x)-(y)$);
\coordinate (v12) at ($0.5*(v1)+0.5*(v2)$);
\coordinate (v34) at ($0.5*(v3)+0.5*(v4)$);

\draw[particle] (v1) node[left]{$s$} -- (v12);
\draw[particle] (v12) -- (v2) node[right]{$u$};
\draw[particle] (v4) node[right]{$u$} -- (v34);
\draw[particle] (v34) -- (v3) node[left]{$d$};

\draw[photon] (v12) -- node[right]{$W$} (v34);

\begin{scope}[xshift = 2cm]
\node{$\to$} (A);
\end{scope}

\begin{scope}[xshift = 4cm]
\coordinate (A) at (0,0);
\coordinate (x) at (1,0);
\coordinate (y) at (0,1);

\coordinate (v1) at ($(A)-(x)+(y)$);
\coordinate (v2) at  ($(A)+(x)+(y)$);
\coordinate (v3) at  ($(A)-(x)-(y)$); 
\coordinate (v4) at ($(A)+(x)-(y)$);
\draw[particle] (v1) node[left]{$s$} -- (A);
\draw[particle] (A) -- (v2) node[right]{$u$};
\draw[particle] (v4) node[right]{$u$} -- (A);
\draw[particle] (A) -- (v3) node[left]{$d$};
\draw[fill] (A) circle (0.1cm);
\end{scope}

\end{tikzpicture} 
}}
\end{equation}
\vskip0.5cm
\noindent That is, when the momentum through the $W$ propagator is negligible
compared to $M_W$
\begin{multline*}
\left(-i\frac{g_2}{\sqrt2}V^*_{ud}\widebar{d}\gamma^\mu P_L u\right)
\left (-i\frac{g_{\mu\nu}-q_\mu q_\nu/M_W^2}{q^2-M_W^2}\right)
\left(-i\frac{g_2}{\sqrt2}V_{us}\widebar{u}\gamma^\nu P_L s\right)\\
\xrightarrow[q\to0]{}
-i\frac{g_2^2}{2M_W^2}V^*_{ud}V^{\phantom{*}}_{us}\widebar{d}\gamma^\mu
P_L u\,
\widebar{u}\gamma_\mu P_L s
\end{multline*}
This corresponds to an effective Hamiltonian density\footnote{Note: since $M_W^2=\frac14 g_2^2 v^2$, the coefficient of the effective
Hamiltonian density is proportional to  $1/v^2$. This is also Fermi's constant. More
precisely, $G_F$ is defined by
$\mathcal{H}_w=G_F/\sqrt2(V-A)\otimes(V-A)$, where $V-A$ stands for a
``vector minus axial'' current, as in
$\widebar{u}\gamma^\mu(1-\gamma_5) d$. Hence $G_F/\sqrt2=1/(2v^2)$,
and the well known value $G_F=1.17\times10^{-5}\text{GeV}^{-2}$ gives
the value of the VEV in EW theory, $v=1/\sqrt{2G_F}=246\,\text{GeV}$.}
\[
\mathcal{H}^{\Delta S=1}_w\approx 
\frac{g_2^2}{2M_W^2}V^*_{ud}V^{\phantom{*}}_{us}\widebar{d}_L\gamma^\mu u_L\,
\widebar{u}_L\gamma_\mu s_L,
\]
and similarly for other terms (with varying external quarks).  

One advantage of using this effective Hamiltonian\footnote{We get
  tired of saying ``effective Hamiltonian density'' so it is standard
  practice to omit ``density.''}  is that for the
$\Delta S=2$ transitions, the formulae for $M_{12}$ and $\Gamma_{12}$,
we first get a contribution at second order, rather than fourth, in
the expansion (that is, order $G_F^2$). However
\begin{enumerate}[(i)]
\item the evaluation of
the second order term, the one with the PP in
\eqref{eq:M12fromTheory}, is cumbersome, and
\item we have not exploited the fact that $m_K\ll m_t$.
\end{enumerate}
In fact, we could also consider the charm quark as so heavy that an
approximation based on $m_K\ll m_c$ is useful. 
All these comments are related: if we can approximate
\begin{center}
\begin{tikzpicture} 
\coordinate (A) at (0,0);
\coordinate (x) at (1,0);
\coordinate (y) at (0,-1);

\coordinate (v1) at ($(A)+(x)$);
\coordinate (v2) at  ($(v1)+(x)$);
\coordinate (v3) at  ($(v2)+(x)$);
\coordinate (v4) at ($(A)+(y)$);
\coordinate (v5) at  ($(v4)+(x)$);
\coordinate (v6) at  ($(v5)+(x)$);
\coordinate (v7) at  ($(v6)+(x)$);

\draw[particle] (A) -- node[above=0.1cm]{$s$} (v1);
\draw[particle] (v1) -- node[above]{$c,t$} (v2);
\draw[particle] (v2) -- node[above=0.1cm]{$d$} (v3);
\draw[particle] (v7) -- node[below=0.1cm]{$s$} (v6);
\draw[particle] (v6) -- node[below]{$c,t$} (v5);
\draw[particle] (v5)  -- node[below=0.0cm]{$d$} (v4);

\draw[photon] (v1) -- node[left]{$W$} (v5);
\draw[photon] (v2) -- node[right]{$W$} (v6);

\begin{scope}[xshift = 4cm,yshift=-0.5cm]
\node{$\to$} (A);
\end{scope}

\begin{scope}[xshift = 6cm, yshift=-0.5cm, scale=0.6]
\coordinate (A) at (0,0);
\coordinate (x) at (1,0);
\coordinate (y) at (0,1);

\coordinate (v1) at ($(A)-(x)+(y)$);
\coordinate (v2) at  ($(A)+(x)+(y)$);
\coordinate (v3) at  ($(A)-(x)-(y)$); 
\coordinate (v4) at ($(A)+(x)-(y)$);
\draw[particle] (v1) node[left]{$s$} -- (A);
\draw[particle] (A) -- (v2) node[right]{$d$};
\draw[particle] (v4) node[right]{$s$} -- (A);
\draw[particle] (A) -- (v3) node[left]{$d$};
\draw[fill] (A) circle (0.1cm);
\end{scope}

\end{tikzpicture} 
 \end{center}
\noindent then we can we can insert the local vertex into the first
order term in the expression for $M_{12}$,
Eq.~\eqref{eq:M12fromTheory}. We call these ``short distance
contributions.'' This is because $\Delta t\sim\hbar /\Delta E\sim
\hbar/(M_X-m_K)$, for $X=$ the least of $M_W, m_c$ and $m_t$, is
much smaller than any time scale associated with the dynamics of
strong interactions. This is, of course, much better for the case of
top-quarks than for charm-quarks, but even for these, the
approximation is pretty good and particles with charm cannot appear as
on-shell intermediate states in the sum in \eqref{eq:M12fromTheory}. 

So we have split the calculation into two pieces, a short distance
contribution, evaluated from the first order term in
\eqref{eq:M12fromTheory} and containing the diagrams where both
internal quarks are heavy ($c$ or $t$), and a ``long distance
contribution,''  evaluated from the sum over states in
\eqref{eq:M12fromTheory}. The latter can involve states that propagate
over long times when their energy is close to $m_K$ (near the energy
pole), {\it e.g.}, for $|n\rangle= |\pi\pi\rangle, |\pi\pi\pi\rangle,\ldots$ 
 
Long distance contributions are difficult to compute. But generally
contributions from NP to the long distance terms are negligible. So we
can happily extract the long distance contributions from data (using $\Delta
M\approx 2\text{Re}M_{12}$) and concentrate on
computing the short distance contributions, where NP may more readily
show up. Note that this is consistent with the semi-empirical approach
adopted above in writing formulae for $\epsilon $ and
$\epsilon'/\epsilon$. Moreover, in some cases the long distance
contributions can be calculated through Monte Carlo simulations of
QCD on the lattice.\footnote{See, {\it e.g.}, Ref.~\cite{Christ:2012se}.} So
from here on we concentrate on the short distance
contribution. Roughly,
\[
\text{Im}M_{12}\approx\text{Im}\left(
\hbox to 5.2cm{\hskip-0.9cm\vbox to 1.3cm {\begin{tikzpicture} 
\coordinate (A) at (0,0);
\coordinate (x) at (1,0);
\coordinate (y) at (0,-1);

\coordinate (v1) at ($(A)+(x)$);
\coordinate (v2) at  ($(v1)+(x)$);
\coordinate (v3) at  ($(v2)+(x)$);
\coordinate (v4) at ($(A)+(y)$);
\coordinate (v5) at  ($(v4)+(x)$);
\coordinate (v6) at  ($(v5)+(x)$);
\coordinate (v7) at  ($(v6)+(x)$);

\draw[particle] (A) -- node[above=0.1cm]{$s$} (v1);
\draw[particle] (v1) -- node[above]{$u,c,t$} (v2);
\draw[particle] (v2) -- node[above=0.1cm]{$d$} (v3);
\draw[particle] (v7) -- node[below=0.1cm]{$s$} (v6);
\draw[particle] (v6) -- node[below]{$u,c,t$} (v5);
\draw[particle] (v5)  -- node[below=0.0cm]{$d$} (v4);

\draw[photon] (v1) -- node[left]{$W$} (v5);
\draw[photon] (v2) -- node[right]{$W$} (v6);

\draw[fill] ($(A)+0.5*(y)$) ellipse (0.1cm and 0.5cm) ;
\draw[fill] ($(A)+0.5*(y)+3*(x)$) ellipse (0.1cm and 0.5cm) ;

\draw[line width=3pt] ($(A)+0.5*(y)$) -- +($-0.5*(x)$)node[left]{$\widebar{K}^0$} ;
\draw[line width=3pt] ($(A)+0.5*(y)+3*(x)$) -- +($0.5*(x)$)node[right]{$K^0$} ;

\end{tikzpicture} 
}}
\right)\sim\hskip4cm
\]
\[
\text{Im}\left[\frac{G_F^2M_W^2}{4\pi^2}\sum_{q,q'=u,c,t}V^*_{qd}V^{\phantom{*}}_{qs}V^*_{q'd}V^{\phantom{*}}_{q's}\,f\!\left(m_q,m_{q'}\right)
\langle K^0|\widebar{d}_L\gamma^\mu s_L\,\widebar{d}_L\gamma_\mu s_L|\widebar{K}^0\rangle\right]
\]
Here $f$ is a dimensionless function that is computed from a Feynman integral of the
box diagram and depends on $M_W$ implicitly. Note that the diagram has a double GIM, one per quark
line. In the second line above, the non-zero imaginary part is from the phase in the
KM-matrix. In the standard parametrization $V_{ud}$ and $V_{us}$ are
real, so we need at least one heavy quark in the Feynman diagram to
get a non-zero imaginary part. I will explain later why the diagram
with one $u$ quark and one heavy, $c$ or $t$, quark is suppressed. We
are left with $c$ and $t$ contributions only. Notice also that
KM-unitarity gives $\sum_q V^*_{qd}V^{\phantom{*}}_{qs}=0$, and since
$\text{Im}V^*_{ud}V^{\phantom{*}}_{us}=0$, we have a single common
coefficient,
$\text{Im}V^*_{cd}V^{\phantom{*}}_{cs}=-\text{Im}V^*_{td}V^{\phantom{*}}_{ts}=A^2\lambda^5\eta$
in terms of the Wolfenstein parametrization. For later use we define
$\lambda_q\equiv V^*_{qd}V_{qs\phantom{d}\!\!\!}^{\phantom{\dagger}}$, and as we just
saw $\text{Im}\lambda_c=-\text{Im}\lambda_t$. We will also need $\text{Re}\lambda_c\approx-\lambda$ and $\text{Re}\lambda_t\approx-A^2\lambda^5(1-\rho)$. 

The matrix element of the four-quark operator between kaon states
requires understanding of non-trivial hadrodynamics. We parametrize
our ignorance as follows,
\begin{equation}
\label{eq:BKdefd}
\langle K^0|\widebar{d}_L\gamma^\mu s_L\,\widebar{d}_L\gamma_\mu s_L|\widebar{K}^0\rangle
=\tfrac23f_K^2m_K^2B_K
\end{equation}
where we have used the standard, relativistic normalization of states. Here the mass $m_K$ and the decay constant $f_K$ are known data, so
$B_K$ is the dimensionless parameter characterizing the value of  the
matrix element.  Monte Carlo simulations of QCD on the lattice suggest
$B_K\approx 0.77$~\cite{Laiho:2013wwa}. The choice of parametrization may seem peculiar. It
is motivated by the following exercise:
\begin{exercises}
\begin{exercise}
$B_K=1$ in the ``vacuum insertion approximation.'' This
consists of summing over all possible insertions of the vacuum,
$|0\rangle\langle0|$,  as an
intermediate state in the matrix element above, including Fierz
rearrangements. For $N_c$ colors show that
in the vacuum insertion approximation  one has 
$\langle K^0|\widebar{d}_L\gamma^\mu s_L\,\widebar{d}_L\gamma_\mu s_L|\widebar{K}^0\rangle
=\tfrac14(2+2/N_c)f_K^2m_K^2$. You will need $\langle 0|\widebar
d\gamma^\mu\gamma_5 s|\widebar{K}^0(p)\rangle = f_Kp^\mu$. 
\end{exercise}
\begin{exercise}
Compute the partial width for $K^+\to e^+\nu$ in terms of $f_K$. This explains why $f_K$ is called the ``kaon decay constant.''
\end{exercise}
\end{exercises}
Assembling all these factors we get
\begin{multline*}
\text{Im}M_{12}=-A^2\lambda^5\eta (\tfrac23B_Km_K^2f_K^2)\frac{G_F^2M_W^2}{4\pi^2}\frac1{2m_K}\\
\times\left[A^2\lambda^5(1-\rho)f(m_t,m_t)-\lambda\big(f(m_c,m_c)-f(m_c,m_t)\big)\right]
\end{multline*}
We have included a factor of $1/2m_K$ to revert to relativistic normalization of states. All that is left to do is to compute the function $f(x,y)$ by performing a Feynman integral. This takes some work, or you can find it in the literature~\cite{Inami:1980fz}, so I will only display the result, given in terms of the ratios $x_q=m_q^2/M_W^2$ using the approximation $x_c\ll 1$:
\begin{align*}
f(m_c,m_t) &= x_c\left[\ln\frac{x_c}{x_t}-\frac{3x_t}{4(1-x_t)}-\frac{3x_t^2\ln x_t}{4(1-x_t)^2}\right]\\
f(m_t,m_t) &= \frac{4x_t-11x_t^2+x_t^3}{4(1-x_t)^2}-\frac{3x_t^3\ln x_t}{2(1-x_t)^3}
\end{align*}

Our final result is (modulo short distance QCD corrections which we will return to below) is
\begin{equation}
\label{eq:epsPreQCDSD}
\epsilon\approx e^{i\pi/4}C_\epsilon A^2\lambda^5\eta\left[A^2\lambda^5(1-\rho)f(m_t,m_t)-\lambda\big(f(m_c,m_c)-f(m_c,m_t)\big)\right]
\end{equation}
where the fixed, uninteresting, data-driven constant is 
\[
C_\epsilon=\frac{G_F^2f_K^2m_KM_W^2B_K}{6\sqrt2\pi^2\Delta M_K}\approx 3\times 10^4.
\]
Let's check the order of magnitude of the contributions of the various terms. The overall coefficient includes a $A^2\lambda^5\eta\sim(0.2)^5\sim3\times10^{-4}$, and the terms in the square bracket in \eqref{eq:epsPreQCDSD} give
\begin{gather*}
A^2\lambda^5(1-\rho)f(m_t,m_t)\sim(0.2)^5\sim3\times10^{-4}\Rightarrow \epsilon \sim 3\times 10^{-3}\\
\lambda f(m_c,m_t)\sim\lambda f(m_c,m_c)\sim (0.2)\left(\frac{1.5}{80}\right)^2\sim 10^{-4}\Rightarrow \epsilon \sim 10^{-3}
\end{gather*}
The result is quite remarkable: all the terms give comparable contributions to $\epsilon$, and all of them are of the correct order of magnitude!

\begin{exercises}
\begin{exercise}
\begin{enumerate}[(i)]
\item Pretend you can compute $\text{Re}M_{12}$ by computing Feynman diagrams and therefore using $\mathcal{H}_W^{\Delta S=2}=\frac1{4\pi^2}G_F^2M_W^2(\cdots)(\widebar s_L \gamma^\mu d_L)(\widebar s_L \gamma_\mu d_L)$, so as to ignore the cumbersome $\sum_n' \text{PP}$. Estimate $\Delta M_K$. Compare with the experimental value. 
\item How does this change if you were to ignore $c,t$ quarks (so there is no GIM mechanism)?
\item Now pretend there are only two generations (ignore the $b$ and $t$ quarks). How large does $m_c$ have to be to account for $\Delta M_K$? Historically this computation is very important: it led to the {\sl prediction} of the existence of the charm quark and of its mass, and it is how the GIM mechanism was discovered~\cite{Glashow:1970gm}.
\end{enumerate}
\end{exercise}
\end{exercises}

\paragraph{Short distance QCD corrections: a precap.\footnote{{\sl Precap} is defined in \protect\url{www.urbandictionary.com} as: Annoying pre-commercial preview of what's to come after the commercial break on the program you are already in the midst of watching. }}
Before we move on:  it may bother you that we made the replacement 
\begin{center}
\begin{tikzpicture} 
\coordinate (A) at (0,0);
\coordinate (x) at (1,0);
\coordinate (y) at (0,-1);

\coordinate (v1) at ($(A)+(x)$);
\coordinate (v2) at  ($(v1)+(x)$);
\coordinate (v3) at  ($(v2)+(x)$);
\coordinate (v4) at ($(A)+(y)$);
\coordinate (v5) at  ($(v4)+(x)$);
\coordinate (v6) at  ($(v5)+(x)$);
\coordinate (v7) at  ($(v6)+(x)$);

\draw[particle] (A) -- (v1);
\draw[particle] (v1) -- (v2);
\draw[particle] (v2) --  (v3);
\draw[particle] (v7) --  (v6);
\draw[particle] (v6) --   (v5);
\draw[particle] (v5)  -- (v4);

\draw[photon] (v1) --  (v5);
\draw[photon] (v2) --  (v6);

\begin{scope}[xshift = 4cm,yshift=-0.5cm]
\node{$\to$} (A);
\end{scope}

\begin{scope}[xshift = 6cm, yshift=-0.5cm, scale=0.6]
\coordinate (A) at (0,0);
\coordinate (x) at (1,0);
\coordinate (y) at (0,1);

\coordinate (v1) at ($(A)-(x)+(y)$);
\coordinate (v2) at  ($(A)+(x)+(y)$);
\coordinate (v3) at  ($(A)-(x)-(y)$); 
\coordinate (v4) at ($(A)+(x)-(y)$);
\draw[particle] (v1) -- (A);
\draw[particle] (A) -- (v2);
\draw[particle] (v4)  -- (A);
\draw[particle] (A) -- (v3) ;
\draw[fill] (A) circle (0.1cm);
\end{scope}

\end{tikzpicture} 
 \end{center}
We have expanded in powers of $G_F$ (perturbation theory in the weak interactions) but kept strong interactions exact:  states are due to strong interactions and the computation of matrix elements is in the strongly interacting theory. But we understand strong interactions as described by QCD and we can easily see we have left some contributions out, for example:
\begin{center}
\begin{tikzpicture}[scale=1.3] 
\coordinate (A) at (0,0);
\coordinate (x) at (1,0);
\coordinate (y) at (0,-1);

\coordinate (v1) at ($(A)+(x)$);
\coordinate (v2) at  ($(v1)+(x)$);
\coordinate (v3) at  ($(v2)+(x)$);
\coordinate (v9) at  ($(v3)+(x)$);
\coordinate (v4) at ($(A)+(y)$);
\coordinate (v5) at  ($(v4)+(x)$);
\coordinate (v6) at  ($(v5)+(x)$);
\coordinate (v7) at  ($(v6)+(x)$);
\coordinate (v8) at  ($(v7)+(x)$);

\draw[particle] (A) -- (v1);
\draw[particle] (v1) -- (v2);
\draw[particle] (v2) --  (v3);
\draw[particle] (v3) --  (v9);
\draw[particle] (v8) --  (v7);
\draw[particle] (v7) --  (v6);
\draw[particle] (v6) --   (v5);
\draw[particle] (v5)  -- (v4);

\draw[photon] (v1) -- node[left]{$W$} (v5);
\draw[photon] (v3) -- node[right]{$W$} (v7);
\draw[gluon] (v2) -- node[right]{$g$} (v6);

\end{tikzpicture} 
\hskip1cm
\begin{tikzpicture}[scale=1.3] 
\coordinate (A) at (0,0);
\coordinate (x) at (1,0);
\coordinate (y) at (0,-1);

\coordinate (v1) at ($(A)+0.5*(x)$); 
\coordinate (v2) at  ($(v1)+0.5*(x)$);
\coordinate (v3) at  ($(v2)+0.5*(x)$); 
\coordinate (v9) at  ($(v3)+0.5*(x)$); 
\coordinate (v10) at  ($(v9)+(x)$); 

\coordinate (v4) at ($(A)+(y)$);
\coordinate (v5) at  ($(v4)+(x)$);
\coordinate (v6) at  ($(v5)+(x)$);
\coordinate (v7) at  ($(v6)+(x)$);

\draw[particle] (A) -- (v1);
\draw[particle] (v1) -- (v2);
\draw[particle] (v2) --  (v3);
\draw[particle] (v3) --  (v9);
\draw[particle] (v9) --  (v10);
\draw[particle] (v7) --  (v6);
\draw[particle] (v6) --   (v5);
\draw[particle] (v5)  -- (v4);

\draw[photon] (v2) -- node[left]{$W$} (v5);
\draw[photon] (v9) -- node[right]{$W$} (v6);
\draw[gluon] (v1) let \p1=($ (v3)-(v2)$) in arc  (180:0:{veclen(\x1,\y1)}) ;
\node at ($(v2)-0.8*(y)$) {$g$}; 

\end{tikzpicture} 
\begin{equation}
\label{emptyForFig}
\end{equation}
\begin{tikzpicture}[scale=1.3] 
\coordinate (A) at (0,0);
\coordinate (x) at (1,0);
\coordinate (y) at (0,-1);

\coordinate (v1) at ($(A)+0.5*(x)$); 
\coordinate (v2) at  ($(v1)+0.5*(x)$);
\coordinate (v3) at  ($(v2)+(x)$); 
\coordinate (v9) at  ($(v3)+0.5*(x)$); 
\coordinate (v10) at  ($(v9)+0.5*(x)$); 

\coordinate (v4) at ($(A)+(y)$);
\coordinate (v5) at  ($(v4)+0.5*(x)$);
\coordinate (v6) at  ($(v5)+2*(x)$);
\coordinate (v7) at  ($(v6)+0.5*(x)$);

\draw[particle] (A) -- (v1);
\draw[particle] (v1) -- (v2);
\draw[particle] (v2) --  (v3);
\draw[particle] (v3) --  (v9);
\draw[particle] (v9) --  (v10);
\draw[particle] (v7) --  (v6);
\draw[particle] (v6) --   (v5);
\draw[particle] (v5)  -- (v4);

\draw[photon] (v1) -- node[left]{$W$} (v5);
\draw[photon] (v9) -- node[right]{$W$} (v6);
\draw[gluon] (v2) let \p1=($0.5*(x)$) in arc  (180:0:{veclen(\x1,\y1)}) ;
\node at ($(v2)+0.5*(x)-0.8*(y)$) {$g$}; 

\end{tikzpicture} 

 \end{center}

 Notice however that graphs with gluons connecting external quarks are
 accounted for already, since they are included in the computation of
 states and matrix elements, {\it e.g.},
\begin{center}
\begin{tikzpicture}[scale=1.3] 
\coordinate (A) at (0,0);
\coordinate (x) at (1,0);
\coordinate (y) at (0,-1);

\coordinate (v1) at ($(A)+(x)$);
\coordinate (v2) at  ($(v1)+(x)$);
\coordinate (v3) at  ($(v2)+(x)$);
\coordinate (v9) at  ($(v3)+(x)$);
\coordinate (v4) at ($(A)+(y)$);
\coordinate (v5) at  ($(v4)+(x)$);
\coordinate (v6) at  ($(v5)+(x)$);
\coordinate (v7) at  ($(v6)+(x)$);
\coordinate (v8) at  ($(v7)+(x)$);

\draw[particle] (A) -- (v1);
\draw[particle] (v1) -- (v2);
\draw[particle] (v2) --  (v3);
\draw[particle] (v3) --  (v9);
\draw[particle] (v8) --  (v7);
\draw[particle] (v7) --  (v6);
\draw[particle] (v6) --   (v5);
\draw[particle] (v5)  -- (v4);

\draw[gluon] (v1) -- node[left]{$g$} (v5);
\draw[photon] (v3) -- node[right]{$W$} (v7);
\draw[photon] (v2) -- node[right]{$W$} (v6);

\node at ($0.5*(v9)+0.5*(v8)+0.7*(x)$) {$\rightarrow$};
\begin{scope}[xshift = 6cm, yshift=-0.5cm, scale=0.8]
\coordinate (A) at (0,0);
\coordinate (x) at (1,0);
\coordinate (y) at (0,1);

\coordinate (v1) at ($(A)-(x)+(y)$);
\coordinate (v11) at($0.4*(A)+0.6*(v1)$); 
\coordinate (v2) at  ($(A)+(x)+(y)$);
\coordinate (v3) at  ($(A)-(x)-(y)$); 
\coordinate (v31) at  ($0.4*(A)+0.6*(v3)$);
\coordinate (v4) at ($(A)+(x)-(y)$);
\draw[particle] (v1) -- (v11);
\draw[particle] (v11) -- (A);
\draw[particle] (A) -- (v2);
\draw[particle] (v4)  -- (A);
\draw[particle] (A) -- (v31) ;
\draw[particle] (v31) -- (v3) ;
\draw[fill] (A) circle (0.1cm);

\draw[gluon] (v31) -- node[left]{$g$} (v11);
\end{scope}

\end{tikzpicture} 
\vskip1cm
\begin{tikzpicture}[scale=1.3] 
\coordinate (A) at (0,0);
\coordinate (x) at (1,0);
\coordinate (y) at (0,-1);

\coordinate (v1) at ($(A)+0.5*(x)$); 
\coordinate (v2) at  ($(v1)+0.5*(x)$);
\coordinate (v3) at  ($(v2)+(x)$); 
\coordinate (v9) at  ($(v3)+0.5*(x)$); 
\coordinate (v10) at  ($(v9)+0.5*(x)$); 

\coordinate (v4) at ($(A)+(y)$);
\coordinate (v5) at  ($(v4)+(x)$);
\coordinate (v6) at  ($(v5)+(x)$);
\coordinate (v7) at  ($(v6)+(x)$);

\draw[particle] (A) -- (v1);
\draw[particle] (v1) -- (v2);
\draw[particle] (v2) --  (v3);
\draw[particle] (v3) --  (v9);
\draw[particle] (v9) --  (v10);
\draw[particle] (v7) --  (v6);
\draw[particle] (v6) --   (v5);
\draw[particle] (v5)  -- (v4);

\draw[photon] (v2) -- node[left]{$W$} (v5);
\draw[photon] (v3) -- node[right]{$W$} (v6);
\draw[gluon] (v1) let \p1=($0.5*(v1)-0.5*(v9)$) in arc  (180:0:{veclen(\x1,\y1)}) ;
\node at ($0.5*(v1)+0.5*(v9)-0.7*(y)$) {$g$}; 

\node at ($0.5*(v9)+0.5*(v8)+0.7*(x)$) {$\rightarrow$};

\begin{scope}[xshift = 6cm, yshift=-0.1cm, scale=0.8]
\coordinate (A) at (0,0);
\coordinate (x) at (1,0);
\coordinate (y) at (0,1);

\coordinate (v1) at ($(A)-(x)+(y)$);
\coordinate (v11) at($0.4*(A)+0.6*(v1)$); 
\coordinate (v2) at  ($(A)+(x)+(y)$);
\coordinate (v21) at  ($0.4*(A)+0.6*(v2)$);
\coordinate (v3) at  ($(A)-(x)-(y)$); 
\coordinate (v4) at ($(A)+(x)-(y)$);
\draw[particle] (v1) -- (v11);
\draw[particle] (v11) -- (A);
\draw[particle] (A) -- (v21);
\draw[particle] (v21) -- (v2);
\draw[particle] (v4)  -- (A);
\draw[particle] (A) -- (v3) ;
\draw[fill] (A) circle (0.1cm);

\draw[gluon] (v11) -- node[above=0.1cm]{$g$} (v21);
\end{scope}
\end{tikzpicture} 
 \end{center}

We will later return to this question and explain  how terms like
those in \eqref{emptyForFig}  give contributions that are
perturbatively computable, of the form $\frac{\alpha_s}{4\pi} \ln m_q/m_K$ or $\alpha_s/4\pi \ln M_W/m_K$ and
we  will find out how to use the renormalization group to sum the
leading logs, {\it i.e.}, the contributions from multi-loop diagrams
that contain $(\alpha_s/4\pi \ln m_q/\mu)^n$. This will also tell
us at what scale to compute $\alpha_s(\mu)$, hence removing this
uncertainty. Numerically the effect is to modify our result,
Eq.~\eqref{eq:epsPreQCDSD}, by some short distance correction factors
$\eta_{1,2,3}$, as follows:
\begin{equation}
\label{eq:epsPostQCDSD}
\epsilon\approx e^{i\pi/4}C_\epsilon A^2\lambda^5\eta\left[\eta_2A^2\lambda^5(1-\rho)f(m_t,m_t)-\lambda\big(\eta_1f(m_c,m_c)-\eta_3f(m_c,m_t)\big)\right]
\end{equation}
with $\eta_{1,2,3}\approx 1.4, 0.6, 0.5$~\cite{Gilman:1980di,Buchalla:1995vs}.

\subsection{ Direct CPV:  $\epsilon'/\epsilon$}
I'll be brief. Most concepts have been introduced. I will concentrate
on new features. 

Recall, from Eq.~\eqref{eq:epsprime}
\[\epsilon'\approx i\frac{\text{Im}A_2}{\sqrt2
  A_0}e^{i(\delta_2-\delta_0)}\qquad \text{($\text{Im}A_0=0$ basis).}
\]
We must look at $K\to\pi\pi$, that is, at $s\to d u \widebar{u}$. We
will need to include underlying transitions at the loop level, else we
will not obtain any CPV. At tree level
\vskip0.2cm
\begin{center}
\begin{tikzpicture}[scale=1.3] 
\coordinate (A) at (0,0);
\coordinate (x) at (1,0);
\coordinate (y) at (0,1);

\coordinate (v1) at ($(A)+0.5*(y)$); 
\coordinate (v2) at  ($(v1)+(x)$);
\coordinate (v3) at  ($(v2)+(x)+0.25*(y)$);
\coordinate (v4) at ($(v3)-0.5*(y)$);
\coordinate (v5) at  ($(A)+(x)$);
\coordinate (v6) at  ($(v4)-0.5*(y)$);
\coordinate (v7) at  ($(v6)-0.5*(y)$);
 
\coordinate (v8) at  ($(A)-0.5*(y)$);

\draw[particle] (v1) -- node[above]{$s$} (v2);
\draw[particle] (v2) --  (v3);
\draw[particle] (v4) --  (v5);
\draw[particle] (v5) --  (v6);
\draw[particle] (v7) .. controls ($(v7)-0.8*(x)+0.2*(y)$) and ($(v8)+(x)$)  ..  (v8);

\draw[photon] (v2) -- node[left]{$W$} (v5);

\draw[fill] ($(A)$) let \p1=($(v1)-(A)$) in ellipse ({0.2*veclen(\x1,\y1)} and {veclen(\x1,\y1)}) node[left=0.1cm]{$\widebar{K}^0$} ;
\draw[fill] ($0.5*(v3)+0.5*(v4)$) let \p1= ($0.5*(v3)-0.5*(v4)$) in ellipse ({0.2*veclen(\x1,\y1)} and {veclen(\x1,\y1)}) node[right]{$\pi$} ;
\draw[fill] ($0.5*(v6)+0.5*(v7)$) let \p1= ($0.5*(v6)-0.5*(v7)$) in ellipse ({0.2*veclen(\x1,\y1)} and {veclen(\x1,\y1)})  node[right]{$\pi$};
\begin{scope}[xshift=4.5cm]
\node at (0,0) {$+$ other tree level};
\end{scope}

\end{tikzpicture} 
\end{center}
and at 1-loop
\vskip0.2cm
\begin{center}
\begin{tikzpicture}[scale=1.3] 
\coordinate (A) at (0,0);
\coordinate (x) at (1,0);
\coordinate (y) at (0,1);

\coordinate (v1) at ($(A)+0.5*(y)$); 
\coordinate (v2) at  ($(v1)+0.5*(x)$);
\coordinate (v3) at  ($(v2)+0.5*(x)$);
\coordinate (v4) at ($(v3)+0.5*(x)$);
\coordinate (v5) at  ($(v4)+(x)+0.5*(y)$);
\coordinate (v6) at  ($(A)-0.5*(y)$);
\coordinate (v7) at  ($(v6)+(x)$);
 \coordinate (v8) at  ($(v7)+1.5*(x)-0.5*(y)$); 

\coordinate (v9) at ($0.7*(v5)+0.3*(v8)$);
\coordinate (v10) at ($0.3*(v5)+0.7*(v8)$);

\draw[particle] (v1) -- (v2);
\draw[photon] (v2) --node[above]{$W$}  (v4);
\draw[particle] (v4) --  (v5);
\draw[particle] (v6) --  (v7);
\draw[particle] (v7) --  (v8);

\draw[style={thick,draw=black, postaction={decorate},
    decoration={markings,mark=at position .55 with {\arrow[black]{triangle 45}}}}] (v9)  let \p1=($0.5*(v10)-0.5*(v9)$) in arc (90:270:{veclen(\x1,\y1)});

\draw[particle] (v2) let \p2=($(v3)-(v2)$) in arc (180:270:{veclen(\x2,\y2)});
\draw[style={thick,draw=black, postaction={decorate},
    decoration={markings,mark=at position .6 with {\arrow[black]{triangle 45}}}}] ($0.5*(v7)+0.5*(v3)$)  let \p1=($(v3)-(v2)$) in arc (270:360:{veclen(\x1,\y1)});
 
\draw[gluon] (v7) -- node[right=0.1cm]{$g$} ($0.5*(v7)+0.5*(v3)$);

\draw[fill] ($(A)$) let \p1=($(v1)-(A)$) in ellipse ({0.2*veclen(\x1,\y1)} and {veclen(\x1,\y1)}) node[left=0.1cm]{$\widebar{K}^0$} ;
\draw[fill] ($0.5*(v5)+0.5*(v9)$) let \p1= ($0.5*(v5)-0.5*(v9)$) in ellipse ({0.2*veclen(\x1,\y1)} and {veclen(\x1,\y1)}) node[right]{$\pi$} ;
\draw[fill] ($0.5*(v8)+0.5*(v10)$) let \p1= ($0.5*(v8)-0.5*(v10)$) in ellipse ({0.2*veclen(\x1,\y1)} and {veclen(\x1,\y1)})  node[right]{$\pi$};

\begin{scope}[xshift=4.5cm]
\node at (0,0) {$+$ other 1-loop level};
\end{scope}

\end{tikzpicture} 
\end{center}

Digression. This one loop contribution is called a ``penguin'' diagram. I do not know why. I have heard many stories. It was certainly first introduced in the context we are studying. Here is a penguin-like depiction of the diagram:

\begin{figure}[H]
\begin{center}
\includegraphics[width=0.4\textwidth]{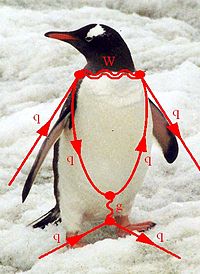}
\end{center}
\caption{Penguin Feynman diagram. }
\end{figure}
\noindent End digression.

Now since gluons have $I=0$, the only isospin change is through the $W$-loop in the figure, transmuting an $s$-quark to a $u$-quark. Hence, penguin diagrams  give 
$\Delta I=1/2$ but no $\Delta I=3/2$. Therefore, importantly
\begin{enumerate}[(a)]
\item Penguins may play a role in understanding the origin of the $\Delta I=1/2$ rule,
\item The computation will lead to a phase only in $K\to\pi\pi(I=0)$. But we chose $\text{Im} A_0=0$ by re-phasing (re-defining the state by a phase) $K^0$. So before re-phasing compute the phase of $A_0'$, say $\xi$. Then rotate $K^0$, $A_0=A_0' e^{-i\xi}$ and $A_2=A_2' e^{-i\xi}$ so that $A_0$ is real. 
\end{enumerate}
Then 
\begin{equation}
\label{eq:epsprimeexplicit}
\epsilon' = i\frac{\text{Im} e^{-i\xi}A_2'}{\sqrt2 A_0} e^{i(\delta_2-\delta_0)}=-i\sin\xi \frac{|A_2|}{\sqrt2 A_0}e^{i(\delta_2-\delta_0)} = -\frac{e^{i\pi/4}}{20\sqrt2}\xi
\end{equation}
where I have used $\xi\ll1$, the $\Delta I=1/2$ rule and the empirical value of  $ \delta_2-\delta_0$. 

Estimating $\epsilon'/\epsilon$ in the SM is more difficult than
computing $\epsilon$. The reason is a combination of two facts. First,
for $\epsilon'$
one needs to compute decay amplitudes, which involve three particles,
one in the initial and two in the final states, as opposed to for
$\epsilon$ which requires 
amplitudes with two particles in the matrix element for mixing. The
second is that there are several competing contributions and there are
some delicate numerical cancellations. I will therefore only say a few
words and refer the interested student to the slightly more complete
treatment in the textbook in Ref.~\cite{Donoghue:1992dd}. As we have
seen the penguin diagram gives only a contribution to $A_0$. However, if
in the penguin diagram we replace a photon or a $Z$ for the gluon the
resulting graph gives also a contribution to $A_2$. The reason is that
neither the photon nor the $Z$ couplings respect isospin, they
transform as a combination of $I=0$ and $I=1$. These digrams are
called ``electroweak penguins'' to distinguish them from the plain
vanilla penguins (sometimes called ``strong'' penguins, to emphasize
the distinction). Why bother? After all these digrams are suppressed
by $\alpha_{1,2}/\alpha_3$ relative the strong penguin. The point is
that the $\Delta I=1/2$ rule acts to amplify the direct contributions
to a phase in $A_2$. More specifically,  the phase $\xi$ in
\eqref{eq:epsprimeexplicit}  gets a contribution from
both $\Delta I=1/2$ and $\Delta I=3/2$ amplitudes,
\[
\xi=\frac{\text{Im}A_0}{\text{Re}A_0}-\frac{\text{Im}A_2}{\text{Re}A_2}\,.
\]
Furthermore, since there are cancellations, it turns out the effects
of isospin breaking by quark masses are non-negligible and have to be
included. A full account is beyond the scope of these lectures.

\section{Time Evolution in $X^0$-$\xoverline{X}^0$ mixing.}
We have looked at processes involving the `physical' states $K_L$ and $K_S$. As these are eigenvectors of $H$ their time evolution is quite simple 
\[
i\frac{d}{dt}|K_{L,S}\rangle = (M_{L,S}- \tfrac{i}2\Gamma_{L,S})|K_{L,S}\rangle\quad\Rightarrow\quad
|K_{L,S}(t)\rangle=e^{-iM_{L,S}t}e^{-\frac12\Gamma_{L,S}t}|K_{L,S}(0)\rangle  . 
\]
Since $|K_{L,S}\rangle $ are eigenvectors of $\mathbf{H}$, they do not mix as they evolve. But often one creates $K^0$ or $\widebar{K}^0$ in the lab. These, of course, mix with each other since they are linear combinations of $K_L$ and $K_S$. 

We'll analyze this in some generality so we may apply the results to $D^0$, $B^0$ and $B_s$ as well. The two mesons system  $X^0$-$\widebar{X}^0$ has effective `hamiltonian'
\[ \mathbf{H} =\mathbf{M}-\tfrac{i}2\mathbf{\Gamma}, \quad \mathbf{M}^\dagger=\mathbf{M}, \mathbf{\Gamma}^\dagger = \mathbf{\Gamma},
\]
and the physical eigenstates are labeled conventionally as Heavy and Light:
\begin{equation}
\label{eq:XHLofX0}
 |X_H\rangle = p|X^0\rangle + q|\widebar{X}^0\rangle, \qquad  |X_L\rangle = p|X^0\rangle - q|\widebar{X}^0\rangle 
\end{equation}
As before, 
\[
\frac{p}{q}=2\frac{M_{12}-\frac{i}2\Gamma_{12}}{\Delta M-\frac{i}2\Delta\Gamma}=
\frac12\frac{\Delta M-\frac{i}2\Delta\Gamma}{M^*_{12}-\frac{i}2\Gamma^*_{12}}
\]
where 
\[
M_{{H\atop L }}-\tfrac{i}2\Gamma_{{H\atop L}}=M-\tfrac{i}2\Gamma\pm\tfrac12(\Delta M-\tfrac{i}2\Delta\Gamma)
\]
The time evolution of $X_{{H\atop L }}$ is as above,
\[
|X_{{H\atop L }}(t)\rangle=e^{-iM_{{H\atop L }}t}e^{-\frac12\Gamma_{{H\atop L}}t}|X_{{H\atop L }}(0)\rangle  . 
\]
Now we can invert, 
\begin{align*}
|X^0\rangle&=\tfrac1{2p}\left(|X_H\rangle+|X_L\rangle\right),\\
|\widebar{X}^0\rangle&=\tfrac1{2q}\left(|X_H\rangle-|X_L\rangle\right).
\end{align*}
Hence, 
\[
|X^0(t)\rangle=\frac1{2p}\left[ e^{-iM_Ht}e^{-\frac12\Gamma_Ht}|X_H (0)\rangle  +
e^{-iM_Lt}e^{-\frac12\Gamma_Lt}|X_H (0)\rangle \right]
\]
and using \eqref{eq:XHLofX0} for the states at $t=0$ we obtain
\begin{equation}
|X^0(t)\rangle=f_+(t)|X^0\rangle+\tfrac{q}{p}f_-(t)|\widebar{X}^0\rangle
\end{equation}
where
\begin{equation}
\begin{aligned}
f_\pm(t)&=\tfrac12\left[e^{-iM_Ht}e^{-\frac12\Gamma_Ht}\pm e^{-iM_Lt}e^{-\frac12\Gamma_Lt}\right]\\
&=\tfrac12 e^{-iM_Ht}e^{-\frac12\Gamma_Ht}\left[1\pm e^{i\Delta Mt}e^{\frac12\Delta\Gamma t}\right]\\
&=\tfrac12 e^{-iM_Lt}e^{-\frac12\Gamma_Lt}\left[ e^{-i\Delta Mt}e^{-\frac12\Delta\Gamma t}\pm1\right]
\end{aligned}
\end{equation}
Similarly,
\begin{equation}
|\widebar{X}^0(t)\rangle=\tfrac{p}{q}f_-(t)|X^0\rangle+f_+(t)|\widebar{X}^0\rangle.
\end{equation}

\begin{figure}
\begin{center}
\includegraphics[width=4in]{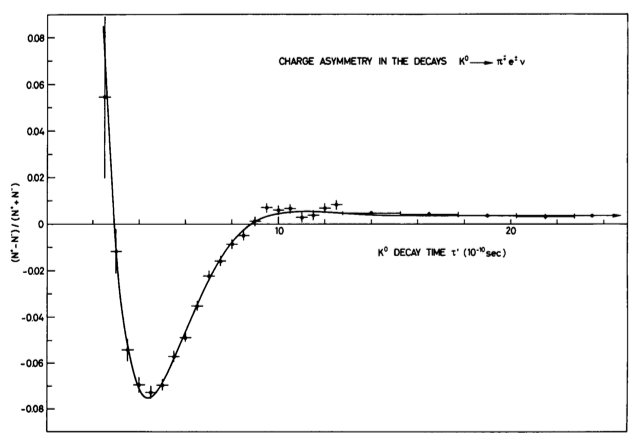}
\end{center}
\caption{\label{fig:deltaKl3} Charge asymmetry in semi-leptonic
  neutral kaon decays, from an experiment by Gjesdal {\it et al},~\cite{Gjesdal:1974yj}.
  The solid curve is a fit to the formula
  \eqref{eq:deltaKl3-timeDep} from which the parameters $\Gamma_S$,
  $\Delta M$, $a$ and $\text{Re}(\epsilon)$ are extracted. }
\end{figure}

\paragraph{Example: Time dependent asymmetry in semileptonic $K$ decay (``$K_{\ell3}$ decay'').}
This is the time dependent analogue of $\delta$ above. The experimental set-up is as follows:

\begin{tikzpicture}[scale=1.3] 
\coordinate (A) at (0,0);
\coordinate (x) at (1,0);
\coordinate (y) at (0,1);

\draw[->,>=stealth,thick] (A) --node[below]{$p$ beam} ++($3*(x)$);
\draw[fill] ($(A)+4*(x)$) ellipse (0.1cm and 0.6cm) node[below=0.7cm]{target};

\node at ($(A)+5.*(x)$) [rectangle, fill=black!50]{};
\node at ($(A)+5.*(x)$) [below=0.1cm]{``magic box''};

\draw[->,>=stealth,thick] ($(A)+6.5*(x)$) -- node[above]{monochromatic beam of $K^0$ and $\widebar{K}^0$} +($3*(x)$);

\draw[->,>=stealth,thick] ($(A)+7.2*(x)-0.1*(y)$) -- +($0.3*(x)-0.5*(y)$) node[below]{$e^-\pi^+\widebar\nu$};
\draw[->,>=stealth,thick] ($(A)+8.2*(x)-0.1*(y)$) -- +($0.3*(x)-0.5*(y)$) node[below]{$e^+\pi^-\nu$};

\node[rectangle,draw,fill=black!10] at  ($(A)+8*(x)-1.3*(y)$){detector array};
\end{tikzpicture} 

The proton beam hits a target, and the magic box produces a clean monochromatic beam of neutral $K$ mesons. These decay in flight and the semileptonic decays are registered in the detector array. Assume there are $N_K^0$ and $N_{\widebar{K}^0}$ $K^0$ and $\widebar{K}^0$'s from the beam, respectively. Measure
\[
\delta(t)=\frac{N^+-N_-}{N^++N_-}
\]
 as a function of distance from the beam (which can be translated into time from production at the magic box). Here $N^\pm$ refers to the total number of $K_{\ell3}$ events observed with charge $\pm$ lepton.  In reality ``$\pi^\pm$'' really stands for ``hadronic stuff'' since only the electrons are detected. We have then,
\begin{equation*}
\delta(t)=\frac{\begin{aligned}N_{K^0}\!\left[\Gamma(K^0(t)\to\pi^-e^+\nu)\right.&-\left.\Gamma(K^0(t)\to\pi^+e^-\widebar\nu)\right]\\
&+N_{\widebar{K}^0}\!\left[\Gamma(\widebar{K}^0(t)\to\pi^-e^+\nu)-\Gamma(\widebar{K}^0(t)\to\pi^+e^-\widebar\nu)\right]\end{aligned}}%
{\begin{aligned}N_{K^0}\!\left[\Gamma(K^0(t)\to\pi^-e^+\nu)\right.&+\left.\Gamma(K^0(t)\to\pi^+e^-\widebar\nu)\right]\\
&+N_{\widebar{K}^0}\!\left[\Gamma(\widebar{K}^0(t)\to\pi^-e^+\nu)+\Gamma(\widebar{K}^0(t)\to\pi^+e^-\widebar\nu)\right]\end{aligned}}
\end{equation*}

\begin{exercises}
\begin{exercise}
Use $\Gamma(K^0(t)\to\pi^-e^+\nu)\propto|\langle
\pi^-e^+\nu|H_W|K^0(t)\rangle|^2$ and the assumptions that
\begin{enumerate}[(i)]
\item $\langle
\pi^-e^+\nu|H_W|\widebar{K}^0(t)\rangle=0=\langle
\pi^+e^-\nu|H_W|{K}^0(t)\rangle$
\item $\langle\pi^-e^+\nu|H_W|K^0(t)\rangle=\langle
\pi^+e^-\nu|H_W|\widebar{K}^0(t)\rangle$
\end{enumerate}
to show that
\[
\delta(t)=\frac{(N_{K^0}-N_{\widebar{K}^0})\left[|f_+(t)|^2-|f_-(t)|^2\tfrac12\left(\left|\frac{q}{p}\right|^2+
      \left|\frac{p}{q}\right|^2\right)\right] 
+\tfrac12 (N_{K^0}+N_{\widebar{K}^0})|f_-(t)|^2\left(\left|\frac{p}{q}\right|^2-
                         \left|\frac{q}{p}\right|^2\right)}%
{(N_{K^0}+N_{\widebar{K}^0})\left[|f_+(t)|^2+|f_-(t)|^2\tfrac12\left(\left|\frac{q}{p}\right|^2+
      \left|\frac{p}{q}\right|^2\right)\right] 
-\tfrac12 (N_{K^0}-N_{\widebar{K}^0})|f_-(t)|^2\left(\left|\frac{p}{q}\right|^2-
                         \left|\frac{q}{p}\right|^2\right)}
\]
Justify assumptions (i) and (ii).
\end{exercise}
\end{exercises}
The formula in the exercise is valid for any $X^0$-$\widebar{X}^0$
system. We can simplify further for kaons, using
$p/q=(1+\epsilon)/(1-\epsilon)$, $a\equiv
(N_{K^0}-N_{\widebar{K}^0})/(N_{K^0}+N_{\widebar{K}^0})$ and
$\Delta\Gamma\approx-\Gamma_S$. Then 
\begin{align}
\delta(t)&=
\frac{a\left[|f_+(t)|^2-|f_-(t)|^2\right]  +4\text{Re}(\epsilon) |f_-(t)|^2}%
{\left[|f_+(t)|^2+|f_-(t)|^2\right] -4a\text{Re}(\epsilon) |f_-(t)|^2}\nonumber\\
\label{eq:deltaKl3-timeDep}
&\approx\frac{2ae^{-\frac12\Gamma_St}\cos(\Delta  Mt)
+\big(1+e^{-\Gamma_St}-2e^{-\frac12\Gamma_St}\cos(\Delta Mt)\big)2\left(1+\tfrac{a}2\right)\text{Re}(\epsilon)}%
{1+e^{-\Gamma_St}}
\end{align}

Figure~\ref{fig:deltaKl3}  shows the experimental measurement of the
asymmetry~\cite{Gjesdal:1974yj}. The solid curve is a fit to the formula
  \eqref{eq:deltaKl3-timeDep} from which the parameters $\Gamma_S$,
  $\Delta M$, $a$ and $\text{Re}(\epsilon)$ are extracted. The fit to
  this figure gives $\Delta M_K=(0.5287\pm0.0040)\times10^{10}\,\text{s}^{-1}$. The current value, from the \href{http://pdg.lbl.gov}{PDG} is $\Delta M_K=(0.5293\pm 0.0009)\times10^{10}\,\text{s}^{-1}$.

\section{CP-Asymmetries: Interference of Mixing and Decay}
Very generically a CP-asymmetry $a$ is defined as
\[
a=\frac{\Gamma(\text{process})-\Gamma(\text{CP conjugate
    process})}{\Gamma(\text{process})+\Gamma(\text{CP conjugate
    process})}\equiv\frac{\Gamma-\widebar{\Gamma}}{\Gamma+\widebar{\Gamma}}.
\]
Under what conditions can this be non-zero? 
$\Gamma\sim|\langle\text{out}|\text{in}\rangle|^2$ and if $\langle\text{out}|\text{in}\rangle$
has definite transformation properties under  CP, {\it e.g.},
$\langle\text{out}|\text{in}\rangle\xrightarrow[CP]{}\pm\langle\text{out}|\text{in}\rangle$,
then $\widebar{\Gamma}=\Gamma$ and $a=0$. We can get a non-zero asymmetry from
interference between two amplitudes with opposite, definite 
 CP properties, one even and one odd under CP: 
$\langle\text{out}|\text{in}\rangle=\langle\text{out}|\text{in}\rangle_++\langle\text{out}|\text{in}\rangle_-
\to \langle\text{out}|\text{in}\rangle_+-\langle\text{out}|\text{in}\rangle_- $. Then 
\[
a=\frac{|\langle\text{out}|\text{in}\rangle_++\langle\text{out}|\text{in}\rangle_-|^2-|\langle\text{out}|\text{in}\rangle_+-\langle\text{out}|\text{in}\rangle_-|^2}{|\langle\text{out}|\text{in}\rangle_++\langle\text{out}|\text{in}\rangle_-|^2+|\langle\text{out}|\text{in}\rangle_+-\langle\text{out}|\text{in}\rangle_-|^2}
=\frac{2\text{Re}(\langle\text{out}|\text{in}\rangle_+\langle\text{out}|\text{in}\rangle_-^*)}%
{|\langle\text{out}|\text{in}\rangle_+|^2+|\langle\text{out}|\text{in}\rangle_-|^2}
\]

One way to get an interference is to have two ``paths'' from
$|\text{in}\rangle$  to $|\text{out}\rangle$. For example, consider an
asymmetry constructed from  $\Gamma=\Gamma(X^0\to
f)$ and $\widebar\Gamma=\Gamma(\widebar X^0\to\widebar f)$,
where $f$ stands for some final state and $\widebar f$ its CP
conjugate. Then $\Gamma$ may get contributions either from a direct
decay $X^0\to f$ or it may first oscillate into $\widebar X^0$ and
then decay $\widebar X^0\to f$. Note that this requires that both
$X^0$ and its antiparticle, $\overline X^0$, decay to the same common
state. Similarly for $\widebar \Gamma$ we may get contributions from
both $\widebar X^0\to \widebar f$  and the oscillation of
$\widebar X^0$ into $X^0$ followed by a decay into $\widebar f$. In
pictures,

\begin{tikzpicture}[scale=1.] 

\node (center) {};
\node (A)  [above= of center] {$X^0$};
\node (B)  [below=of center] {$\widebar X^0$};
\node (C)  [left=of center]{$X^0$} ;
\node (D) [right=of center]{$f$} ;

\draw[->,>=stealth,thick] (C) -- (A);
\draw[->,>=stealth,thick] (C) -- (B);\draw[->,>=stealth,thick] (A) -- (D);\draw[->,>=stealth,thick] (B) -- (D);

\begin{scope}[xshift=5cm]
\node (center) {};
\node (A)  [above= of center] {$X^0$};
\node (B)  [below=of center] {$\widebar X^0$};
\node (C)  [left=of center]{$\widebar X^0$} ;
\node (D) [right=of center]{$\widebar f$} ;

\draw[->,>=stealth,thick] (C) -- (A);
\draw[->,>=stealth,thick] (C) -- (B);\draw[->,>=stealth,thick] (A) -- (D);\draw[->,>=stealth,thick] (B) -- (D);

\end{scope}
\end{tikzpicture}

Concretely,
\begin{align*}
\Gamma(X^0(t)\to f)&\propto |f_+(t)\langle f|H_w|X^0\rangle
+f_-(t)\tfrac{q}{p}\langle f|H_w|\widebar X^0\rangle|^2\\
&\equiv|f_+(t)A_f+f_-(t)\tfrac{q}{p}\widebar A_f|^2\\
\Gamma(\widebar X^0(t)\to \widebar f)&\propto
|f_-(t)\tfrac{p}{q}\langle \widebar f|H_w|X^0\rangle
+f_+(t)\langle \widebar f|H_w|\widebar X^0\rangle|^2\\
&\equiv|\tfrac{p}{q}f_-(t)A_{\overline f}+f_+(t)\widebar
A_{\overline f}|^2
\end{align*}
I hope the notation, which is pretty standard, is not just
self-explanatory, but fairly explicit. The bar over an amplitude $A$ refers to the
decaying state being $\widebar X^0$, while the decay product is
explicitly given by the subscript, {\it e.g.},  $\widebar
A_{\overline f} = \langle \widebar f|H_w|\widebar X^0\rangle$. 

\begin{exercises}
\begin{exercise}
\label{ex:CPTeigenstates}
If $f$ is an eigenstate of the strong interactions, show that CPT
implies $|A_f|^2=|\widebar A_{\overline f}|^2$ and $|A_{\overline f} |^2=|\widebar A_{ f}|^2$ 
\end{exercise}
\end{exercises}

The time dependent asymmetry is 
\[
a(t)=\frac{\Gamma(X^0(t)\to f)-\Gamma(\widebar X^0(t)\to \widebar f)}{\Gamma(X^0(t)\to f)+\Gamma(\widebar X^0(t)\to \widebar f)}
\]
and the time integrated asymmetry is
\[
a=\frac{\Gamma(X^0\to f)-\Gamma(\widebar X^0\to \widebar
  f)}{\Gamma(X^0\to f)+\Gamma(\widebar X^0\to \widebar f)}
\]
where $\Gamma(X^0\to f)\equiv\int_0^\infty dt\,\Gamma(X^0(t)\to f)$,
and likewise for the CP conjugate. These are the generalizations of
the quantities we called $\delta$ and $\delta(t)$ we studied for
kaons. 

For the rest of this section we will make the approximation that $\Delta\Gamma$ is negligible. For the case of $B^0$, for example, $\Delta\Gamma/\Gamma\sim10^{-2}$, while for $B_s$ the ratio is about 10\%. This simplifies matters because in this approximation
\[
f_\pm(t)\approx e^{-i\Delta M t}e^{-\frac12\Gamma t}\begin{cases}\cos(\tfrac12\Delta M t)\\-i\sin(\tfrac12\Delta M t)\end{cases}
\]
If one further approximates $|q/p|=1$, which is the analog of $\text{Re}(\epsilon)\ll1$ for $K^0$, then one finds (see exercises below):
\begin{equation}
\label{eq:t-int-a}
a=\frac{\frac{\Delta M}{\Gamma}\text{Im}\left(\rho\frac{q}{p}-\widebar\rho\frac{p}{q}\right)}%
{2\left(1+\frac12\left(\frac{\Delta M}{\Gamma}\right)^2\right)+ 
\frac{\Delta M}{\Gamma}\text{Im}\left(\rho\frac{q}{p}+\widebar\rho\frac{p}{q}\right)
+\left(\frac{\Delta M}{\Gamma}\right)^2|\rho|^2}
\end{equation}
where\footnote{It is unfortunate that the standard notation for
  $\widebar A_f/A_f$ uses the same symbol, $\rho$, as the parameter of
  the Wolfenstein parametrization of the CKM matrix elements. It will
  hopefully be clear from the context which one we are referring to.}
\[
\rho\equiv \frac{\widebar A_f}{A_f}\qquad\text{and}\qquad \widebar\rho\equiv \frac{\widebar A_{\overline f}}{A_{\overline f}}.
\]
This formula will be our workhorse. I will leave it up to the student, guided by specific exercises at the end of the section, to go through the same analysis in the time dependent case. In fact, the time dependent case is simpler to analyze\footnote{So the choice of presentation must seem non-pedagogical, but I did want to have the student have the opportunity to work out the time dependent case.}
but the central observations are the same in both time dependent and time independent asymmetries. We consider several special cases.

\paragraph{Case I: Self-conjugate final state.} Assume $\widebar
f=\pm f$. Such self-conjugate states are easy to come by. For example
$D^+D^-$ or, to good approximation, $J/\psi K_S$. Now, in this case we
have $A_{\widebar f}=\pm A_f$
 and $\widebar  A_{\overline f}=\pm \widebar A_f$, so that $\widebar\rho=1/\rho$. Since these final states are eigenstates of the strong interactions, using the result of exercise \ref{ex:CPTeigenstates} we have $|\widebar\rho|=|\rho|$ and therefore $|\rho|=1$. We already assumed  $|q/p|=1$, so we have $\text{Im} \rho\tfrac{q}{p}=-\text{Im}\widebar\rho\tfrac{p}{q}$. So one has
\[
a=\frac{\frac{\Delta M}{\Gamma}}{1+\left(\frac{\Delta M}{\Gamma}\right)^2}\text{Im}\left(\rho\frac{q}{p}\right)
\]
Here is what is amazing about this formula, for which Bigi and Sanda~\cite{Bigi:1981qs} were awarded the Sakurai Prize for Theoretical Particle Physics: the pre-factor (the stuff multiplying the ``Im'' part), depending only on $\Delta M/\Gamma$ can be determined from independent measurements (mixing and lifetime), and then what is left is independent of non-computable, non-perturbative corrections. The point is that what most often frustrates us in extracting fundamental parameters from experiment is our inability to calculate, that is, make a prediction that depends on the parameter to be measured. I now explain this claim.

The leading contributions to the processes $B^0\to f$ and $\widebar{B}^0\to f$ in the case $f=D^+D^-$ are shown in the following figures:
\begin{center}
\begin{tikzpicture}[scale=1.3] 
\coordinate (A) at (0,0);
\coordinate (x) at (1,0);
\coordinate (y) at (0,1);

\coordinate (v1) at ($(A)+0.5*(y)$); 
\coordinate (v2) at  ($(v1)+1.3*(x)$); 
\coordinate (v3) at  ($(v2)+(x)+0.25*(y)$); 
\coordinate (v4) at ($(v3)+(x)+0.5*(y)$); 
\coordinate (v5) at  ($(v3)+(x)-0.5*(y)$); 
\coordinate (v6) at  ($(A)-0.5*(y)$);
\coordinate (v7) at  ($(v6)+2.3*(x)-0.5*(y)$);
\coordinate (v8) at  ($(v7)+(y)$); 

\draw[particle] (v2) -- node[above]{$b$} (v1);
\draw[particle] (v3) -- node[above=0.1cm]{$c$} (v4);
\draw[particle] (v5) -- node[below]{$d$} (v3);
\draw[particle] (v8) -- node[above right]{$c$}  (v2);
\draw[particle] (v6) .. controls ($(v6)+(x)$) and ($(v7)-(x)+0.2*(y)$)  .. node[below]{$d$} (v7);

\draw[photon] (v2) -- node[above]{$W$} (v3);

\draw[fill] ($(A)$) let \p1=($(v1)-(A)$) in ellipse ({0.2*veclen(\x1,\y1)} and {veclen(\x1,\y1)}) node[left=0.1cm]{${B}^0$} ;
\draw[fill] ($0.5*(v5)+0.5*(v4)$) let \p1= ($0.5*(v5)-0.5*(v4)$) in ellipse ({0.2*veclen(\x1,\y1)} and {veclen(\x1,\y1)}) node[right=0.1cm]{$D^+$} ;
\draw[fill] ($0.5*(v8)+0.5*(v7)$) let \p1= ($0.5*(v8)-0.5*(v7)$) in ellipse ({0.2*veclen(\x1,\y1)} and {veclen(\x1,\y1)})  node[right=0.1cm]{$D^-$};
\begin{scope}[yshift=-1.5cm,xshift=1cm]
\node at (0,0) {$A_{D^+D^-}\propto V_{cb}^*V_{cd}$};
\end{scope}

\end{tikzpicture} 
\begin{tikzpicture}[scale=1.3] 
\coordinate (A) at (0,0);
\coordinate (x) at (1,0);
\coordinate (y) at (0,1);

\coordinate (v1) at ($(A)+0.5*(y)$); 
\coordinate (v2) at  ($(v1)+1.3*(x)$); 
\coordinate (v3) at  ($(v2)+(x)+0.25*(y)$); 
\coordinate (v4) at ($(v3)+(x)+0.5*(y)$); 
\coordinate (v5) at  ($(v3)+(x)-0.5*(y)$); 
\coordinate (v6) at  ($(A)-0.5*(y)$);
\coordinate (v7) at  ($(v6)+2.3*(x)-0.5*(y)$);
\coordinate (v8) at  ($(v7)+(y)$); 

\draw[particle] (v1) -- node[above]{$b$} (v2);
\draw[particle] (v4) -- node[above=0.1cm]{$c$} (v3);
\draw[particle] (v3) -- node[below]{$d$} (v5);
\draw[particle] (v2) -- node[above right]{$c$}  (v8);
\draw[particle] (v7) .. controls ($(v7)-(x)+0.2*(y)$)  and  ($(v6)+(x)$)   .. node[below]{$d$} (v6);

\draw[photon] (v2) -- node[above]{$W$} (v3);

\draw[fill] ($(A)$) let \p1=($(v1)-(A)$) in ellipse ({0.2*veclen(\x1,\y1)} and {veclen(\x1,\y1)}) node[left=0.1cm]{$\widebar{B}^0$} ;
\draw[fill] ($0.5*(v5)+0.5*(v4)$) let \p1= ($0.5*(v5)-0.5*(v4)$) in ellipse ({0.2*veclen(\x1,\y1)} and {veclen(\x1,\y1)}) node[right=0.1cm]{$D^-$} ;
\draw[fill] ($0.5*(v8)+0.5*(v7)$) let \p1= ($0.5*(v8)-0.5*(v7)$) in ellipse ({0.2*veclen(\x1,\y1)} and {veclen(\x1,\y1)})  node[right=0.1cm]{$D^+$};
\begin{scope}[yshift=-1.5cm,xshift=1cm]
\node at (0,0) {$\widebar{A}_{D^+D^-}\propto V_{cb}V_{cd}^*$};
\end{scope}

\end{tikzpicture} 
\end{center}
\noindent Either using CPT or noting that as far as the strong
interactions are concerned the two diagrams are identical, we have 
\[
\rho=\frac{V_{cb}^{\phantom{*}}V_{cd}^*}{V_{cb}^*V_{cd}^{\phantom{*}}}.
\]
Since $|\rho|=1$ we know it is a pure phase, and we see that the phase is given purely in terms of KM elements.

To complete the argument we need $q/p$. To this end we analyze mixing in the case of $B^0$ mesons. Now, the diagram that gives mixing is just like in the neutral kaon system:
\begin{center}
\begin{tikzpicture} 
\coordinate (A) at (0,0);
\coordinate (x) at (1,0);
\coordinate (y) at (0,-1);

\coordinate (v1) at ($(A)+(x)$);
\coordinate (v2) at  ($(v1)+(x)$);
\coordinate (v3) at  ($(v2)+(x)$);
\coordinate (v4) at ($(A)+(y)$);
\coordinate (v5) at  ($(v4)+(x)$);
\coordinate (v6) at  ($(v5)+(x)$);
\coordinate (v7) at  ($(v6)+(x)$);

\draw[particle] (A) -- node[above=0.1cm]{$b$} (v1);
\draw[particle] (v1) -- node[above]{$u,c,t$} (v2);
\draw[particle] (v2) -- node[above=0.1cm]{$d$} (v3);
\draw[particle] (v7) -- node[below=0.1cm]{$b$} (v6);
\draw[particle] (v6) -- node[below]{$u,c,t$} (v5);
\draw[particle] (v5)  -- node[below=0.0cm]{$d$} (v4);

\draw[photon] (v1) -- node[left]{$W$} (v5);
\draw[photon] (v2) -- node[right]{$W$} (v6);

\draw[fill] ($(A)+0.5*(y)$) ellipse (0.1cm and 0.5cm) ;
\draw[fill] ($(A)+0.5*(y)+3*(x)$) ellipse (0.1cm and 0.5cm) ;

\draw[line width=3pt] ($(A)+0.5*(y)$) -- +($-0.5*(x)$)node[left]{$\widebar{B}^0$} ;
\draw[line width=3pt] ($(A)+0.5*(y)+3*(x)$) -- +($0.5*(x)$)node[right]{$B^0$} ;

\end{tikzpicture} 
\end{center}
and just like in the $K^0$-$\widebar{K}^0$ case it involves a factor
$\sum_{q.q'} V_{qb}^*V_{qd}^{\phantom{*}}\,V_{q'b}^*V_{q'd}^{\phantom{*}}\,f(m_q,m_{q'})$. But
now $V_{qb}^*V_{qd}^{\phantom{*}}$ are sides of a fat triangle so that
$f(m_t,m_t)$ dominates the contributions of the virtual
quarks. Therefore $M_{12}$ is short distance dominated, $\Gamma_{12}$
is negligible (explaining why $\Delta\Gamma$ is negligible), and the
phase in $M_{12}$ is dominantly from $M_{12}\propto
(V_{tb}^*V_{td}^{\phantom{*}})^2$. Neglecting $\Gamma_{12}$ we have
then
\[
\frac{p}{q}=\frac{2M_{12}}{\Delta M}=\frac{\Delta
  M}{2M_{12}^*}=\frac{M_{12}}{|M_{12}|}
=\frac{V_{tb}^*V_{td}^{\phantom{*}}}{V_{tb}^{\phantom{*}}V_{td}^*}.
\]

\begin{figure}[H]
\begin{minipage}[c]{0.5\textwidth}\includegraphics[width=2.4in]{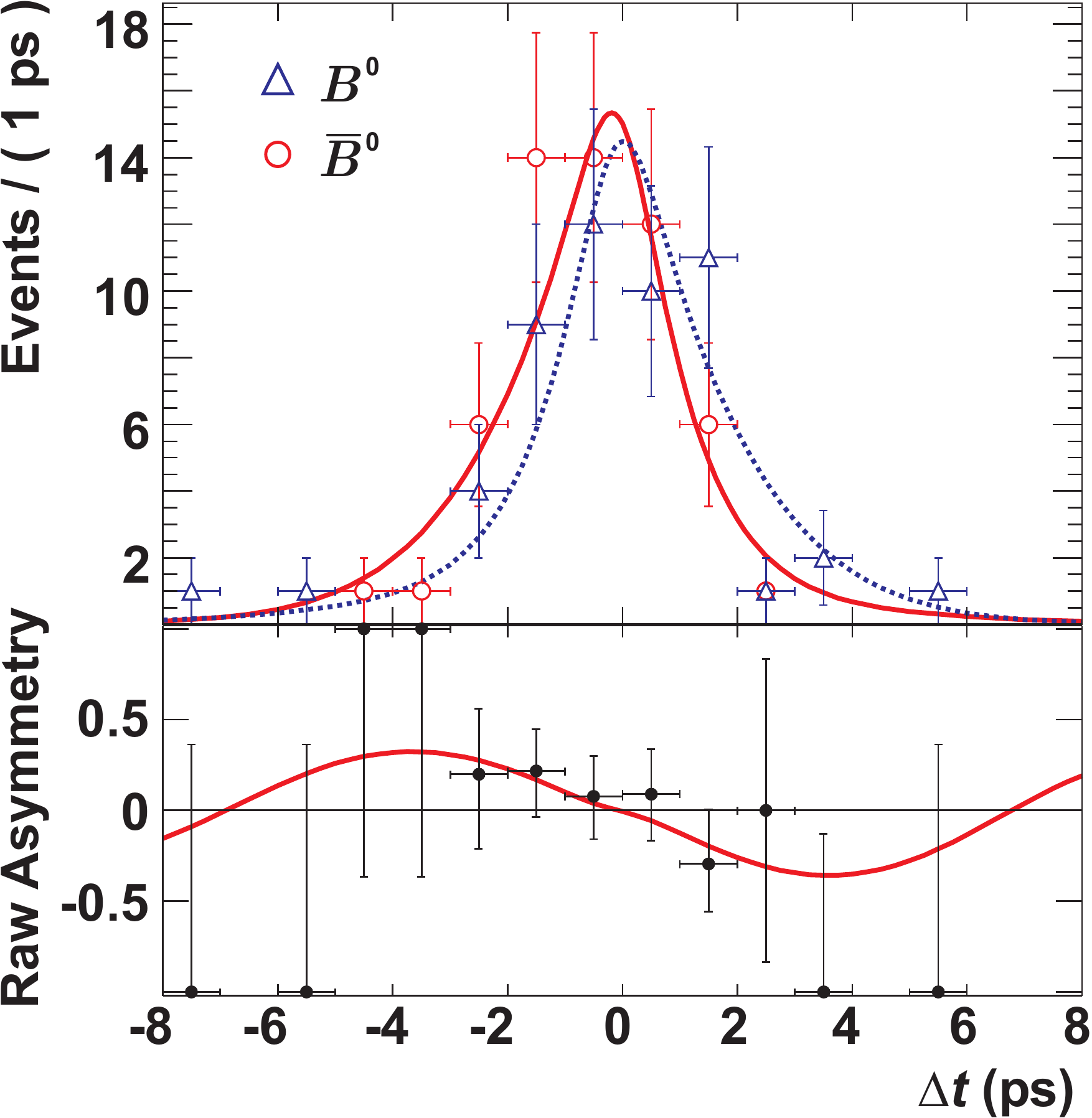}
\end{minipage}
\begin{minipage}[c]{0.3\textwidth}
\includegraphics[width=3in]{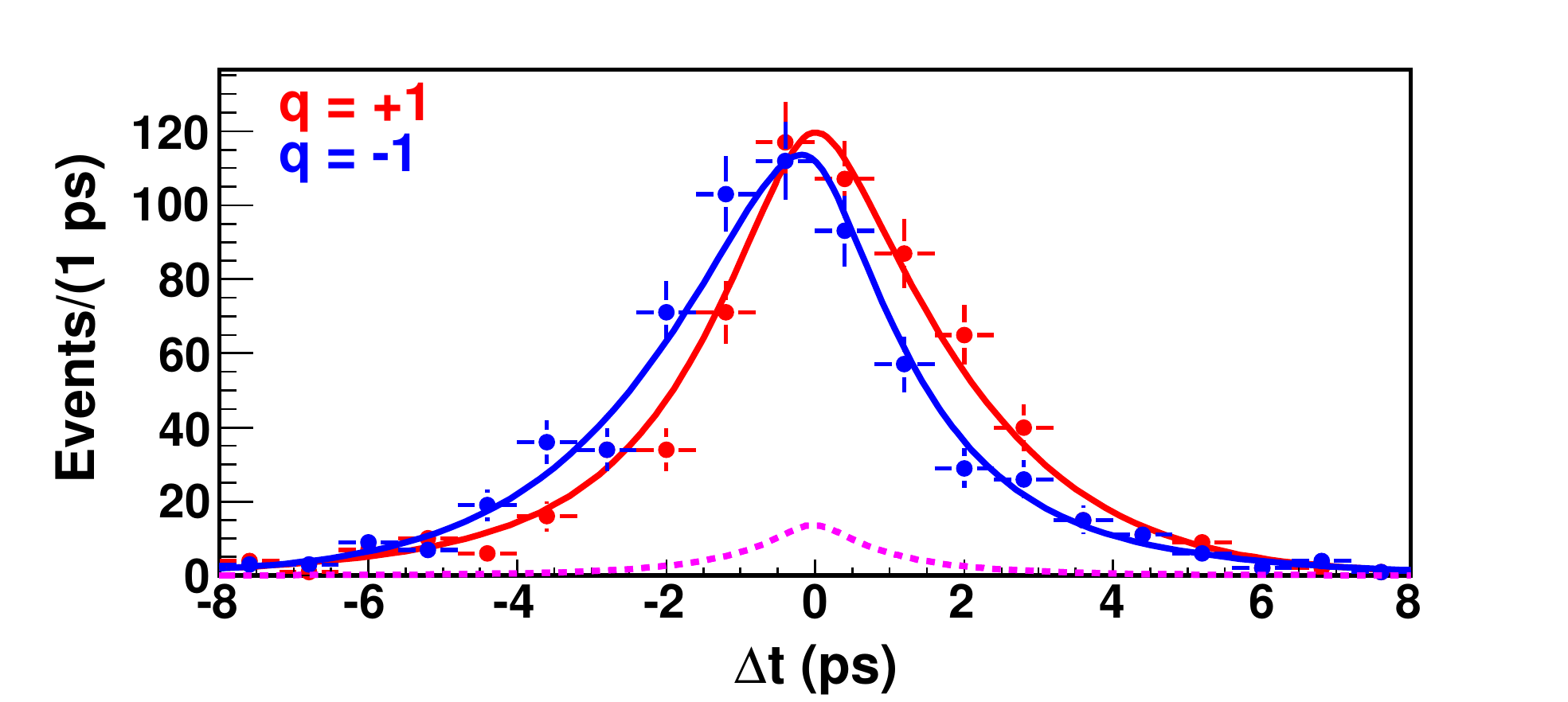}
\includegraphics[width=3in]{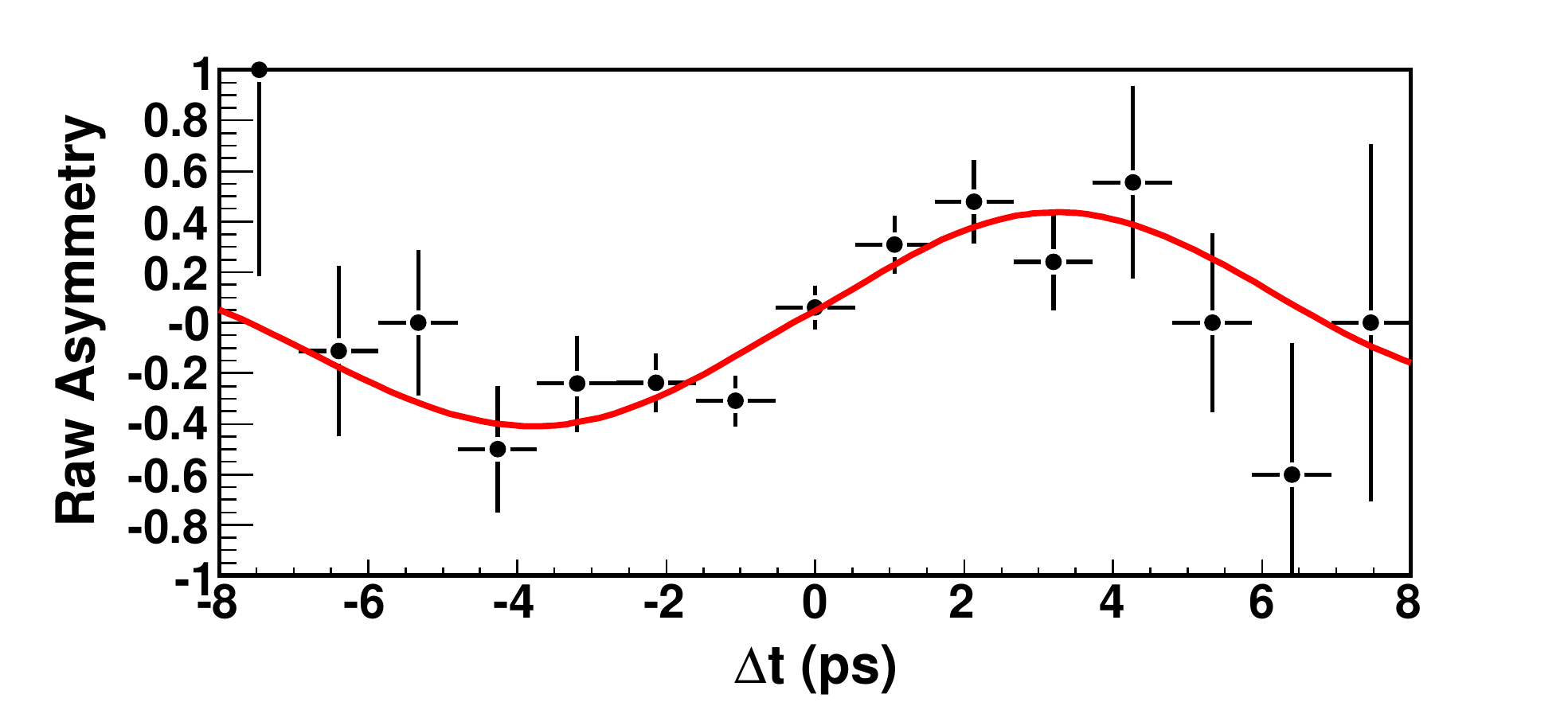}
\end{minipage}
\caption{\label{fig:expDD}  Experimental evidence for time
  dependent CP-asymmetry in $b\to c\protect\widebar c s$ decays. The left
  pannel shows results from BaBar for  $B^0(\protect\widebar B^0)\to
  D^+D^-$~\protect\cite{Aubert:2008ah} while the right pannel is Belle's
  $B^0(\protect\widebar B^0)\to \psi(2S) K_S$~\protect\cite{Abe:2007gj}.}
\end{figure}

Collecting results we have
\[
\text{Im}\left(\rho\frac{q}{p}\right) =
\text{Im}\left(\frac{V_{cb}^{\phantom{*}}V_{cd}^*}{V_{cb}^*V_{cd}^{\phantom{*}}}
\frac{V_{tb}^*V_{td}^{\phantom{*}}}{V_{tb}^{\phantom{*}}V_{td}^*}\right)
=\text{Im}(e^{2i\beta})=\sin(2\beta)
\]
and the asymmetry is $a=[(\Delta M/\Gamma)/(1+(\Delta
M/\Gamma)^2)]\sin(2\beta)$. Measurements of the asymmetry and the
mixing parameters give (twice the sine of) one of the angles of the unitarity triangle
without hadronic uncertainties!

\begin{exercises}

\begin{exercise}
Calculate the  time integrated asymmetry without assuming
$|q/p|=1$. Then take $|q/p|\to1$ to verify \eqref{eq:t-int-a}.
\end{exercise}

\begin{solution}
\[a = \frac{\frac12\left(\frac{\Delta M}{\Gamma}\right)^2\left(\left|\frac{q}{p}\right|^2-\left|\frac{p}{q}\right|^2\right)|\rho|^2
+\frac{\Delta M}{\Gamma}\text{Im}\left(\rho\frac{q}{p}-\widebar\rho\frac{p}{q}\right)}%
{2\left(1+\frac12\left(\frac{\Delta M}{\Gamma}\right)^2\right)+ 
\frac12\left(\frac{\Delta M}{\Gamma}\right)^2\left(\left|\frac{q}{p}\right|^2+\left|\frac{p}{q}\right|^2\right)|\rho|^2
+\frac{\Delta M}{\Gamma}\text{Im}\left(\rho\frac{q}{p}+\widebar\rho\frac{p}{q}\right)}\]
\end{solution}

\begin{exercise}
With the same assumptions as in the analysis above for the
time-integrated asymmetry ($|\rho|=|q/p|=1$), show that the
time-dependent asymmetry in $B^0\to D^+D^-$ is $a=\sin(\Delta M t)\sin(2\beta)$.
\end{exercise}

\begin{solution}
The numerator and denominator in $a(t)$ involve
\[f_++\tfrac{q}{p}\rho f_-|^2\pm \tfrac{p}{q}f_-+\rho f_+|^2\]
Since this is a ratio we can simply replace $c=\cos(\frac12 \Delta M
t)$ and $s=\sin(\frac12 \Delta M t)$ for $f_+$ and $if_-$,
respectively. So the line above is
\[(1+|\rho|^2)c^2+
\left(\left|\rho\frac{q}{p}\right|^2\pm\left|\frac{p}{q}\right|^2\right)s^2+2sc\text{Im}\left(
  \frac{q}{p}\rho\pm \frac{p}{q}\rho^*\right)\]
If $\rho$=|q/p|=1, then 
\[a=\frac{4sc \text{Im}\left(
  \frac{q}{p}\rho\right)}{2c^2+2s^2}=\sin(\Delta M t)\text{Im}\left(
  \frac{q}{p}\rho\right)\]
\end{solution}

\begin{exercise}
Above we assumed that $B^0\to D^+D^-$ (and $\widebar B^0\to D^+D^-$) is dominated by a single diagram, and from this we showed that $\rho$ is a pure CKM phase. 
\begin{enumerate}[(i)]
\item Exhibit other diagrams (other topologies) that contribute to
  this decay that have different CKM matrix  dependence than this leading, ``tree level'' one.
\item Show one can always write $A=e^{i\alpha}T + e^{i\beta}P$ where $\alpha$ and $\beta$ are weak phases (NOT necessarily  the phases in the unitarity triangle) and $T$ and $P$ are the rest, including the magnitude of the CKM and the strong interaction matrix elements. Moreover, show that $\pm \widebar A=e^{-i\alpha}T + e^{-i\beta}P$.
\item Assuming $ |P/T|\ll 1$ show that now
\begin{equation}
\label{eq:timedepA}
a(t)=\cos(\Delta M t) C + \sin(\Delta M t) S
\end{equation}
where $C= \mathcal{O}(|P/T|)$ and $S=\sin(2\beta)+\mathcal{O}(|P/T|)$. Exhibit these relations explicitly.
\item More generally, for any $\widebar f=\pm f$ write
  $\lambda_f=\frac{q}{p}\rho_f$ with $|q/p|=1$ to show that 
\[
C=\frac{1-|\lambda_f|^2}{1+|\lambda_f|^2}, \qquad  S=\frac{2\text{Im}\lambda_f}{1+|\lambda_f|^2}
\]
\end{enumerate}
Take a look at Fig.~\ref{fig:expDD} where the experimental measurement
of the rates for $B^0(\protect\widebar B^0)\to
  D^+D^-$~\cite{Aubert:2008ah} and 
  $B^0(\protect\widebar B^0)\to \psi(2S) K_S$~\cite{Abe:2007gj}
  display a clearly visible asymmetry, and  Eq.~\eqref{eq:timedepA} is
  a good fit to the time dependent
  asymmetry. Nice!
\end{exercise}
\begin{solution}
sol here
\end{solution}
\begin{exercise}
The most celebrated case is $B\to J/\psi K_S$. Here are the leading diagrams:
\begin{center}
\begin{tikzpicture}[scale=1.3] 
\coordinate (A) at (0,0);
\coordinate (x) at (1,0);
\coordinate (y) at (0,1);

\coordinate (v1) at ($(A)+0.5*(y)$); 
\coordinate (v2) at  ($(v1)+1.3*(x)$); 
\coordinate (v3) at  ($(v2)+(x)+0.25*(y)$); 
\coordinate (v4) at ($(v3)+(x)+0.5*(y)$); 
\coordinate (v5) at  ($(v3)+(x)-0.5*(y)$); 
\coordinate (v6) at  ($(A)-0.5*(y)$);
\coordinate (v7) at  ($(v6)+2.3*(x)-0.5*(y)$);
\coordinate (v8) at  ($(v7)+(y)$); 


\draw[particle] (v2) -- node[above]{$b$} (v1);
\draw[particle] (v3) -- node[above=0.1cm]{$c$} (v4);
\draw[particle] (v5) -- node[below]{$c$} (v3);
\draw[particle] (v8) -- node[above right]{$s$}  (v2);
\draw[particle] (v6) .. controls ($(v6)+(x)$) and ($(v7)-(x)+0.2*(y)$)  .. node[below]{$d$} (v7);

\draw[photon] (v2) -- node[above]{$W$} (v3);

\draw[fill] ($(A)$) let \p1=($(v1)-(A)$) in ellipse ({0.2*veclen(\x1,\y1)} and {veclen(\x1,\y1)}) node[left=0.1cm]{${B}^0$} ;
\draw[fill] ($0.5*(v5)+0.5*(v4)$) let \p1= ($0.5*(v5)-0.5*(v4)$) in ellipse ({0.2*veclen(\x1,\y1)} and {veclen(\x1,\y1)}) node[right=0.1cm]{$J/\psi$} ;
\draw[fill] ($0.5*(v8)+0.5*(v7)$) let \p1= ($0.5*(v8)-0.5*(v7)$) in ellipse ({0.2*veclen(\x1,\y1)} and {veclen(\x1,\y1)})  node[right=0.1cm]{$K^0(K_S)$};

\end{tikzpicture} 
\begin{tikzpicture}[scale=1.3] 
\coordinate (A) at (0,0);
\coordinate (x) at (1,0);
\coordinate (y) at (0,1);

\coordinate (v1) at ($(A)+0.5*(y)$); 
\coordinate (v2) at  ($(v1)+1.3*(x)$); 
\coordinate (v3) at  ($(v2)+(x)+0.25*(y)$); 
\coordinate (v4) at ($(v3)+(x)+0.5*(y)$); 
\coordinate (v5) at  ($(v3)+(x)-0.5*(y)$); 
\coordinate (v6) at  ($(A)-0.5*(y)$);
\coordinate (v7) at  ($(v6)+2.3*(x)-0.5*(y)$);
\coordinate (v8) at  ($(v7)+(y)$); 


\draw[particle] (v1) -- node[above]{$b$} (v2);
\draw[particle] (v4) -- node[above=0.1cm]{$c$} (v3);
\draw[particle] (v3) -- node[below]{$c$} (v5);
\draw[particle] (v2) -- node[above right]{$s$}  (v8);
\draw[particle] (v7) .. controls ($(v7)-(x)+0.2*(y)$)  and  ($(v6)+(x)$)   .. node[below]{$d$} (v6);

\draw[photon] (v2) -- node[above]{$W$} (v3);

\draw[fill] ($(A)$) let \p1=($(v1)-(A)$) in ellipse ({0.2*veclen(\x1,\y1)} and {veclen(\x1,\y1)}) node[left=0.1cm]{$\widebar{B}^0$} ;
\draw[fill] ($0.5*(v5)+0.5*(v4)$) let \p1= ($0.5*(v5)-0.5*(v4)$) in ellipse ({0.2*veclen(\x1,\y1)} and {veclen(\x1,\y1)}) node[right=0.1cm]{$J/\psi$} ;
\draw[fill] ($0.5*(v8)+0.5*(v7)$) let \p1= ($0.5*(v8)-0.5*(v7)$) in ellipse ({0.2*veclen(\x1,\y1)} and {veclen(\x1,\y1)})  node[right=0.1cm]{$\widebar K^0(K_S)$};
\begin{scope}[yshift=-1.5cm,xshift=1cm]
\node at (0,0) {$\widebar{A}_{D^+D^-}\propto V_{cb}V_{cd}^*$};
\end{scope}

\end{tikzpicture} 
\end{center}
\begin{enumerate}[(i)]
\item Show that other ({\it e.g.}, ``penguin'') contributions are loop plus CKM suppressed.
\item Neglecting those corrections determine $a$ and $a(t)$. Be sure
  to include the $q/p$ factors for the component of $K^0(\widebar
  K^0)$ in  $K_S$. 
\end{enumerate}
\end{exercise}
\begin{solution}
sol here
\end{solution}
\begin{exercise}
Another case of much interest is $B_d\to\pi^+\pi^-$ (and its cousin,
$B_d\to\pi^0\pi^0$). 
\begin{enumerate}[(i)]
\item Display the diagrams that contribute to these decays.
\item Assuming tree level dominance, what angle of the unitarity
  triangle determines the decay asymmetry?
\item Argue that in this case, however, tree level dominance is
  questionable at best. 
\end{enumerate}
Once it was realized that the tree diagrams are not
dominant~\cite{Grinstein:1989df,London:1989ph}, a method was proposed that combines
several decays using isospin symmetry, for 
a clean determination of an angle of the unitarity triangle~\cite{Gronau:1990ka}.
\end{exercise}
\begin{solution}
sol here
\end{solution}
\begin{exercise}
We've discussed $K^0$ and $B^0$ mixing within the SM, but not $B_s$ nor $D^0$. For $B_s$ we need replace $V_{qs}$ for $V_{qd}$ everywhere. To see that you understand this show that the CP asymmetry for $B_s\to J/\psi \phi$ (to good approximation $\phi$ is a pure $s\widebar s$ state) is no more than a few per-cent in the SM.
\end{exercise}
\begin{solution}
sol here
\end{solution}
\end{exercises}

\chapter{Effective Field Theory}
\section{Introduction}
Recall that we left open the question of how to include
\begin{center}
\begin{tikzpicture}[scale=1.3] 
\coordinate (A) at (0,0);
\coordinate (x) at (1,0);
\coordinate (y) at (0,-1);

\coordinate (v1) at ($(A)+(x)$);
\coordinate (v2) at  ($(v1)+(x)$);
\coordinate (v3) at  ($(v2)+(x)$);
\coordinate (v9) at  ($(v3)+(x)$);
\coordinate (v4) at ($(A)+(y)$);
\coordinate (v5) at  ($(v4)+(x)$);
\coordinate (v6) at  ($(v5)+(x)$);
\coordinate (v7) at  ($(v6)+(x)$);
\coordinate (v8) at  ($(v7)+(x)$);

\draw[particle] (A) -- (v1);
\draw[particle] (v1) -- (v2);
\draw[particle] (v2) --  (v3);
\draw[particle] (v3) --  (v9);
\draw[particle] (v8) --  (v7);
\draw[particle] (v7) --  (v6);
\draw[particle] (v6) --   (v5);
\draw[particle] (v5)  -- (v4);

\draw[photon] (v1) -- node[left]{$W$} (v5);
\draw[photon] (v3) -- node[right]{$W$} (v7);
\draw[gluon] (v2) -- node[right]{$g$} (v6);

\end{tikzpicture} 
\hskip1cm
\begin{tikzpicture}[scale=1.3] 
\coordinate (A) at (0,0);
\coordinate (x) at (1,0);
\coordinate (y) at (0,-1);

\coordinate (v1) at ($(A)+0.5*(x)$); 
\coordinate (v2) at  ($(v1)+0.5*(x)$);
\coordinate (v3) at  ($(v2)+0.5*(x)$); 
\coordinate (v9) at  ($(v3)+0.5*(x)$); 
\coordinate (v10) at  ($(v9)+(x)$); 

\coordinate (v4) at ($(A)+(y)$);
\coordinate (v5) at  ($(v4)+(x)$);
\coordinate (v6) at  ($(v5)+(x)$);
\coordinate (v7) at  ($(v6)+(x)$);

\draw[particle] (A) -- (v1);
\draw[particle] (v1) -- (v2);
\draw[particle] (v2) --  (v3);
\draw[particle] (v3) --  (v9);
\draw[particle] (v9) --  (v10);
\draw[particle] (v7) --  (v6);
\draw[particle] (v6) --   (v5);
\draw[particle] (v5)  -- (v4);

\draw[photon] (v2) -- node[left]{$W$} (v5);
\draw[photon] (v9) -- node[right]{$W$} (v6);
\draw[gluon] (v1) let \p1=($ (v3)-(v2)$) in arc  (180:0:{veclen(\x1,\y1)}) ;
\node at ($(v2)-0.8*(y)$) {$g$}; 

\end{tikzpicture} 
\end{center}
in $X^0$-$\widebar X^0$ mixing, since we computed in terms of the
matrix element $\langle X^0|\mathcal{O}|\widebar X^0\rangle$ of a
local operator $\mathcal{O}$. In fact, we never quite justified
properly why we can use a local operator instead of a time ordered
product of interactions in the electro-weak Lagrangian. We also
mentioned a related problem, how to deal with the scale uncertainty,
that is, how to choose between, say, $\alpha_s(M_W)$ and
$\alpha_S(m_K)$. We'll address these problems in this chapter. We will
get into the guts of how it all works; the price we'll pay is limited
time for explicit examples. I think it is the correct emphasis.

The scale uncertainty problem derives form having disparate scales. The technique we'll utilize to address this is the effective field theory (EFT). It allows one to look at the physics of the shortest distance/time scales ignoring the longer ones, and then move sequentially to longer distances/times.

The problems we are facing are artifacts of perturbation theory. For example, if we could compute non-perturbatively, or at least perturbatively to all orders, we would use $\alpha_s(\mu)$ for the coupling, with an arbitrary renormalization scale $\mu$, together with $\lambda_{U,D}(\mu)$, $g_{1,2}(\mu)$. Then the physical amplitudes would actually be $\mu$-independent. Of course this is the content of the renormalization group equation (RGE), which we will use extensively. 

There is a related point worth mentioning. Disparate scales often result in possible breakdown of perturbation theory. The best example is in grand unified theories (GUTs) for which $M_{\text{GUT}}$ can be enormously larger that the electroweak scale, of order $10^{15} v$ (with $v=246~\text{GeV}$). If you compute, say, $e^+e^-\to\mu^+\mu^-$ at a CM energy of the order of current accelerators in a GUT in terms of its sole coupling constant, $g_{\text{GUT}}$, you'll find to 1-loop that
\[
\mathcal{A}=\mathcal{A}_{\text{Born}}\left(1+c\frac{\alpha_{\text{GUT}}}{\pi}\ln\frac{M^2_{\text{GUT}}}{v^2}+\cdots\right)
\]
Here $c=$ some number of $\mathcal{O}(1)$ and I have omitted terms that do not contain the large enhancement factor $\ln\frac{M^2_{\text{GUT}}}{v^2}\approx 30$. Now $\alpha_{\text{GUT}}\sim1/40$ is fairly typical and $c$ can easily be of order $\pi$, if not for this process for some of the great many low energy processes in the PDG book. Not only is the 1-loop correction large, $\mathcal{O}(100\%)$, but at $n$-loops there will be a correction of order $(\frac{\alpha_{\text{GUT}}}{\pi}\ln\frac{M^2_{\text{GUT}}}{v^2})^n$. 

If you can account for all of the terms of the form $(\frac{\alpha_{\text{GUT}}}{\pi}\ln\frac{M^2_{\text{GUT}}}{v^2})^n$, say by summing the corresponding $\sum_n c_n (\frac{\alpha_{\text{GUT}}}{\pi}\ln\frac{M^2_{\text{GUT}}}{v^2})^n$, then the next order gives corrections of the form  $\sum_n c'_n \frac{\alpha_{\text{GUT}}}{\pi}(\frac{\alpha_{\text{GUT}}}{\pi}\ln\frac{M^2_{\text{GUT}}}{v^2})^n$. If 
 $\frac{\alpha_{\text{GUT}}}{\pi}\ln\frac{M^2_{\text{GUT}}}{v^2}\sim1$, then these subleading corrections are of order $\frac{\alpha_{\text{GUT}}}{\pi}\sim 1/\ln\frac{M^2_{\text{GUT}}}{v^2}\sim1/30$. Nice. All we need to do to get per-cent accuracy is to sum those ``leading-logs.'' But failing to do so we incur in 100\% errors. 

The EFT technique takes advantage of the simpler form of the RGE when
there is only one relevant scale (one at a time!) in the problem, to
sum the leading-logs (LL) and if needed the next to leading-logs
(NLL), {\it i.e.},  $\frac{\alpha_{\text{GUT}}}{\pi}(\frac{\alpha_{\text{GUT}}}{\pi}\ln\frac{M^2_{\text{GUT}}}{v^2})^n$, etc. 

\section{Intuitive EFT}
We can get a good preview of what EFT is about from general
considerations. Suppose you have some particles whose interactions you
are studying in a particle accelerator with CM energy $E$. Obviously
all these particles, including the colliding particles and those
produced in the collisions, have masses less than $E$. Let's call these
particles ``light.''  Suppose further
that you know of the existence of a particle that interacts
with the light particles, but has   mass $M$ much larger
than $E$; it's a ``heavy'' particle.  You also know that the
interactions among all these particles are well described by a
local,  Lorentz invariant, renormalizable QFT. We call this the ``full'' theory. 

While you could take the full theory to calculate  reaction rates among the
light particles, using the full QFT including the heavy particle may strike you as overkill. Since the heavy particle
cannot be used in the collider, nor can it be produced by the
collisions, can we just ignore its presence? That is, delete it from
the theory? 

Clearly the heavy particle can affect collision rates among light particles through its virtual effects. The space-time separation between the point at which the virtual heavy particle is created and the one at which it is annihilated is of order $1/M$. To be sure, the exchange of the heavy particle produces what appears as a non-local interaction among the light particles. But the scale of the non-locality is very short, of order $1/M$, and we can Taylor expand in this parameter so that on distance scales $L\gg 1/M$, the interaction appears local: the expansion is in powers of  $1/(ML)$.  So if we are prepared to say that such a separation is undetectable in low energy collisions we can model the effects of the virtual exchange by local operators. Some of the local operators are of dimension 4 and would come into the Lagrangian with dimensionless coefficients. Some of these may already exist in the Lagrangian of our full QFT. In fact, if the full Lagrangian contained all possible dimension 4 operators there are no new operators of this dimension to add. But are the coefficients modified, and if so,  how significantly? 

By dimensional analysis the effects of the particle of mass $M$ should come into scattering amplitudes only through the logarithm of this mass, say $\ln(M/\mu)$, where $\mu$ is a renormalization scale. Powers of $E/M$ are also allowed, but powers of  $E$  translate into powers of momenta, hence derivatives on local operators, and this means higher dimension operators. So the Lagrangian is modified by allowing the dimensionless 
coefficients of dimension 4 operators to be functions of $\ln(M/\mu)$. Even if we do not know these functions, we can model the low energy interactions by writing the most general renormalizable Lagrangian for interactions among the light particles. Knowledge of the full QFT, including the heavy particle, simply gives us additional constraints on the dimensionless couplings of this Lagrangian. This is fine provided we do not need accuracy 
better than $E/M$ in our predictions of  scattering amplitudes. 

Sometimes, however, there are processes that are simply not allowed by the renormalizable Lagrangian of the light particles, but are allowed if we include operators of dimension higher than 4. This can happen also as a result of a virtual exchange of a heavy particle. The effect is captured in the low energy Lagrangian by including higher dimension operators (that is, of dimension higher than 4) with dimensional coefficients made up with inverse powers of $M$. Hence, our ``effective'' Lagrangian for interactions among the light particles is the most general renormalizable Lagrangian for the light particles, consistent with Lorentz and gauge symmetries, supplemented by higher dimensional operators accompanied by inverse powers of $M$.

There is one more insight that is worth dwelling on. Suppose you want to compute the scattering amplitude for some light particles to accuracy $E/M$. To make our discussion simpler let's assume that the scattering amplitude vanishes identically if only renormalizable interactions are retained in your effective theory. That is, it vanishes in  order $(E/M)^0$. That is, it vanishes as $M\to\infty$. Then to get a non-zero amplitude you must use the interaction terms that include one power of $1/M$. And any one of these interactions may come in  only once, else you will get more powers of $1/M$. Think about it this way: you do first order perturbation theory in these interactions, but compute to all orders in the renormalizable interactions. Then there are many-loop contributions that probe the $1/M$ vertex at arbitrarily short distances. It would seem we have a problem, since the premise of our construction is that we could expand the non-local interactions on long distances $L$ in powers of $1/(ML)$. And here is the crux of the matter. In these loops, when the virtual (light) particles probe long distances, the approximations we have made work well by design. And while the approximation fails when the virtual particles probe short distances, the failure can be modeled by a local interaction. But we already have all possible local interactions in our effective Lagrangian. All we need to do is to choose their interaction strengths (a.k.a. operator coefficients, or coupling constants) appropriately so as to reproduce the scattering amplitude of the full theory.

In fact, you can go one step further. In the absence of detailed knowledge of the full QFT, we have argued on general grounds that the physics of the light particles is described by a non-renormalizable Lagrangian, the most general sum of  local operators consistent with symmetries with the  coefficients of operators of dimension $4+n$ including a factor of $1/M^n$. This should approximate extremely well any amplitude provided the energies involved are small compared to $M$. The mass scale $M$ acts, in effect, as an energy/momentum cut-off.\footnote{The alert reader will notice that if we apply this prescription to operators of dimension less than 4 we will render the light particles as heavy as $M$, and the whole procedure meaningless. This is the hierarchy problem, that the Higgs mass term ought to be as large as the scale of the cut-off.}

Let' s make these ideas somewhat more precise.

\section{The Appelquist-Carazzone Decoupling Theorem}
(Note: there is no wikipedia page for this: here is your chance to make a mark!)\\
While we call this a `theorem' neither them nor I are mathematicians,
so do not expect a `proof.'  Instead we will explain what it is and how it works~\cite{Appelquist:1974tg}. Consider a model with a renormalizable Lagrangian
\[
\mathcal{L}_{\text{full}}=\mathcal{L}_{\text{light}}+\tfrac12\left[\left(\partial_\mu\phi\right)^2-M^2\phi^2\right]+\mathcal{L}_{\phi-\text{light}}
\]
$\mathcal{L}_{\text{light}}$ may include many fields but all with
mass~$\ll M$. It depends on coupling constants, which we take to be
dimensionless parameters, $g_i$,  and possibly also masses~$m_j$. $\mathcal{L}_{\phi-\text{light}}$ has the interactions between $\phi$ and the light-fields and depends on $g_i$ and possibly on additional couplings $h_i$. 

Consider Green functions $G^{(n)}(p_1,\ldots, p_n)$ (or, better yet, amplitudes) of the  {\it light} fields (associated with light particles) restricted to $|p_i|\ll M$ (all components of all momenta). Then 
\[
G^{(n)}(p_1,\ldots, p_n)=Z^{\frac{n}2}\tilde G^{(n)}(p_1,\ldots, p_n)(1+\mathcal{O}(1/M))
\]
where $\tilde G^{(n)}(p_1,\ldots, p_n)$ is computed from
$\mathcal{L}_{\text{eff}}=\tilde{\mathcal{L}}_{\text{light}}$, where
$\tilde{\mathcal{L}}_{\text{light}}$ is a renormalizable Lagrangian
constructed out of the fields in $\mathcal{L}_{\text{light}}$, with
new (possibly more) couplings $\tilde g_i$. The $\tilde g_i=\tilde
g_i(\{g_i\},M)$ and $Z=Z(\{g_i\},M)$ are not functions of momenta,
only of the indicated arguments, and are {\it universal}. That is, the
same $\tilde g$ and $Z$ appear in any Green function. To be sure,
there is a different $Z$ for each light field, and the factor
$Z^{\frac n2}$ really stands for a $Z^{\frac12}$ for each field in the
Green function, but as we will see shortly this is of little
consequence (so I will continue to be sloppy in the way the factor is included). 

The meaning is clear, heavy particles appear in $G^{(n)}$ only through virtual effects, by construction. At large $M$ ($M\gg m, |p|$) the effects of $M$ decrease as powers of $1/M$, except possibly when $M$ appears in logarithms. The content of the decoupling theorem is that (i) there are no positive powers of $M$, and (ii) the $\ln(M)$ terms can all be absorbed into $\tilde g$ and $Z$.

For the theorem to work you have to be able to take $M$ arbitrarily
large holding $g_i$ constant. It fails when $M=gv$, the case where the
field $\phi$ and some of the light fields acquire their mass solely from spontaneous symmetry breaking, because either $v\to \infty$ and all particles get heavy, or 
$g_i\to\infty$ together with $M$, so that the order $1/M$ corrections can go as $g/M=1/v=$~fixed. 

Concretely, lets consider GUTs. Say $SU(5)$. We can apply decoupling to the $M_{\text{GUT}}$-heavy fields. Then by construction $\mathcal{L}_{\text{eff}}$ is just the SM with couplings $\tilde g_{1,2,3}=\tilde g_{1,2,3}(g_{\text{GUT}},M_{\text{GUT}})$. To understand what is going on consider again $e^+e^-\to\mu^+\mu^-$:
\begin{center}
\begin{tikzpicture}[scale=0.9] 
\coordinate (A) at (0,0);
\coordinate (x) at (1,0);
\coordinate (y) at (0,-1);

\coordinate (v1) at ($(A)+0.5*(y)$); 
\coordinate (v2) at  ($(A)-0.5*(y)$);
\coordinate (v3) at  ($(A)+(x)$); 
\coordinate (v4) at  ($(v3)+(x)$); 
\coordinate (v5) at  ($(v4)+(x)+0.5*(y)$);
\coordinate (v6) at  ($(v4)+(x)-0.5*(y)$);

\draw[particle] (v1) node[left]{$e^-$} -- (v3);
\draw[particle] (v3) --  (v2) node [left]{$e^+$};
\draw[particle] (v4) --  (v5) node[right]{$\mu^-$};
\draw[particle] (v6) node[right]{$\mu^+$}--  (v4);

\draw[photon] (v3) -- (v4);
\begin{scope}[xshift=4cm]
\node at (0,0) {$+$};
\end{scope}

\end{tikzpicture} 
\begin{tikzpicture}[scale=0.9] 
\coordinate (A) at (0,0);
\coordinate (x) at (1,0);
\coordinate (y) at (0,-1);

\coordinate (v1) at ($(A)+0.5*(y)$); 
\coordinate (v2) at  ($(A)-0.5*(y)$);
\coordinate (v3) at  ($(A)+(x)$); 
\coordinate (v4) at  ($(v3)+3*(x)$); 
\coordinate (v5) at  ($(v4)+(x)+0.5*(y)$);
\coordinate (v6) at  ($(v4)+(x)-0.5*(y)$);

\coordinate (v7) at  ($(v3)+(x)$);
\coordinate (v8) at  ($(v4)-(x)$);

\draw[particle] (v1) node[left]{$e^-$} -- (v3);
\draw[particle] (v3) --  (v2) node [left]{$e^+$};
\draw[particle] (v4) --  (v5) node[right]{$\mu^-$};
\draw[particle] (v6) node[right]{$\mu^+$}--  (v4);

\draw[photon] (v3) -- (v7);
\draw[photon] (v8) -- (v4);

\draw[photon] ($0.5*(v7)+0.5*(v8)$) let \p1=($0.5*(v8)-0.5*(v7)$) in  circle ({veclen(\x1,\y1)}); 

\begin{scope}[xshift=7cm]
\node at (0,0) {$+\cdots$};
\end{scope}

\end{tikzpicture} 
\end{center}
 The effect of the loop of GUT-heavy particles differentiates the photon and $Z$ from gluons by differences in $\ln(M_{\text{GUT}}^2/\mu^2)$, because the photon and $Z$ couple differently than gluons to GUT-heavy particles. Of course $\gamma/Z$ couple differently than gluons  to light particles, but at this point that difference is not from the coupling constant, which is the common $g_{\text{GUT}}$.  The  photon (or $Z$, or gluon) self-energy, $p^2\Pi(p^2)$ has two contributions, roughly
\[
\Pi(p^2)\sim\frac{\alpha_{\text{GUT}}}{\pi}
\ln\frac{M_{\text{GUT}}^2}{\mu^2}+
\frac{\alpha_{\text{GUT}}}{\pi}\ln\frac{\mu^2}{p^2}+
\mathcal{O}\left(\frac{p^2}{M_{\text{GUT}}^2}\right).
\]
The first term is from the GUT heavy particles while the second, which I call $\tilde \Pi$ below, is from effectively massless particles. 
So, neglecting powers of $p^2/M_{\text{GUT}}^2$ the new coupling has been shifted by $\ln\frac{M_{\text{GUT}}^2}{\mu^2}$. More explicitly, if the tree-level (Born) amplitude is $\mathcal{A}^{(0)}=g^2_{\text{GUT}}A^{(0)}/p^2$, then the self-energy correction to the one-loop amplitude is
\begin{equation}
\label{eq:GUTeff1}
\frac{g^2_{\text{GUT}}}{p^2(1+\Pi)}A^{(0)}\sim
\left(\frac{g^2_{\text{GUT}}}{ 1+
\frac{\alpha_{\text{GUT}}}{\pi}\ln\frac{M_{\text{GUT}}}{\mu}}\right)
\left( \frac{1}{p^2(1+\tilde\Pi)}\right)A^{(0)}
\end{equation}
We have used the fact that the product of the GUT-heavy  and massless particles contributions to $\Pi$ is higher order in $\alpha_{\text{GUT}}/\pi$ and hence can be neglected to the order we are working. Now in \eqref{eq:GUTeff1} the first term effectively changes the value of the coupling constant, 
\[
g^2_{\text{GUT}}\to {\tilde g}^2(\mu)\sim 
\frac{g^2_{\text{GUT}}}{ 1+
\frac{\alpha_{\text{GUT}}}{\pi}\ln\frac{M_{\text{GUT}}}{\mu}}
\]
while the second term is the contribution of light particles to the self energy. Both terms depend on the renormalization scale, but in such a way that once we account for self-energies of the external states (electrons and muons) the amplitude is $\mu$-independent. We see that the amplitude could be computed from the effective Lagrangian, $\mathcal{\tilde L}_{\text{light}}$, which is just the Lagrangian of the SM provided the coupling constant used is $\tilde g$. We still have to account for the effect of self-energies on the external particles. This can also be broken into a contribution from GUT-heavy particles, that goes into the factor of $Z^{\frac12}$, and a contribution from light ones, and the latter is produced by  $\mathcal{\tilde L}_{\text{light}}$:\\
\begin{tikzpicture}[scale=1,gaugeZ/.style={decorate, draw=black,
    decoration={coil,aspect=0, segment length=4,amplitude=2}},] 
\coordinate (A) at (0,0);
\coordinate (x) at (1,0);

\coordinate (v1) at ($(A)+(x)$); 

\draw[particle] (A) --  (v1);

\begin{scope}[xshift=1.5cm]
\node at (0,0) {$+$};
\end{scope}

\begin{scope}[xshift=2cm]
\coordinate (A) at (0,0);
\coordinate (x) at (1,0);
\coordinate (y) at (0,0.5);

\coordinate (v1) at ($(A)+0.5*(x)$); 
\coordinate (v2) at ($(v1)+(x)$); 
\coordinate (v3) at ($(v2)+0.5*(x)$); 

\draw[particle] (A) -- (v1);
\draw[particle] (v1) -- (v2);
\draw[particle] (v2) -- (v3);

\draw[gaugeZ] (v1) let \p1=($0.5*(v2)-0.5*(v1)$) in  arc  (180:0:{veclen(\x1,\y1)}); 
\node[above] at ($0.5*(v2)+0.5*(v1)+(y)$) {$\gamma,Z$};

\end{scope}
\begin{scope}[xshift=4.5cm]
\node at (0,0) {$+$};
\end{scope}

\begin{scope}[xshift=5cm]
\coordinate (A) at (0,0);
\coordinate (x) at (1,0);
\coordinate (y) at (0,0.5);

\coordinate (v1) at ($(A)+0.5*(x)$); 
\coordinate (v2) at ($(v1)+(x)$); 
\coordinate (v3) at ($(v2)+0.5*(x)$); 

\draw[particle] (A) -- (v1);
\draw[particle] (v1) -- (v2);
\draw[particle] (v2) -- (v3);

\draw[gaugeZ] (v1) let \p1=($0.5*(v2)-0.5*(v1)$) in  arc  (180:0:{veclen(\x1,\y1)}); 
\node[above] at ($0.5*(v2)+0.5*(v1)+(y)$) {GUT};

\end{scope}

\end{tikzpicture} 
\[
\sim\left(1+\frac{\alpha_{\text{GUT}}}{\pi}\ln\frac{M_{\text{GUT}}}{\mu}\right)
\left(
\hbox to 4.7cm{\hskip-0.3cm\vbox to 0.8cm {\begin{tikzpicture}[scale=1,gaugeZ/.style={decorate, draw=black,
    decoration={coil,aspect=0, segment length=4,amplitude=2}}] 
\coordinate (A) at (0,0);
\coordinate (x) at (1,0);

\coordinate (v1) at ($(A)+(x)$); 

\draw[particle] (A) --  (v1);

\begin{scope}[xshift=1.5cm]
\node at (0,0) {$+$};
\end{scope}

\begin{scope}[xshift=2cm]
\coordinate (A) at (0,0);
\coordinate (x) at (1,0);
\coordinate (y) at (0,0.5);

\coordinate (v1) at ($(A)+0.5*(x)$); 
\coordinate (v2) at ($(v1)+(x)$); 
\coordinate (v3) at ($(v2)+0.5*(x)$); 

\draw[particle] (A) -- (v1);
\draw[particle] (v1) -- (v2);
\draw[particle] (v2) -- (v3);

\draw[gaugeZ] (v1) let \p1=($0.5*(v2)-0.5*(v1)$) in  arc  (180:0:{veclen(\x1,\y1)}); 
\node[above] at ($0.5*(v2)+0.5*(v1)+(y)$) {$\gamma,Z$};

\end{scope}
\end{tikzpicture} 
}}
\right)+\cdots
\]
Thus, we see, the EFT formalism is much like  a factorization theorem, except that not really, because the efective couplings, $\tilde g$, appear everywhere. 

\section{Beyond $\mathcal{\tilde L}_{\text{light}}$.}
$\mathcal{\tilde L}_{\text{light}}$ is the most general renormalizable Lagrangian of the light fields. It may be less general if exact symmetries of the full Lagrangian, $\mathcal{L}_{\text{full}}$,  forbid some terms in the effective Lagrangian.  But there are processes described by $\mathcal{ L}_{\text{full}}$ involving only external light fields that may be absent from $\mathcal{\tilde L}_{\text{light}}$. For example, 
\begin{center}
\begin{tikzpicture}[scale=0.9] 
\coordinate (A) at (0,0);
\coordinate (x) at (1,0);
\coordinate (y) at (0,-1);

\coordinate (v1) at ($(A)+0.5*(y)$); 
\coordinate (v2) at  ($(A)-0.5*(y)$);
\coordinate (v3) at  ($(A)+(x)$); 
\coordinate (v4) at  ($(v3)+1.5*(x)$); 
\coordinate (v5) at  ($(v4)+(x)+0.5*(y)$);
\coordinate (v6) at  ($(v4)+(x)-0.5*(y)$);

\draw[particle] (v1) node[left]{$\ell$} -- (v3);
\draw[particle] (v3) --  (v2) node [left]{$d^c$};
\draw[particle] (v4) --  (v5) node[right]{$u^c$};
\draw[particle] (v6) node[right]{$q$}--  (v4);

\draw[photon] (v3) -- node[above]{GUT} (v4);
\begin{scope}[xshift=4cm]
\node at (0,0) {or};
\end{scope}

\end{tikzpicture} 
\begin{tikzpicture}[scale=0.9] 
\coordinate (A) at (0,0);
\coordinate (x) at (1,0);
\coordinate (y) at (0,-1);

\coordinate (v1) at ($(A)+0.5*(y)$); 
\coordinate (v2) at  ($(A)-0.5*(y)$);
\coordinate (v3) at  ($(A)+(x)$); 
\coordinate (v4) at  ($(v3)+1.5*(x)$); 
\coordinate (v5) at  ($(v4)+(x)+0.5*(y)$);
\coordinate (v6) at  ($(v4)+(x)-0.5*(y)$);

\draw[particle] (v1) node[left]{$q$} -- (v3);
\draw[particle] (v3) --  (v2) node [left]{$u^c$};
\draw[particle] (v4) --  (v5) node[right]{$e^c$};
\draw[particle] (v6) node[right]{$q$}--  (v4);

\draw[photon] (v3) -- node[above]{GUT}   (v4);
\end{tikzpicture} 
\end{center}
There are no terms in $\mathcal{\tilde L}_{\text{light}}$ one can
write, consistent with gauge symmetry and renormalizability, that reproduce these. In fact, the SM Lagrangian respects baryon and lepton number separately while these processes require  breaking of both symmetries. To obtain this we must supplement  the effective Lagrangian with additional non-renormalizable terms, 
\begin{equation}
\label{eq:Leff-introduced}
\mathcal{L}_{\text{eff}}=\mathcal{\tilde L}_{\text{light}}+\frac1{M_{\text{GUT}}}\mathcal{L}^{(5)}+
\frac1{M^2_{\text{GUT}}}\mathcal{L}^{(6)}+\cdots
\end{equation}
where $\mathcal{L}^{(n)}$ consists of a sum of  operators of dimension $n$ constructed out of light fields and coefficients that depend on the couplings $\tilde g_i$. 

It is easy to see how this works at tree level:
\[
\hbox to 3.5cm{\hskip-0.5cm\vbox to 0.6cm {\begin{tikzpicture}[scale=0.9] 
\coordinate (A) at (0,0);
\coordinate (x) at (1,0);
\coordinate (y) at (0,-1);

\coordinate (v1) at ($(A)+0.5*(y)$); 
\coordinate (v2) at  ($(A)-0.5*(y)$);
\coordinate (v3) at  ($(A)+(x)$); 
\coordinate (v4) at  ($(v3)+1.5*(x)$); 
\coordinate (v5) at  ($(v4)+(x)+0.5*(y)$);
\coordinate (v6) at  ($(v4)+(x)-0.5*(y)$);

\draw[particle] (v1) -- (v3);
\draw[particle] (v3) --  (v2) ;
\draw[particle] (v4) --  (v5) ;
\draw[particle] (v6) --  (v4);

\draw[photon] (v3) --  (v4);

\end{tikzpicture} 
}}
=-g_{\text{GUT}}^2J_1^\mu J_2^\nu\left(-i\frac{g_{\mu\nu}-p_\mu p_\nu/M_{\text{GUT}}^2}{p^2-M_{\text{GUT}}^2}\right)
\rightarrow-i\frac{g_{\text{GUT}}^2}{M_{\text{GUT}}^2}J_1^\mu J_2^\nu g_{\mu\nu}
\]
where $J_i^\mu$ stand for currents that couple to the GUT-heavy gauge bosons and in the last step we have used  $|p|\ll M_{\text{GUT}}$. So we have found a tree level contribution, $\mathcal{L}^{(6)}=-g_{\text{GUT}}^2J_1^\mu J_{2\mu}$.

\section{Beyond Tree Level\footnote{Much of this section is based on Ref.~\cite{Witten:1976kx}}}
\label{sec:loops}
We have established that in GUT for some $\Delta B=\Delta L  \ne0$ 4-pt functions,
\[
G^{(4)}(p_1,\ldots,p_4)=(1)\tilde G^{(4)}(p_1,\ldots,p_4)+\cdots
\]
where the right hand side is computed in an effective theory with light fields only, with Lagrangian $\mathcal{L}_{\text{eff}}$. The same applies to the weak interactions if the exchange particle is a $W/Z$ and the external states are quarks and leptons (except for the very heavy top quark) and the energies involved are all small compared to the $W$ and $Z$ masses. Diagrammatically 
\begin{center}
\begin{tikzpicture}[scale=0.7] 
\coordinate (A) at (0,0);
\coordinate (x) at (1,0);
\coordinate (y) at (0,1);

\coordinate (v1) at ($(A)-(x)+(y)$);
\coordinate (v2) at  ($(A)+(x)+(y)$);
\coordinate (v3) at  ($(A)-(x)-(y)$); 
\coordinate (v4) at ($(A)+(x)-(y)$);
\coordinate (v12) at ($0.5*(v1)+0.5*(v2)$);
\coordinate (v34) at ($0.5*(v3)+0.5*(v4)$);

\draw[particle] (v1)  -- (v12);
\draw[particle] (v12) -- (v2) ;
\draw[particle] (v4)  -- (v34);
\draw[particle] (v34) -- (v3);

\draw[photon] (v12) -- node[right]{$W$} (v34);

\begin{scope}[xshift = 2cm]
\node{$=$} (A);
\end{scope}

\begin{scope}[xshift = 4cm]
\coordinate (A) at (0,0);
\coordinate (x) at (1,0);
\coordinate (y) at (0,1);

\coordinate (v1) at ($(A)-(x)+(y)$);
\coordinate (v2) at  ($(A)+(x)+(y)$);
\coordinate (v3) at  ($(A)-(x)-(y)$); 
\coordinate (v4) at ($(A)+(x)-(y)$);
\draw[particle] (v1) -- (A);
\draw[particle] (A) -- (v2) ;
\draw[particle] (v4)   -- (A);
\draw[particle] (A) -- (v3);
\draw[fill] (A) circle (0.1cm);
\end{scope}

\end{tikzpicture} 
\end{center}
 We have seen this before in \eqref{eq:fig4vertex}, albeit the argument there was somewhat informal. Of course, the equation has corrections of relative order $p^2/M^2$ (I will use $M$ for the heavy mass from here on, which can be a $W$ or a $Z$ mass). So we can make it formally correct by writing:
\[
\lim_{M\to\infty}\left[M^2\left(
\hbox to 4.5cm{\hskip-0.4cm\vbox to 0.9cm {%
\begin{tikzpicture}[scale=0.7] 
\coordinate (A) at (0,0);
\coordinate (x) at (1,0);
\coordinate (y) at (0,1);

\coordinate (v1) at ($(A)-(x)+(y)$);
\coordinate (v2) at  ($(A)+(x)+(y)$);
\coordinate (v3) at  ($(A)-(x)-(y)$); 
\coordinate (v4) at ($(A)+(x)-(y)$);
\coordinate (v12) at ($0.5*(v1)+0.5*(v2)$);
\coordinate (v34) at ($0.5*(v3)+0.5*(v4)$);

\draw[particle] (v1)  -- (v12);
\draw[particle] (v12) -- (v2) ;
\draw[particle] (v4)  -- (v34);
\draw[particle] (v34) -- (v3);

\draw[photon] (v12) -- node[right]{$W$} (v34);

\begin{scope}[xshift = 2cm]
\node{$-$} (A);
\end{scope}

\begin{scope}[xshift = 4cm]
\coordinate (A) at (0,0);
\coordinate (x) at (1,0);
\coordinate (y) at (0,1);

\coordinate (v1) at ($(A)-(x)+(y)$);
\coordinate (v2) at  ($(A)+(x)+(y)$);
\coordinate (v3) at  ($(A)-(x)-(y)$); 
\coordinate (v4) at ($(A)+(x)-(y)$);
\draw[particle] (v1) -- (A);
\draw[particle] (A) -- (v2) ;
\draw[particle] (v4)   -- (A);
\draw[particle] (A) -- (v3);
\draw[fill] (A) circle (0.1cm);
\end{scope}

\end{tikzpicture} 
}}
\right)\right]=0
\]

Our task presently is to extend this argument beyond tree level. The
1-loop corrections we are concerned about are from gluon exchange
among the quarks, since the strong coupling constant is the largest
among the couplings in the SM. Moreover, the coupling constant  grows larger the lower the energy and we want to use this technique for low energy processes, {\it e.g.}, $B$-meson decays.
The treatment of photon mediated loops is entirely analogous. We will discuss the effect of loops with $W/Z$'s below. We want to show\\[0.5cm]
\begin{tikzpicture}[scale=0.8,gaugeZ/.style={decorate, draw=black,
    decoration={coil,aspect=0, segment length=5,amplitude=3}},sgluon/.style={decorate, draw=black,
    decoration={coil,aspect=0.5,segment length=3pt,amplitude=2pt}}
] 
\coordinate (A) at (0,0);
\coordinate (x) at (1,0);
\coordinate (y) at (0,1);

\coordinate (v1) at ($(A)+0.5*(y)$);
\coordinate (v2) at  ($(A)-0.5*(y)$);
\coordinate (v3) at  ($(A)+(x)$); 
\coordinate (v4) at ($(v3)+(x)$); 
\coordinate (v5) at ($(v4)+(x)+0.5*(y)$);
\coordinate (v6) at  ($(v4)+(x)-0.5*(y)$);

\draw (v1)  -- (v3);
\draw (v2) -- (v3) ;
\draw (v4)  -- (v5);
\draw (v4) -- (v6);

\draw[gaugeZ] (v3) -- (v4);

\begin{scope}[xshift = 4cm]
\node{$+$} (A);
\end{scope}

\begin{scope}[xshift = 5cm]
\coordinate (A) at (0,0);
\coordinate (x) at (1,0);
\coordinate (y) at (0,1);

\coordinate (v1) at ($(A)+0.5*(y)$);
\coordinate (v2) at  ($(A)-0.5*(y)$);
\coordinate (v3) at  ($(A)+(x)$); 
\coordinate (v4) at ($(v3)+(x)$); 
\coordinate (v5) at ($(v4)+(x)+0.5*(y)$);
\coordinate (v6) at  ($(v4)+(x)-0.5*(y)$);

\draw (v1)  -- (v3);
\draw (v2) -- (v3) ;
\draw (v4)  -- (v5);
\draw (v4) -- (v6);

\draw[gaugeZ] (v3) -- (v4);
\draw[sgluon] ($0.6*(v2)+0.4*(v3)$) -- node[left]{$g$} ($0.6*(v1)+0.4*(v3)$);

\end{scope}
\begin{scope}[xshift = 9cm]
\node{$+$} (A);
\end{scope}

\begin{scope}[xshift = 10cm]
\coordinate (A) at (0,0);
\coordinate (x) at (1,0);
\coordinate (y) at (0,1);

\coordinate (v1) at ($(A)+0.5*(y)$);
\coordinate (v2) at  ($(A)-0.5*(y)$);
\coordinate (v3) at  ($(A)+(x)$); 
\coordinate (v4) at ($(v3)+(x)$); 
\coordinate (v5) at ($(v4)+(x)+0.5*(y)$);
\coordinate (v6) at  ($(v4)+(x)-0.5*(y)$);

\draw (v1)  -- (v3);
\draw (v2) -- (v3) ;
\draw (v4)  -- (v5);
\draw (v4) -- (v6);

\draw[gaugeZ] (v3) -- (v4);
\draw[{decorate, draw=black,
    decoration={coil,aspect=0.4,segment length=5pt,amplitude=2pt}}] ($0.6*(v1)+0.4*(v3)$) -- 
                node[above]{$g$} ($0.4*(v4)+0.6*(v5)$);

\end{scope}

\begin{scope}[xshift = 14cm]
\node{$+\cdots$} (A);
\end{scope}

\end{tikzpicture}

\[= \left(Z^{\frac12}\right)^4\left[
\hbox to 9.5cm{\hskip-0.4cm\vbox to 0.8cm {%
\begin{tikzpicture}[scale=0.7,gaugeZ/.style={decorate, draw=black,
    decoration={coil,aspect=0, segment length=5,amplitude=3}}] 

\coordinate (A) at (0,0);
\coordinate (x) at (1,0);
\coordinate (y) at (0,1);

\coordinate (v1) at ($(A)-(x)+0.5*(y)$);
\coordinate (v2) at  ($(A)+(x)+0.5*(y)$);
\coordinate (v3) at  ($(A)-(x)-0.5*(y)$); 
\coordinate (v4) at ($(A)+(x)-0.5*(y)$);
\draw (v1) -- (A);
\draw (A) -- (v2) ;
\draw (v4)   -- (A);
\draw (A) -- (v3);
\draw[fill] (A) circle (0.1cm);

\begin{scope}[xshift = 2cm]
\node{$+$} (A);
\end{scope}

\begin{scope}[xshift = 4cm]
\coordinate (A) at (0,0);
\coordinate (x) at (1,0);
\coordinate (y) at (0,1);

\coordinate (v1) at ($(A)-(x)+0.5*(y)$);
\coordinate (v2) at  ($(A)+(x)+0.5*(y)$);
\coordinate (v3) at  ($(A)-(x)-0.5*(y)$); 
\coordinate (v4) at ($(A)+(x)-0.5*(y)$);
\draw (v1) -- (A);
\draw (A) -- (v2) ;
\draw (v4)   -- (A);
\draw (A) -- (v3);
\draw[fill] (A) circle (0.1cm);

\draw[{decorate, draw=black,
    decoration={coil,aspect=0.5,segment length=3pt,amplitude=2pt}}]
  ($0.6*(v3)+0.4*(A)$) -- node[left]{$g$} ($0.6*(v1)+0.4*(A)$); 
\end{scope}
\begin{scope}[xshift = 6cm]
\node{$+$} (A);
\end{scope}

\begin{scope}[xshift = 8cm]
\coordinate (A) at (0,0);
\coordinate (x) at (1,0);
\coordinate (y) at (0,1);

\coordinate (v1) at ($(A)-(x)+0.5*(y)$);
\coordinate (v2) at  ($(A)+(x)+0.5*(y)$);
\coordinate (v3) at  ($(A)-(x)-0.5*(y)$); 
\coordinate (v4) at ($(A)+(x)-0.5*(y)$);
\draw (v1) -- (A);
\draw (A) -- (v2) ;
\draw (v4)   -- (A);
\draw (A) -- (v3);
\draw[fill] (A) circle (0.1cm);

\draw[{decorate, draw=black,
    decoration={coil,aspect=0.4,segment length=4pt,amplitude=2pt}}]
  ($0.6*(v1)+0.4*(A)$) -- node[above]{$g$}  ($0.6*(v2)+0.4*(A)$); 
\end{scope}
\begin{scope}[xshift = 10cm]
\node{$+\cdots$} (A);
\end{scope}

\end{tikzpicture} 
}}
\right]
\]

At least for now graphs on the right  hand side (RHS) are in one-to-one correspondance with those on the LHS. So let's compare one at a time. We ask first
\begin{center}

\begin{tikzpicture}[scale=1,gaugeZ/.style={decorate, draw=black,
    decoration={coil,aspect=0, segment length=5,amplitude=3}},sgluon/.style={decorate, draw=black,
    decoration={coil,aspect=0.5,segment length=3pt,amplitude=2pt}}
] 

\coordinate (A) at (0,0);
\coordinate (x) at (1,0);
\coordinate (y) at (0,1);

\coordinate (v1) at ($(A)+0.5*(y)$);
\coordinate (v2) at  ($(A)-0.5*(y)$);
\coordinate (v3) at  ($(A)+(x)$); 
\coordinate (v4) at ($(v3)+(x)$); 
\coordinate (v5) at ($(v4)+(x)+0.5*(y)$);
\coordinate (v6) at  ($(v4)+(x)-0.5*(y)$);

\draw (v1)  -- (v3);
\draw (v2) -- (v3) ;
\draw (v4)  -- (v5);
\draw (v4) -- (v6);

\draw[gaugeZ] (v3) -- (v4);
\draw[{decorate, draw=black,
    decoration={coil,aspect=0.4,segment length=5pt,amplitude=2pt}}] ($0.6*(v1)+0.4*(v3)$) -- ($0.4*(v4)+0.6*(v5)$);

\begin{scope}[xshift = 4cm]
\node{$? \atop =$} (A);
\end{scope}

\begin{scope}[xshift = 6cm]
\coordinate (A) at (0,0);
\coordinate (x) at (1,0);
\coordinate (y) at (0,1);

\coordinate (v1) at ($(A)-(x)+0.5*(y)$);
\coordinate (v2) at  ($(A)+(x)+0.5*(y)$);
\coordinate (v3) at  ($(A)-(x)-0.5*(y)$); 
\coordinate (v4) at ($(A)+(x)-0.5*(y)$);
\draw (v1) -- (A);
\draw (A) -- (v2) ;
\draw (v4)   -- (A);
\draw (A) -- (v3);
\draw[fill] (A) circle (0.1cm);

\draw[{decorate, draw=black,
    decoration={coil,aspect=0.4,segment length=4pt,amplitude=2pt}}]
  ($0.6*(v1)+0.4*(A)$) --  ($0.6*(v2)+0.4*(A)$); 
\end{scope}
\begin{scope}[xshift = 8cm]
\node{$+~~\mathcal{O}\left(\frac1{M^3}\right)$} (A);
\end{scope}

\end{tikzpicture} 
\end{center}
Keep in mind the first term on the RHS has an explicit factor of  $1/M^2$.  It is easy to see that this equation is non-sensical. The left hand side, sketchily, is
\[
\text{LHS}\sim \int \!\! d^4k\,\left(\frac{1}{\slashed{k}}\right)^2\left(\frac{g_{\mu\nu}}{k^2}\right) \left(\frac{g_{\mu\nu}-\cdots}{k^2-M^2}\right).
\]
This is indeed a sketch. We left out the interaction vertices and set all external momentum  to zero. The point we are trying to make is simple. This integral is clearly UV convergent. As $k\to\infty$ the integrand of the 4-dimensional integral is $\sim1/k^6$. It's a different story for the  RHS,
\[
\text{RHS}\sim \int \!\! d^4k\,\left(\frac{1}{\slashed{k}}\right)^2\left(\frac{g_{\mu\nu}}{k^2}\right) \left(\frac{g_{\mu\nu}-\cdots}{-M^2}\right).
\]
The crucial difference (the only difference!) is that the $W$-propagator has been replaced by a constant, which now gives,  as $k\to\infty$, an integrand  $\sim1/k^4$, {\it i.e.}, logarithmically divergent. The equation makes no sense, one side is a number the other is formally infinite. 

Oops. We need to renormalize. But before that we can see if the finite part of the infinite RHS has the correct dependence on external momenta, $p_i$, up to some trivial $p_i$ dependence in the infinite part. By ``correct dependence'' I mean, of course, that it reproduces the $p_i$ dependence of the LHS. 

To this end take $\partial/\partial p_i^\mu$ on both sides of the would-be-equation (obviously now restoring the external momentum and, if need be, the small masses of the light fields). Now, when the external momentum $p_i$ appears in the propagator of some internal light-field inside the loop (an internal line), then the action of $\partial/\partial p_i^\mu$  on this propagator increases the degree of convergence. For example, for a scalar field,
\[
\frac{\partial}{\partial p_\mu}\frac1{(k+p)^2-m^2}=-2\frac{k^\mu+p^\mu}{\left[(k+p)^2-m^2\right]^2}\sim\frac1{k^3}
\]
Diagrammatically,
\begin{equation}
\label{eq:diag-deriv-insert}
\frac{\partial}{\partial p_\mu}\Bigg(
\begin{tikzpicture}[scale=1]
\coordinate (A) at (0,0);
\coordinate (x) at (1,0);
\coordinate (y) at (0,.2);

\coordinate (v3) at  ($(A)+(x)$);

\draw (A)  -- (v3);
\draw[->,>=stealth,thick] ($(A)+0.3*(x)+(y)$) -- node[above]{$k+p$} +($0.3*(x)$);
\end{tikzpicture} 
\Bigg)
=
\hbox to 2.5cm{\hskip-0.5cm\vbox to 0.85cm {%
\begin{tikzpicture}[scale=1]
\coordinate (A) at (0,0);
\coordinate (x) at (1,0);
\coordinate (y) at (0,.2);

\coordinate (v3) at  ($(A)+(x)$); 
\coordinate (v4) at  ($(v3)+(x)$);

\draw (A)  -- (v3) node {$\times$} -- (v4) ;
\draw[->,>=stealth,thick] ($(A)+0.2*(x)+(y)$) -- node[above]{$k+p$} +($0.3*(x)$);
\end{tikzpicture} 
}}
\end{equation}
where the vertex, indicated as a cross,  stands for  $2i(k+p)^\mu$. The same is true for other propagators. 

\begin{exercises}
\begin{exercise}
Show this.  That is, that $\partial/\partial p_i^\mu$  on an internal
quark or gluon propagator changes the loop momentum UV asymptotic
scaling of the propagator from $1/k$ and $1/k^2$ to $1/k^2$ and
$1/k^3$, respectively. Show that the result can be expressed
diagrammatically just as in \eqref{eq:diag-deriv-insert}, with the
appropriate choice for the vertex indicated as a cross. 
\end{exercise}
\end{exercises}
Returning to the question at hand, we take a derivative and ask whether the following holds:

\begin{center}

\begin{tikzpicture}[scale=1.3,gaugeZ/.style={decorate, draw=black,
    decoration={coil,aspect=0, segment length=5,amplitude=3}},sgluon/.style={decorate, draw=black,
    decoration={coil,aspect=0.5,segment length=3pt,amplitude=2pt}}
] 

\coordinate (A) at (0,0);
\coordinate (x) at (1,0);
\coordinate (y) at (0,1);

\coordinate (v1) at ($(A)+0.5*(y)$);
\coordinate (v2) at  ($(A)-0.5*(y)$);
\coordinate (v3) at  ($(A)+(x)$); 
\coordinate (v4) at ($(v3)+(x)$); 
\coordinate (v5) at ($(v4)+(x)+0.5*(y)$);
\coordinate (v6) at  ($(v4)+(x)-0.5*(y)$);

\draw (v1)  -- (v3);
\draw (v2) -- (v3) ;
\draw (v4)  -- (v5);
\draw (v4) -- (v6);

\draw[gaugeZ] (v3) -- (v4);
\draw[{decorate, draw=black,
    decoration={coil,aspect=0.4,segment length=5pt,amplitude=2pt}}] ($0.7*(v1)+0.3*(v3)$) -- ($0.3*(v4)+0.7*(v5)$);

\node[rotate=30] at  ($0.3*(v1)+0.7*(v3)$) {$+$}; 
\begin{scope}[xshift = 4cm]
\node[scale=2] {$? \atop =$} (A);
\end{scope}

\begin{scope}[xshift = 6cm]
\coordinate (A) at (0,0);
\coordinate (x) at (1,0);
\coordinate (y) at (0,1);

\coordinate (v1) at ($(A)-(x)+0.5*(y)$);
\coordinate (v2) at  ($(A)+(x)+0.5*(y)$);
\coordinate (v3) at  ($(A)-(x)-0.5*(y)$); 
\coordinate (v4) at ($(A)+(x)-0.5*(y)$);
\draw (v1) -- (A);
\draw (A) -- (v2) ;
\draw (v4)   -- (A);
\draw (A) -- (v3);
\draw[fill] (A) circle (0.1cm);

\draw[{decorate, draw=black,
    decoration={coil,aspect=0.4,segment length=4pt,amplitude=2pt}}]
  ($0.7*(v1)+0.3*(A)$) --  ($0.7*(v2)+0.3*(A)$); 
\node[rotate=30]  at  ($0.3*(v1)+0.7*(A)$) {$+$}; 

\end{scope}
\begin{scope}[xshift = 8cm]
\node[scale=1.4]{$+~~\mathcal{O}\left(\frac1{M^3}\right)$} (A);
\end{scope}

\end{tikzpicture} 
\end{center}
To see that this actually does hold, consider 
\[\lim_{M\to\infty}\left(M^2\cdot 
\hbox to 4.5cm{\hskip-0.5cm\vbox to 0.8cm {%
\begin{tikzpicture}[scale=1.3,gaugeZ/.style={decorate, draw=black,
    decoration={coil,aspect=0, segment length=5,amplitude=3}},sgluon/.style={decorate, draw=black,
    decoration={coil,aspect=0.5,segment length=3pt,amplitude=2pt}}
] 

\coordinate (A) at (0,0);
\coordinate (x) at (1,0);
\coordinate (y) at (0,1);

\coordinate (v1) at ($(A)+0.5*(y)$);
\coordinate (v2) at  ($(A)-0.5*(y)$);
\coordinate (v3) at  ($(A)+(x)$); 
\coordinate (v4) at ($(v3)+(x)$); 
\coordinate (v5) at ($(v4)+(x)+0.5*(y)$);
\coordinate (v6) at  ($(v4)+(x)-0.5*(y)$);

\draw (v1)  -- (v3);
\draw (v2) -- (v3) ;
\draw (v4)  -- (v5);
\draw (v4) -- (v6);

\draw[gaugeZ] (v3) -- (v4);
\draw[{decorate, draw=black,
    decoration={coil,aspect=0.4,segment length=5pt,amplitude=2pt}}] ($0.7*(v1)+0.3*(v3)$) -- ($0.3*(v4)+0.7*(v5)$);

\node[rotate=30] at  ($0.3*(v1)+0.7*(v3)$) {$+$}; 

\end{tikzpicture} 
}}
\right)=\lim_{M\to\infty}\left (M^2\int \!\! d^4k (\cdots) \frac1{k^2-M^2}\right).
\]
For simplicity I have chosen the routing of the momentum through the loop so that the momentum of the $W$ is the loop momentum $k$. I have indicated the rest of the factors in the diagram by ``$(\cdots)$.''  All we need to know about this factor is that (i) it is common to the two sides of the equation we are studying and (ii) it contains sufficient inverse  powers of $k$ to render them finite. Now, we are taking two limits. One is the explicit one, and the other one is in the limits of integration of the integral over momenta. We can ask whether the order of taking the limits matter. Well, 
\(\lim_{M\to\infty}\left (\int \!\! d^4k  M^2 (\cdots) \frac1{k^2-M^2}\right)\) converges uniformly, and so does
\(\int \!\! d^4k  \lim_{M\to\infty} \left (M^2 (\cdots) \frac1{k^2-M^2}\right)\), and by standard mathematical non-sense both converge to the same limit. But the latter is simply the result of the effective theory calculation since  \( \lim_{M\to\infty} \left (M^2 (\cdots) \frac1{k^2-M^2}\right)=-(\cdots)\) or, diagrammatically,

\[\lim_{M\to\infty}\left[M^2\left( 
\hbox to 4.5cm{\hskip-0.5cm\vbox to 0.8cm {%
\begin{tikzpicture}[scale=1.3,gaugeZ/.style={decorate, draw=black,
    decoration={coil,aspect=0, segment length=5,amplitude=3}},sgluon/.style={decorate, draw=black,
    decoration={coil,aspect=0.5,segment length=3pt,amplitude=2pt}}
] 

\coordinate (A) at (0,0);
\coordinate (x) at (1,0);
\coordinate (y) at (0,1);

\coordinate (v1) at ($(A)+0.5*(y)$);
\coordinate (v2) at  ($(A)-0.5*(y)$);
\coordinate (v3) at  ($(A)+(x)$); 
\coordinate (v4) at ($(v3)+(x)$); 
\coordinate (v5) at ($(v4)+(x)+0.5*(y)$);
\coordinate (v6) at  ($(v4)+(x)-0.5*(y)$);

\draw (v1)  -- (v3);
\draw (v2) -- (v3) ;
\draw (v4)  -- (v5);
\draw (v4) -- (v6);

\draw[gaugeZ] (v3) -- (v4);
\draw[{decorate, draw=black,
    decoration={coil,aspect=0.4,segment length=5pt,amplitude=2pt}}] ($0.7*(v1)+0.3*(v3)$) -- ($0.3*(v4)+0.7*(v5)$);

\node[rotate=30] at  ($0.3*(v1)+0.7*(v3)$) {$+$}; 

\end{tikzpicture} 
}}
-
\hbox to 3cm{\hskip-0.5cm\vbox to 0.8cm {%
\begin{tikzpicture}[scale=1.3,gaugeZ/.style={decorate, draw=black,
    decoration={coil,aspect=0, segment length=5,amplitude=3}},sgluon/.style={decorate, draw=black,
    decoration={coil,aspect=0.5,segment length=3pt,amplitude=2pt}}
] 
\coordinate (A) at (0,0);
\coordinate (x) at (1,0);
\coordinate (y) at (0,1);

\coordinate (v1) at ($(A)-(x)+0.5*(y)$);
\coordinate (v2) at  ($(A)+(x)+0.5*(y)$);
\coordinate (v3) at  ($(A)-(x)-0.5*(y)$); 
\coordinate (v4) at ($(A)+(x)-0.5*(y)$);
\draw (v1) -- (A);
\draw (A) -- (v2) ;
\draw (v4)   -- (A);
\draw (A) -- (v3);
\draw[fill] (A) circle (0.1cm);

\draw[{decorate, draw=black,
    decoration={coil,aspect=0.4,segment length=4pt,amplitude=2pt}}]
  ($0.7*(v1)+0.3*(A)$) --  ($0.7*(v2)+0.3*(A)$); 
\node[rotate=30]  at  ($0.3*(v1)+0.7*(A)$) {$+$}; 

\end{tikzpicture} 
}}
\right)\right]=0
\]
As we saw earlier, this is what we mean formally by equality up to corrections that vanish as $1/M$  of the full and effective (once-differentiated) graphs. 

Next, do this for each propagator on which $\partial/\partial p_i$ acts and for each momenta: think of these as different graphs with one insertion of the ``cross'' vertex. Do it also for each non-vanishing light mass that may appear in loop propagators. The same argument goes through, graph by graph,  in establishing the equality between full and effective theories.\footnote{Maybe it's worth emphasizing that we are only considering insertion of the ``cross'' in internal propagators. On external propagators the degree of divergence of the graph is not reduced. You can alternatively consider only amputated Green functions.} We conclude then that  
\[
\frac{\partial}{\partial p_i}\left(
\hbox to 3.2cm{\hskip-0.5cm\vbox to 0.6cm {%
\begin{tikzpicture}[scale=1,gaugeZ/.style={decorate, draw=black,
    decoration={coil,aspect=0, segment length=5,amplitude=3}},sgluon/.style={decorate, draw=black,
    decoration={coil,aspect=0.5,segment length=3pt,amplitude=2pt}}
] 

\coordinate (A) at (0,0);
\coordinate (x) at (1,0);
\coordinate (y) at (0,1);

\coordinate (v1) at ($(A)+0.5*(y)$);
\coordinate (v2) at  ($(A)-0.5*(y)$);
\coordinate (v3) at  ($(A)+(x)$); 
\coordinate (v4) at ($(v3)+(x)$); 
\coordinate (v5) at ($(v4)+(x)+0.5*(y)$);
\coordinate (v6) at  ($(v4)+(x)-0.5*(y)$);

\draw (v1)  -- (v3);
\draw (v2) -- (v3) ;
\draw (v4)  -- (v5);
\draw (v4) -- (v6);

\draw[gaugeZ] (v3) -- (v4);
\draw[{decorate, draw=black,
    decoration={coil,aspect=0.4,segment length=5pt,amplitude=2pt}}] ($0.6*(v1)+0.4*(v3)$) -- ($0.4*(v4)+0.6*(v5)$);

\end{tikzpicture}
}}\right)
=\frac{\partial}{\partial p_i}\left(
\hbox to 2.2cm{\hskip-0.5cm\vbox to 0.6cm {%
\begin{tikzpicture}[scale=1,gaugeZ/.style={decorate, draw=black,
    decoration={coil,aspect=0, segment length=5,amplitude=3}},sgluon/.style={decorate, draw=black,
    decoration={coil,aspect=0.5,segment length=3pt,amplitude=2pt}}
] 
\coordinate (A) at (0,0);
\coordinate (x) at (1,0);
\coordinate (y) at (0,1);

\coordinate (v1) at ($(A)-(x)+0.5*(y)$);
\coordinate (v2) at  ($(A)+(x)+0.5*(y)$);
\coordinate (v3) at  ($(A)-(x)-0.5*(y)$); 
\coordinate (v4) at ($(A)+(x)-0.5*(y)$);
\draw (v1) -- (A);
\draw (A) -- (v2) ;
\draw (v4)   -- (A);
\draw (A) -- (v3);
\draw[fill] (A) circle (0.1cm);

\draw[{decorate, draw=black,
    decoration={coil,aspect=0.4,segment length=4pt,amplitude=2pt}}]
  ($0.6*(v1)+0.4*(A)$) --  ($0.6*(v2)+0.4*(A)$); 

\end{tikzpicture} 
}}
\right)
\]
and similarly for light field mass derivatives. Integrating we get
\[
\hbox to 3.2cm{\hskip-0.5cm\vbox to 0.6cm {%
\begin{tikzpicture}[scale=1,gaugeZ/.style={decorate, draw=black,
    decoration={coil,aspect=0, segment length=5,amplitude=3}},sgluon/.style={decorate, draw=black,
    decoration={coil,aspect=0.5,segment length=3pt,amplitude=2pt}}
] 

\coordinate (A) at (0,0);
\coordinate (x) at (1,0);
\coordinate (y) at (0,1);

\coordinate (v1) at ($(A)+0.5*(y)$);
\coordinate (v2) at  ($(A)-0.5*(y)$);
\coordinate (v3) at  ($(A)+(x)$); 
\coordinate (v4) at ($(v3)+(x)$); 
\coordinate (v5) at ($(v4)+(x)+0.5*(y)$);
\coordinate (v6) at  ($(v4)+(x)-0.5*(y)$);

\draw (v1)  -- (v3);
\draw (v2) -- (v3) ;
\draw (v4)  -- (v5);
\draw (v4) -- (v6);

\draw[gaugeZ] (v3) -- (v4);
\draw[{decorate, draw=black,
    decoration={coil,aspect=0.4,segment length=5pt,amplitude=2pt}}] ($0.6*(v1)+0.4*(v3)$) -- ($0.4*(v4)+0.6*(v5)$);

\end{tikzpicture}
}}
=
\hbox to 2.2cm{\hskip-0.5cm\vbox to 0.6cm {%
\begin{tikzpicture}[scale=1,gaugeZ/.style={decorate, draw=black,
    decoration={coil,aspect=0, segment length=5,amplitude=3}},sgluon/.style={decorate, draw=black,
    decoration={coil,aspect=0.5,segment length=3pt,amplitude=2pt}}
] 
\coordinate (A) at (0,0);
\coordinate (x) at (1,0);
\coordinate (y) at (0,1);

\coordinate (v1) at ($(A)-(x)+0.5*(y)$);
\coordinate (v2) at  ($(A)+(x)+0.5*(y)$);
\coordinate (v3) at  ($(A)-(x)-0.5*(y)$); 
\coordinate (v4) at ($(A)+(x)-0.5*(y)$);
\draw (v1) -- (A);
\draw (A) -- (v2) ;
\draw (v4)   -- (A);
\draw (A) -- (v3);
\draw[fill] (A) circle (0.1cm);

\draw[{decorate, draw=black,
    decoration={coil,aspect=0.4,segment length=4pt,amplitude=2pt}}]
  ($0.6*(v1)+0.4*(A)$) --  ($0.6*(v2)+0.4*(A)$); 

\end{tikzpicture} 
}}
+C
\] 
The difference $C$ in the amputated Green functions, which we know is infinite,  must vanish upon differentiation with respect to the external momenta or the light masses. $C$ may depend on $g_s$, $g_w$ and $M$ (really $g_2$ and $M_W$). And it must have the same chiral structure as the other terms, $\gamma^\mu(1-\gamma_5)\otimes\gamma_\mu(1-\gamma_5)$ which, up to a numerical factor is just the tree level amputated Green function in the effective theory. So  writing
\[ C=c
\hbox to 9.5cm{\hskip-0.25cm\vbox to 0.5cm {%
\begin{tikzpicture}[scale=0.7,gaugeZ/.style={decorate, draw=black,
    decoration={coil,aspect=0, segment length=5,amplitude=3}}] 

\coordinate (A) at (0,0);
\coordinate (x) at (1,0);
\coordinate (y) at (0,1);

\coordinate (v1) at ($(A)-(x)+0.5*(y)$);
\coordinate (v2) at  ($(A)+(x)+0.5*(y)$);
\coordinate (v3) at  ($(A)-(x)-0.5*(y)$); 
\coordinate (v4) at ($(A)+(x)-0.5*(y)$);
\draw (v1) -- (A);
\draw (A) -- (v2) ;
\draw (v4)   -- (A);
\draw (A) -- (v3);
\draw[fill] (A) circle (0.1cm);

\end{tikzpicture}
}}
\]
and noting that $c$ is oder $\alpha_s/\pi$, we have
\[
\hbox to 7cm{\hskip-0.4cm\vbox to 0.8cm {%
\begin{tikzpicture}[scale=0.8,gaugeZ/.style={decorate, draw=black,
    decoration={coil,aspect=0, segment length=5,amplitude=3}},sgluon/.style={decorate, draw=black,
    decoration={coil,aspect=0.5,segment length=3pt,amplitude=2pt}}
] 
\coordinate (A) at (0,0);
\coordinate (x) at (1,0);
\coordinate (y) at (0,1);

\coordinate (v1) at ($(A)+0.5*(y)$);
\coordinate (v2) at  ($(A)-0.5*(y)$);
\coordinate (v3) at  ($(A)+(x)$); 
\coordinate (v4) at ($(v3)+(x)$); 
\coordinate (v5) at ($(v4)+(x)+0.5*(y)$);
\coordinate (v6) at  ($(v4)+(x)-0.5*(y)$);

\draw (v1)  -- (v3);
\draw (v2) -- (v3) ;
\draw (v4)  -- (v5);
\draw (v4) -- (v6);

\draw[gaugeZ] (v3) -- node[below]{$W$} (v4);

\begin{scope}[xshift = 4cm]
\node{$+$} (A);
\end{scope}

\begin{scope}[xshift = 5cm]
\coordinate (A) at (0,0);
\coordinate (x) at (1,0);
\coordinate (y) at (0,1);

\coordinate (v1) at ($(A)+0.5*(y)$);
\coordinate (v2) at  ($(A)-0.5*(y)$);
\coordinate (v3) at  ($(A)+(x)$); 
\coordinate (v4) at ($(v3)+(x)$); 
\coordinate (v5) at ($(v4)+(x)+0.5*(y)$);
\coordinate (v6) at  ($(v4)+(x)-0.5*(y)$);

\draw (v1)  -- (v3);
\draw (v2) -- (v3) ;
\draw (v4)  -- (v5);
\draw (v4) -- (v6);

\draw[gaugeZ] (v3) -- node[below]{$W$} (v4);
\draw[{decorate, draw=black,
    decoration={coil,aspect=0.4,segment length=5pt,amplitude=2pt}}] ($0.6*(v1)+0.4*(v3)$) -- 
                node[above]{$g$} ($0.4*(v4)+0.6*(v5)$);

\end{scope}
\end{tikzpicture}
}}
= \left(1+c\right)\left[
\hbox to 4.5cm{\hskip-0.4cm\vbox to 0.8cm {%
\begin{tikzpicture}[scale=0.7,gaugeZ/.style={decorate, draw=black,
    decoration={coil,aspect=0, segment length=5,amplitude=3}}] 

\coordinate (A) at (0,0);
\coordinate (x) at (1,0);
\coordinate (y) at (0,1);

\coordinate (v1) at ($(A)-(x)+0.5*(y)$);
\coordinate (v2) at  ($(A)+(x)+0.5*(y)$);
\coordinate (v3) at  ($(A)-(x)-0.5*(y)$); 
\coordinate (v4) at ($(A)+(x)-0.5*(y)$);
\draw (v1) -- (A);
\draw (A) -- (v2) ;
\draw (v4)   -- (A);
\draw (A) -- (v3);
\draw[fill] (A) circle (0.1cm);

\begin{scope}[xshift = 2cm]
\node{$+$} (A);
\end{scope}

\begin{scope}[xshift = 4cm]
\coordinate (A) at (0,0);
\coordinate (x) at (1,0);
\coordinate (y) at (0,1);

\coordinate (v1) at ($(A)-(x)+0.5*(y)$);
\coordinate (v2) at  ($(A)+(x)+0.5*(y)$);
\coordinate (v3) at  ($(A)-(x)-0.5*(y)$); 
\coordinate (v4) at ($(A)+(x)-0.5*(y)$);
\draw (v1) -- (A);
\draw (A) -- (v2) ;
\draw (v4)   -- (A);
\draw (A) -- (v3);
\draw[fill] (A) circle (0.1cm);

\draw[{decorate, draw=black,
    decoration={coil,aspect=0.4,segment length=4pt,amplitude=2pt}}]
  ($0.6*(v1)+0.4*(A)$) -- node[above]{$g$}  ($0.6*(v2)+0.4*(A)$); 
\end{scope}

\end{tikzpicture} 
}}
\right]
\]
Note that $c$ is infinite and this equation holds because it cancels the infinity in the 1-loop diagram on the RHS. That is, it is the counter-term to that 1-loop diagram.

This goes through when  we include all 1-loop graphs that contribute to the amputated Green functions. So we obtain, at least at 1-loop,
\[
\Gamma^{(4)}(p_1,\ldots, p_4)=\frac1{M^2}\,D \,\widetilde{\Gamma}^{(4)}_{\mathcal{O}}(p_1,\ldots, p_4)+\cdots
\]
where
\begin{align*}
\Gamma &=\text{amputated Green function, renormalized}\\
\widetilde\Gamma &=\text{idem in the EFT with Lagrangian $\mathcal{\tilde L}_{\text{light}}$}\\
\widetilde\Gamma_{\mathcal{O}} &=\text{idem with a zero momentum insertion of the 4-quark operator $\mathcal{O}$}\\
D&=\text{finite coefficient, of order $\alpha_s/\pi$}\\
\cdots &=\text{higher order in $1/M^2$}
\end{align*}
I hope you recognize this is precisely the statement that $\mathcal{L}_{\text{eff}}$   of Eq.~\ref{eq:Leff-introduced}
gives the desired approximation to the Green function. 

Comments:
\begin{enumerate}[]
\item $D$ has an expansion, $D=1+d_1\frac{\alpha_s}{\pi}+d_2\left(\frac{\alpha_s}{\pi}\right)^2+\cdots$; it may depend on $M$ but not on $p_1,\ldots,p_4$ nor $m_j$.
\item The dependence on $p_1,\ldots,p_4$ and on $m_j$ of the full and EFT are the same. They have the same analytic structure (same cuts, poles, residues, and what not), provided the approximations we have made (such as $|p_i|\ll M$) are valid. But the full and EFT may differ badly when the approximation fails,  for example, for $|p_i|\sim M$.
\item The above result, we already stated, is summarized by Eq.~\ref{eq:Leff-introduced}, or more specifically, $\mathcal{L}_{\text{eff}}=\mathcal{\tilde L}_{\text{light}}+
\frac1{M^2_{\text{GUT}}}D \mathcal{O}+\cdots$. It should be clear that one is to compute in the EFT to all orders in the interactions in $ \mathcal{\tilde L}_{\text{light}}$ but only to first order in  $\frac1{M^2_{\text{GUT}}}D \mathcal{O}$. This is the statement  that the Green function with {\it one} insertion of the operator is  equal in the full and effective theories. So for may applications we find we need
\[\text{amplitude}\propto\langle\psi_\text{final}|\mathcal{L}_{\text{int}}|\psi_{\text{initial}}\rangle=
\frac1{M^2}\,D\,\langle\psi_\text{final}|\mathcal{O}|\psi_{\text{initial}}\rangle\]
Many of the difficulties in obtaining good predictions in flavor physics arise from the need to calculate the matrix elements of operators. 
\item One has to analyze and consider separately the case with two (or more) virtual heavy particles, as is the case, for example, for the box diagram that gives rise to neutral meson mixing. 
\end{enumerate}

\section{RGE improvement}
Let's inspect the function $D=D(M,g_s)$ more closely. It is dimensionless, so it depends on $M$ only through the ratio $M/\mu$,  with $\mu$ the renormalization scale. Of course, $D$ depends on $\mu$ but it is customary to leave this as implicit dependence. Now, the fact that amplitudes are $\mu$-independent immediately gives us
\begin{equation}
\label{eq:pre-RGE}
\mu\frac{d}{d\mu}
\langle\psi_\text{final}|\frac1{M^2}\,D\,\mathcal{O}|\psi_{\text{initial}}\rangle=0.
\end{equation}
Therefore, if
\[\mu\frac{d}{d\mu}\mathcal{O} =\gamma_{\mathcal{O}}\mathcal{O} \qquad\text{then}
\qquad \mu\frac{d}{d\mu}D =-\gamma_{\mathcal{O}}D.\]
Slow down a little. $\gamma_{\mathcal{O}}$ is called the anomalous dimension of the operator
$\mathcal{O}$. In a mass independent renormalization scheme, like dimensional regularization with MS (subtracting only the $\epsilon=d-4$ poles), the anomalous dimension is only a function of the dimensionless couplings, in this case $g_s$. 

The amazing thing about this is that the whole $M$-dependence of the coefficient function $D$ is fixed by the renormalization group! Let's see how this works in some detail. In order to do so we solve the renormalization group equation (RGE). Since $D=D(M/\mu,g_s)$ the dependence on $\mu$ is either through the ratio of scales, or its logarithm, $t=\ln(\mu/M)$, or implicitly through the coupling $g_s=g_s(\mu)$. Then  the RGE for $D$, Eq.~\ref{eq:pre-RGE}, is explicitly
\begin{equation}
\label{eq:RGE}
\left(\frac{\partial}{\partial t}+\beta(g)\frac{\partial}{\partial g}\right)D(t,g)=-\gamma_{\mathcal{O}}D(t,g)
\end{equation}
We have used $g$ for the strong coupling $g_s$ to streamline notation, and $\beta(g)$ is the beta function for the strong coupling constant. The solution to this equation is well known, but we review it here. We first introduce an auxiliary function, the running coupling constant $\widebar g=\widebar g(t,g)$, defined as the solution to
\[
\frac{d\widebar g(t)}{dt}=\beta\left(\widebar g(t)\right)\quad\text{with boundary condition}\quad \widebar g(0,g)=g.
\]

\begin{exercises}
\begin{exercise}
Show that 
\[ \frac{\partial \widebar g}{\partial g}=\frac{\beta(\widebar g)}{\beta(g)}.\]
Use this to show that the solution to the homogeneous RGE
\[
\left(\frac{\partial}{\partial t}+\beta(g)\frac{\partial}{\partial g}\right)F(t,g)=0
\]
is $F(t,g)=f(\widebar g(-t,g))$. That is, $t$ and $g$ can come into
the arbitrary function $f$ only in the combination $\widebar
g(-t,g)$. Make sure you get that sign in the argument of $\widebar g$
right! Verify that the function $f$ can be thought of as the boundary condition $f(g)=F(0,g)$.
\end{exercise}
\end{exercises}

If you solved the previous exercise then it is not difficult to verify that the solution to the RGE for $D$ is
\[
D(t,g)=D(0,\widebar g(-t,g))\exp\left({\int_g^{\widebar g(-t,g)}dg'\frac{\gamma_{\mathcal{O}}(g')}{\beta(g')}}\right).
\]
For example, at 1-loop one has 
\[
\gamma_{\mathcal{O}}(g)=a_1\frac{g^2}{16\pi^2}\qquad\text{and}\qquad
\beta(g)=-b_0\frac{g^3}{16\pi^2}
\]
Then, to this order, 
\[
\frac{d\widebar g}{dt}=-b_0\frac{\widebar{g}^3}{16\pi^2} \quad\Rightarrow
\frac{1}{\widebar\alpha(t)}=\frac{1}{\widebar \alpha(0)}+\frac{b_0}{2\pi}t
\]
where, as usual, $\alpha=g^2/4\pi$, and then
\[
\int_g^{\widebar g(-t)} dg'\, \frac{\gamma_{\mathcal{O}}(g')}{\beta(g')}
=-\int_g^{\widebar g(-t)} \frac{dg'}{g'}\, \frac{a_1}{b_0}
=-\frac{a_1}{2b_0}\ln\frac{\widebar\alpha(-t)}{\widebar\alpha(0)}
\]
so that
\[
D(t,g)=D(0,\widebar g(-t))\left(\frac{\widebar\alpha(-t)}{\widebar\alpha(0)}\right)^{-\frac{a_1}{2b_0}}
\]

There is some standard language that goes with this. We say we obtain
the EFT by ``integrating out'' the heavy fields (or sometimes, the
heavy ``degrees of freedom'') --- in this case the $W$ vector
boson. The factor
$\left(\frac{\widebar\alpha(-t)}{\widebar\alpha(0)}\right)^{-\frac{a_1}{2b_0}}$
is obtained by ``running'' while the computation of the pre-factor $D(0,\widebar g(-t))$ is
called ``matching.''  The matching computation can be performed by taking $t=0$,
as an expansion in the coupling constant at short distances,
\begin{equation}
\label{eq:EFTmatch}
D(0,\widebar g(0))=D(0, g)=D_0+\frac{\alpha}{4\pi}D_1+\cdots
\end{equation}
Notice that this is computed at $t=0$, that is, at $\mu=M$. This means that in the calculation one may encounter large logs of $p_i$ or $m_j$ over $\mu$. But the coefficient function $D$ is independent of these low energy variables. By taking $t=0$ the matching calculation consists of computing functions of one variable only, namely $\alpha$, and as indicated in \eqref{eq:EFTmatch} this can be done perturbatively. The point is that ``matching'' calculations ensure that the full and effective theories give the same answer, while ``running'' gives the full dependence on the renormalization scale $\mu$ and, as we will see, re-sums logs. For example, the coefficient $D_0$ is computed from comparing the leading order contribution to the amputated Green functions, as in 
\[
\hbox to 3cm{\hskip-0.4cm\vbox to 0.5cm {%
\begin{tikzpicture}[scale=0.9,gaugeZ/.style={decorate, draw=black,
    decoration={coil,aspect=0, segment length=5,amplitude=3}},sgluon/.style={decorate, draw=black,
    decoration={coil,aspect=0.5,segment length=3pt,amplitude=2pt}}
] 
\coordinate (A) at (0,0);
\coordinate (x) at (1,0);
\coordinate (y) at (0,1);

\coordinate (v1) at ($(A)+0.5*(y)$);
\coordinate (v2) at  ($(A)-0.5*(y)$);
\coordinate (v3) at  ($(A)+(x)$); 
\coordinate (v4) at ($(v3)+(x)$); 
\coordinate (v5) at ($(v4)+(x)+0.5*(y)$);
\coordinate (v6) at  ($(v4)+(x)-0.5*(y)$);

\draw (v1)  -- (v3);
\draw (v2) -- (v3) ;
\draw (v4)  -- (v5);
\draw (v4) -- (v6);

\draw[gaugeZ] (v3) -- node[below=0.1cm]{$W$} (v4);

\end{tikzpicture}
}}
= D_0\left[
\hbox to 2.cm{\hskip-0.4cm\vbox to 0.5cm {%
\begin{tikzpicture}[scale=0.7,gaugeZ/.style={decorate, draw=black,
    decoration={coil,aspect=0, segment length=5,amplitude=3}}] 

\coordinate (A) at (0,0);
\coordinate (x) at (1,0);
\coordinate (y) at (0,1);

\coordinate (v1) at ($(A)-(x)+0.5*(y)$);
\coordinate (v2) at  ($(A)+(x)+0.5*(y)$);
\coordinate (v3) at  ($(A)-(x)-0.5*(y)$); 
\coordinate (v4) at ($(A)+(x)-0.5*(y)$);
\draw (v1) -- (A);
\draw (A) -- (v2) ;
\draw (v4)   -- (A);
\draw (A) -- (v3);
\draw[fill] (A) circle (0.1cm);

\end{tikzpicture} 
}}
\right]
\]
and the next order coefficient, $D_1$, is computed form comparing 1-loop graphs,
\[
\hbox to 4.3cm{\hskip-0.4cm\vbox to 0.8cm {%
\begin{tikzpicture}[scale=0.9,gaugeZ/.style={decorate, draw=black,
    decoration={coil,aspect=0, segment length=5,amplitude=3}}] 
\coordinate (A) at (0,0);
\coordinate (x) at (1,0);
\coordinate (y) at (0,1);

\coordinate (v1) at ($(A)+0.5*(y)$);
\coordinate (v2) at  ($(A)-0.5*(y)$);
\coordinate (v3) at  ($(A)+(x)$); 
\coordinate (v4) at ($(v3)+(x)$); 
\coordinate (v5) at ($(v4)+(x)+0.5*(y)$);
\coordinate (v6) at  ($(v4)+(x)-0.5*(y)$);

\draw (v1)  -- (v3);
\draw (v2) -- (v3) ;
\draw (v4)  -- (v5);
\draw (v4) -- (v6);

\draw[gaugeZ] (v3) -- node[below]{$W$} (v4);
\draw[{decorate, draw=black,
    decoration={coil,aspect=0.4,segment length=5pt,amplitude=2pt}}] ($0.6*(v1)+0.4*(v3)$) --   node[above]{$g$} ($0.4*(v4)+0.6*(v5)$);

\begin{scope}[xshift = 3.8cm]
\node{$+\cdots$} (A);
\end{scope}
\end{tikzpicture}
}}
= D_1\frac{\widebar\alpha(0)}{4\pi}
\hbox to 6.5cm{\hskip-0.4cm\vbox to 0.7cm {%
\begin{tikzpicture}[scale=0.9,gaugeZ/.style={decorate, draw=black,
    decoration={coil,aspect=0, segment length=5,amplitude=3}}] 

\coordinate (A) at (0,0);
\coordinate (x) at (1,0);
\coordinate (y) at (0,1);

\coordinate (v1) at ($(A)-(x)+0.5*(y)$);
\coordinate (v2) at  ($(A)+(x)+0.5*(y)$);
\coordinate (v3) at  ($(A)-(x)-0.5*(y)$); 
\coordinate (v4) at ($(A)+(x)-0.5*(y)$);
\draw (v1) -- (A);
\draw (A) -- (v2) ;
\draw (v4)   -- (A);
\draw (A) -- (v3);
\draw[fill] (A) circle (0.1cm);

\begin{scope}[xshift = 1.5cm]
\node{$+$} (A);
\end{scope}

\begin{scope}[xshift = 3cm]
\coordinate (A) at (0,0);
\coordinate (x) at (1,0);
\coordinate (y) at (0,1);

\coordinate (v1) at ($(A)-(x)+0.5*(y)$);
\coordinate (v2) at  ($(A)+(x)+0.5*(y)$);
\coordinate (v3) at  ($(A)-(x)-0.5*(y)$); 
\coordinate (v4) at ($(A)+(x)-0.5*(y)$);
\draw (v1) -- (A);
\draw (A) -- (v2) ;
\draw (v4)   -- (A);
\draw (A) -- (v3);
\draw[fill] (A) circle (0.1cm);

\draw[{decorate, draw=black,
    decoration={coil,aspect=0.4,segment length=4pt,amplitude=2pt}}]
  ($0.6*(v1)+0.4*(A)$) -- node[above]{$g$}  ($0.6*(v2)+0.4*(A)$); 
\end{scope}
\begin{scope}[xshift = 4.8cm]
\node{$+\cdots$} (A);
\end{scope}
\end{tikzpicture} 
}}
\]
where the ellipsis stand for all other one-gluon exchange 1-loop
diagrams (on both sides of the equation). Here by  $D_1$ we mean  the
finite part of the counter-term,  well defined in any specific renormalization scheme, {\it e.g.}, MS in dimensional regularization.

Comments:
\begin{enumerate}[]
\item From 
\[\frac{\widebar\alpha(0)}{\widebar\alpha(-t)}=1+\frac{\alpha}{2\pi}b_0\ln\frac{M}{\mu}
\]
we have
\begin{equation}
\label{eq:LL}
\left(\frac{\widebar\alpha(-t)}{\widebar\alpha(0)}\right)^{-\frac{a_1}{2b_0}}=
\left(1+\frac{\alpha}{2\pi}b_0\ln\frac{M}{\mu}\right)^{\frac{a_1}{2b_0}}
=1+a_1\frac{\alpha}{4\pi}\ln\frac{M}{\mu}+\cdots
\end{equation}
The ellipsis stand for an infinite sum of terms of the form $\left(\frac{\alpha}{2\pi}b_0\ln{M}/{\mu}\right)^n$. This is why we say the running gives a re-summation of the leading logs. Also, the $n=1$ term, displayed explicitly, is precisely what determines the anomalous dimension $\gamma_{\mathcal{O}}$ at 1-loop. Note, finally, that in the expansion we cannot say whether $\alpha$ is evaluated at $M$ or at $\mu$, and we have therefore simply denoted it as $\alpha$. The EFT method has allowed us to make sense of the scale at which $\widebar\alpha$ is evaluated. 
\item In
  $D\,\langle\psi_\text{final}|\mathcal{O}|\psi_\text{initial}\rangle$
  we have cleanly separated the physical scales. This is why we
  introduced the EFT to begin with. The coefficient $D$ depends on
  $\ln M/\mu$, while the matrix element
  $\langle\psi|\mathcal{O}|\psi\rangle$ is independent of $M$ and 
contains $\ln\mu$ only through $\ln \mu/E$ where $E$ stands for any of the  low energy scales associated with the physical process. This shows that we have to deal with large logs: either in $D$ if we take $\mu\sim E$ or in the matrix element if we take $\mu\sim M$. We have explained above that we can compute $D$ even in the presence of large $\ln M/E$ perturbatively, provided $\widebar\alpha(0)$ is sufficiently small for perturbation theory to work. This is clearly the case for $M=$ the mass of the $W/Z$ vector bosons.   
The calculation of $D$ at low $\mu$ is valid provided the leading log
expansion continues to make sense, so that  $\widebar\alpha(-t)$
remains sufficiently small, although possibly very different form
$\widebar\alpha(0)$. In practice $\mu$ is commonly taken to be as low
as 2~GeV (or sometimes even 1~GeV), and we let the non-perturbative
method that determines the matrix element deal with the not so large
$\ln \mu/E$. Even for  $E=\Lambda_{\text{QCD}}\approx 300$~MeV, this
is a reasonably small ratio of scales.
\item Operator mixing: The discussion above has been very sketchy. In most cases we have to deal with more than one operator in the EFT Lagrangian at the time. A simple example is in the EFT for charm decay. Let's take
\begin{equation}
\label{eq:mix1}
\mathcal{O}_1=\widebar c_L\gamma^\mu s_L\,\widebar d_L\gamma_\mu u_L =
\hbox to 2.cm{\hskip0.1cm\vbox to 0.85cm {%
\begin{tikzpicture}[scale=0.9,gaugeZ/.style={decorate, draw=black,
    decoration={coil,aspect=0, segment length=5,amplitude=3}}] 

\coordinate (A) at (0,0);
\coordinate (x) at (1,0);
\coordinate (y) at (0,1.3);
\coordinate (z) at (0,0.05);

\coordinate (v1) at ($(A)-(x)+0.5*(y)$);
\coordinate (v2) at  ($(A)+(x)+0.5*(y)$);
\coordinate (v3) at  ($(A)-(x)-0.5*(y)$); 
\coordinate (v4) at ($(A)+(x)-0.5*(y)$);
\coordinate (vu) at ($(A)+(z)$);
\coordinate (vd) at ($(A)-(z)$);

\draw[particle] (v1) node[left]{$s$}-- (vu);
\draw[particle] (vu) -- (v2)node[right]{$c$} ;
\draw[particle] (v4) node[right]{$u$}  -- (vd);
\draw[particle] (vd) -- (v3) node[left]{$d$};

\end{tikzpicture} 
}}
\end{equation}
\vskip0.5cm
\noindent I have drawn the figure with the local vertex slightly separated as if
there were two distinct vertices, when they are not. It is simply
a device to indicate color flow (which also corresponds in this case to
flow of the indices of the Dirac spinors). What I mean by this is that
we have kept the color indices implicit. To be sure, the operator with
explicit color indices indicated is 
  $\mathcal{O}_1=\widebar c^i_L\gamma^\mu s_{Li}\,\widebar{d}^j_L\gamma_\mu u_{Lj}$
with the sum over $i$ and $j$ from 1 to $N_c=3$ understood. This
operator is the one you get from tree-level matching by integrating
out the $W$ vector-boson. Now, at
1-loop the graph
\begin{center}
\begin{tikzpicture}[scale=0.9]

\coordinate (A) at (0,0);
\coordinate (x) at (1,0);
\coordinate (y) at (0,1.3);
\coordinate (z) at (0,0.05);

\coordinate (v1) at ($(A)-(x)+0.5*(y)$);
\coordinate (v2) at  ($(A)+(x)+0.5*(y)$);
\coordinate (v3) at  ($(A)-(x)-0.5*(y)$); 
\coordinate (v4) at ($(A)+(x)-0.5*(y)$);
\coordinate (vu) at ($(A)+(z)$);
\coordinate (vd) at ($(A)-(z)$);

\draw[particle] (v1) node[left]{$s$}-- (vu);
\draw (vu) -- (v2)node[right]{$c$} ;
\draw (v4) node[right]{$u$}  -- (vd);
\draw[particle] (vd) -- (v3) node[left]{$d$};

\draw[{decorate, draw=black,
    decoration={coil,aspect=0.4,segment length=4pt,amplitude=2pt}}] ($0.3*(vu)+0.7*(v2)$) -- ($0.3*(vd)+0.7*(v4)$) ;
\end{tikzpicture} 
\end{center}
requires a counter-term that, even without computing can be seen to be
an operator 
\begin{equation}
\label{eq:mix2}
\mathcal{O}_2=\widebar c_L\gamma^\mu T^a s_L\,\widebar d_L\gamma_\mu T^a u_L
\end{equation}
where $T^a$ are the generators of $SU(N_c)$  and
the quark (and $T^a$) color-indices and the sum over $a$ are implicit. This means that when we
compute ``running'' a term with $\mathcal{O}_2$ will be generate in
the EFT Lagrangian. More explicitly, we write 
\[\mathcal{L}^{(6)}= \sum_{i=1,2} D_i \mathcal{O}_i\]
and the requirement that amplitudes be $\mu$ independent is the same
as $\mu d\mathcal{L}^{(6)}/d\mu =\mu dD_i \mathcal{O}_i/d\mu =0$ (sum on
repeated $i$-indices understood from here on). Then the RGE for the
operators is 
\[
\mu\frac{d}{d\mu}\mathcal{O}_i
=\gamma_{ij}\mathcal{O}_j\qquad\text{so that}
\qquad\mu\frac{d}{d\mu}D_i
=-D_j\gamma_{ji},
\]
where $\gamma$ is now a $2\times2$  matrix function of $g_s$. 
The non-vanishing counter-term of the form $\mathcal{O}_2$ gives
$\gamma_{12}\ne0$, and the RGE then contains a term $\mu dD_2/d\mu\propto
D_1$ which produces a non-vanishing $D_2(t)$ even if $D_2(0)=0$.
\item Tie some loose ends. In our discussion of $K^0$-$\widebar K^0$
  mixing we postponed discussion of how to handle the two loop graphs
  with gluons attaching to the internal $W$-box diagram. Consider, for
  example, the first Feynman diagram in \eqref{emptyForFig}. We can
  think physically of what is going on here. When all the loop momenta
  is high (meaning, of order of $M$ or larger) the whole diagram is
  effectively a point-like 4-quark interaction. This gives then a
  contribution to the 4-quark operator by matching,
\[
\hbox to 7.cm{\hskip0.1cm\vbox to 0.95cm {%
\begin{tikzpicture}[scale=1.3] 
\coordinate (A) at (0,0);
\coordinate (x) at (1,0);
\coordinate (y) at (0,-1);

\coordinate (v1) at ($(A)+(x)$);
\coordinate (v2) at  ($(v1)+(x)$);
\coordinate (v3) at  ($(v2)+(x)$);
\coordinate (v9) at  ($(v3)+(x)$);
\coordinate (v4) at ($(A)+(y)$);
\coordinate (v5) at  ($(v4)+(x)$);
\coordinate (v6) at  ($(v5)+(x)$);
\coordinate (v7) at  ($(v6)+(x)$);
\coordinate (v8) at  ($(v7)+(x)$);

\draw[particle] (A) node[left]{$s$} -- (v1) ;
\draw[particle] (v1) -- (v2);
\draw[particle] (v2) --  (v3);
\draw[particle] (v3) --  (v9) node[right]{$d$};
\draw[particle] (v8) node[right]{$s$} --  (v7);
\draw[particle] (v7) --  (v6);
\draw[particle] (v6) --   (v5);
\draw[particle] (v5)  -- (v4) node[left]{$d$};

\draw[photon] (v1) -- node[left]{$W$} (v5);
\draw[photon] (v3) -- node[right]{$W$} (v7);
\draw[gluon] (v2) -- node[right]{$g$} (v6);
\end{tikzpicture} 
}}
\rightarrow \quad\frac{\alpha_s}{4\pi}D_1
\hbox to 2.cm{\hskip-0.5cm\vbox to 1.1cm {%
\begin{tikzpicture}[scale=0.8] 
\coordinate (A) at (0,0);
\coordinate (x) at (1,0);
\coordinate (y) at (0,1);

\coordinate (v1) at ($(A)-(x)+(y)$);
\coordinate (v2) at  ($(A)+(x)+(y)$);
\coordinate (v3) at  ($(A)-(x)-(y)$); 
\coordinate (v4) at ($(A)+(x)-(y)$);

\draw[particle] (v1) node[left]{$s$} -- (A);
\draw[particle] (A) -- (v2) node[right]{$d$} ;
\draw[particle] (v4) node[right]{$s$}   -- (A);
\draw[particle] (A) -- (v3) node[left]{$d$} ;
\draw[fill] (A) circle (0.1cm);

\end{tikzpicture} 
}}
\]
\vskip0.5cm
\noindent What about the case when the gluon in the loop carries low momentum?
We can split the computation into two parts. First, compute the
matching to the operator that results from cutting open  the gluon
line:
\[
\hbox to 7.cm{\hskip0.1cm\vbox to 1.9cm {%
\begin{tikzpicture}[scale=1.3] 
\coordinate (A) at (0,0);
\coordinate (x) at (1,0);
\coordinate (y) at (0,-1);

\coordinate (v1) at ($(A)+(x)$);
\coordinate (v2) at  ($(v1)+(x)$);
\coordinate (v3) at  ($(v2)+(x)$); 
\coordinate (v9) at  ($(v3)+(x)$);
\coordinate (v4) at ($(A)+(y)$);
\coordinate (v5) at  ($(v4)+(x)$);
\coordinate (v6) at  ($(v5)+(x)$);
\coordinate (v7) at  ($(v6)+(x)$);
\coordinate (v8) at  ($(v7)+(x)$);

\draw[particle] (A) node[left]{$s$} -- (v1) ;
\draw[particle] (v1) -- (v2);
\draw[particle] (v2) --  (v3);
\draw[particle] (v3) --  (v9) node[right]{$d$};
\draw[particle] (v8) node[right]{$s$} --  (v7);
\draw[particle] (v7) --  (v6);
\draw[particle] (v6) --   (v5);
\draw[particle] (v5)  -- (v4) node[left]{$d$};

\draw[photon] (v1) -- node[left]{$W$} (v5);
\draw[photon] (v3) -- node[right]{$W$} (v7);
\draw[gluon] (v2) -- node[right]{$g$} ($(v2)-(y)$);
\draw[gluon] (v6) -- node[right]{$g$} ($(v6)+(y)$);

\end{tikzpicture} 
}}
\rightarrow \quad\frac{\alpha_s}{4\pi}D'_1
\hbox to 2.cm{\hskip-0.5cm\vbox to 1.1cm {%
\begin{tikzpicture}[scale=0.8] 
\coordinate (A) at (0,0);
\coordinate (x) at (1,0);
\coordinate (y) at (0,1);

\coordinate (v1) at ($(A)-(x)+(y)$);
\coordinate (v2) at  ($(A)+(x)+(y)$);
\coordinate (v3) at  ($(A)-(x)-(y)$); 
\coordinate (v4) at ($(A)+(x)-(y)$);

\draw[particle] (v1) node[left]{$s$} -- (A);
\draw[particle] (A) -- (v2) node[right]{$d$} ;
\draw[particle] (v4) node[right]{$s$}   -- (A);
\draw[particle] (A) -- (v3) node[left]{$d$} ;
\draw[fill] (A) circle (0.1cm);

\draw[gluon] (A) -- node[below right]{$g$} ($(A)-1.5*(y)$);
\draw[gluon] (A) -- node[above right]{$g$} ($(A)+1.5*(y)$);

\end{tikzpicture} 
}}
\]
\vskip1.25cm
\noindent This graph matches to an operator with 4-quarks and two gluons, which
is a dimension-8 operator (at least). The evaluation of the matrix
element of this operator will include a contribution in which the
gluons reconnect. At low energies this precisely reproduces the low
energy part of the first diagram in \eqref{emptyForFig}. At high
energies it gives again a local operator which is properly accounted
for in the 4-quark operator if we have done the complete matching
calculation correctly. Most importantly, since this is a dimension-8
operator it appears in the EFT Lagrangian with a coefficient
$1/M_W^4$, and is therefore suppressed by $E^2/M^2\ll 1$ relative to
the contribution of the 4-quark operator. Notice, incidentally, that 
I labeled the coefficient with a prime to make sure we distinguish
this from the  4-quark operator.
\end{enumerate}

\begin{exercises}
\begin{exercise}
Consider the operator that occurs in
$K^0$-$\widebar K^0$ mixing, namely, 
$\mathcal{O}=\widebar s_L\gamma^\mu d_L\,\widebar s_L\gamma^\mu
d_L$. 
Argue, without an explicit  computation, that it does not mix with
other operators under renormalization, so that  $\gamma_{\mathcal{O}}$
is a $1\times1$ matrix. 
Find $\gamma_{\mathcal{O}}$. Note: this  is not   an EFT calculation, but an exercise in
perturbative QCD. You must compute the renormalized Green function
with an insertion of the operator and then take the logarithmic $\mu$
derivative. If you have never seen this, you should first consult a
text on Quantum Field Theory that covers this (for example, Peskin and
Schroeder~\cite{Peskin:1995ev}, section 12.4).
\end{exercise}

\begin{exercise}
The operator in $\mathcal{O}_2$ in \eqref{eq:mix2} was introduced out
of necessity: it was found to be a counter-termcounter-termcounter-termcounter-term to the operator
produced by the weak interactions at tree level, $\mathcal{O}_1$ of
\eqref{eq:mix1}. 
\begin{enumerate}[(i)]
\item Show that this set of operators closes under
  renormalization. That is, no new operators need be introduced to
  renormalize this pair. This justifies using a $2\times 2$ matrix of
  anomalous dimensions.  You should be able to argue from symmetry and
  power counting, and therefore construct an all orders
  argument. Note: we are assuming we are working in a mass independent
  subtraction scheme, like MS in dimensional regularization, so that operators only mix
  with operators of the same dimension.
\item Show that the set of operators $\mathcal{Q}_1=\mathcal{O}_1$ and
  $\mathcal{Q}_2=\widebar c_L\gamma^\mu u_L\,\widebar
  d_L\gamma_\mu  s_L$ is equivalent to the set $\mathcal{O}_{1,2}$ and
  give the explicit relation. We sometimes speak of these as different
  ``basis'' of operators (we can think of the space of operators as a
  linear vector space, since we can form linear combinations of
  them, and the computation of Green functions with one insertion
  respects linearity). 
\end{enumerate}
\end{exercise}

\begin{exercise}
Our discussion of EFT used as a prototype the exchange of a heavy
field (a $W$) to produce a dimension-greater-than-4 operator in the
effective Lagrangian (a dimension-6 4-quark operator in the case of
$W$ exchange). For $K^0$-$\widebar K^0$ mixing, or more generally for
any neutral meson mixing, the full process in the full theory is from
a loop diagram containing two $W$ propagators; see
the ``box'' diagram in \eqref{eq:fig:KKbarmix}.  
\begin{enumerate}[(i)]
\item Consider the box diagram for fixed light internal quarks; for
  concreteness take both internal quark lines to be charm quarks. Use
  the methods of Sec.~\ref{sec:loops} to argue that even though the
  graph contains two $W$ propagators, it produces only one factor of
  $1/M_W^2$ in the coefficient of the EFT's 4-quark  operator.
\item Now include the other quarks. Work in the fictitious case that
  the top quark is much lighter than the $W$ so that the old-fashion
  GIM  mechanism applies; see Sec.\ref{ssec:oldgim}. Show that now the
  coefficient of the 4-quark operator has a $1/M_W^4$. What makes up
  the dimensions so the EFT Lagrangian is still of dimension-4, as it must?   
\end{enumerate}
\end{exercise}
\begin{exercise}
In the comments above we showed how to resolve the problem of light
degrees of freedom inside loops. In particular we analyzed the first
Feynman diagram in \eqref{emptyForFig}. 
\begin{enumerate}[(i)]
\item Analyze the other two diagrams in \eqref{emptyForFig} and
  explain how to account for them in the EFT.
\item The alert student may realize there are other light degrees of
  freedom in the loop: the light quarks! Consider for simplicity  the
  box diagram \eqref{eq:fig:KKbarmix} for the case that one internal
  quark is light and the other is the top quark. It makes sense to
  integrate out the top quark simultaneously with the $W$ (we incur in
  errors from failing to distinguish $\widebar\alpha(M_W)$ from
  $\widebar\alpha(m_t)$; we can live with that for now!). Analyze this
  case.  What is the dimension of the operator that needs to be
  included to account for low momentum in the light quark in the loop? 
\end{enumerate}
\end{exercise}

\end{exercises}


\end{document}